\documentclass[twocolumn,amsmath,amssymb,eqsecnum,nofootinbib,aps,prd,10pt]{revtex4-1}
\usepackage{graphicx}
\graphicspath{{./images/}}
\usepackage{hyperref}
\usepackage[noabbrev]{cleveref}            
\usepackage{mathtools}
\usepackage[scr=esstix]{mathalfa}          

\crefname{figure}{Fig.}{Figs.}
\crefname{section}{Sec.}{Secs.}

\makeatletter
\newcommand{\superimpose}[3][\mathord]{#1{\mathpalette\superimpose@{{#2}{#3}}}}
\newcommand{\superimpose@}[2]{\superimpose@@{#1}#2}
\newcommand{\superimpose@@}[3]{%
  \ooalign{%
    \hfil$\m@th#1#2$\hfil\cr
    \hfil$\m@th#1#3$\hfil\cr
  }%
}

\makeatother
\newcommand{\doublewedge}{\superimpose[\mathbin]{\wedge}{\scriptscriptstyle\wedge}}

\DeclareMathOperator{\Tr}{Tr}
\DeclareMathOperator{\arctanh}{arctanh}

\DeclareMathOperator{\arcsinh}{arcsinh}

\DeclareMathOperator{\arcsech}{arcsech}

\newcommand{\dom}{\mathcal}

\newcommand{\ts}[1]{{\boldsymbol{#1}}}

\newcommand{\dif}{\ts{d}}

\newcommand{\Riem}{\mathbf{Riem}}
\newcommand{\Ric}{\mathbf{Ric}}
\newcommand{\Ein}{\mathbf{Ein}}

\newcommand{\scR}{\mathcal{R}}

\newcommand{\Terg}{\ts{T}}

\newcommand{\spcd}{{\pmb{\triangledown}}}

\newcommand{\spsc}{{\mathscr{r}}}
\newcommand{\spgc}{{\mathscr{k}}}
\newcommand{\tnrm}{{\ts{u}}}

\newcommand{\scur}{\alpha}

\newcommand{\ph}{\varphi}

\newcommand{\eps}{\varepsilon}
\newcommand{\kap}{\varkappa}

\newcommand{\bph}{{\overline\ph}}
\newcommand{\bchi}{{\overline\chi}}
\newcommand{\btau}{{\overline\tau}}
\newcommand{\bphi}{{\bar\phi}}
\newcommand{\bR}{{\bar{R}}}
\newcommand{\bT}{{\bar{T}}}

\newcommand{\const}{\text{const}}

\newcommand{\bKV}{\ts{\xi}}
\newcommand{\Soax}{\Sigma_\oi}
\newcommand{\Sbax}{\Sigma_\bst}
\newcommand{\bst}{*}

\newcommand{\AdS}{{\mathrm{AdS}}}
\newcommand{\BTZ}{{\mathrm{BTZ}}}
\newcommand{\Cmt}{{\mathrm{C}}}

\newcommand{\oi}{{\mathrm{o}}}
\newcommand{\iix}{{\mathrm{in}}}
\newcommand{\oix}{{\mathrm{out}}}
\newcommand{\ai}{{\mathrm{a}}}
\newcommand{\asA}{{\mathrm{I}}}
\newcommand{\asB}{{\mathrm{II}}}
\newcommand{\hor}{{\mathrm{hor}}}
\newcommand{\tot}{{\mathrm{tot}}}
\newcommand{\itr}{{\mathrm{int}}}
\newcommand{\con}{{\mathrm{con}}}
\newcommand{\ren}{{\mathrm{ren}}}
\newcommand{\can}{{\mathrm{can}}}
\newcommand{\nor}{{\mathrm{nor}}}
\newcommand{\wh}{{\mathrm{wh}}}
\newcommand{\tst}{{\mathrm{test}}}

\newcommand{\mBTZ}{M}
\newcommand{\MBTZ}{M_\BTZ}
\newcommand{\MKV}{\mathscr{M}}
\newcommand{\locmass}{\mathscr{m}}

\newcommand{\intn}{\mathbb{Z}}
\newcommand{\realn}{\mathbb{R}}

\begin{document}


    \title{Point particles on the string in a three-dimensional AdS universe}

    \author{Petr Luke\v{s}}
    \email{petr.lukes934@student.cuni.cz}
    \author{Pavel Krtou\v{s}}%
    \email{pavel.krtous@utf.mff.cuni.cz}
    \affiliation{%
        Institute of Theoretical Physics,\\
        Faculty of Mathematics and Physics, Charles University\\
        V Hole\v{s}ovi\v{c}k\'{a}ch 2, 180 00 Prague 8, Czech Republic
    }%

    \date{\today}

    \begin{abstract}
        BTZ spacetime is a long-known locally AdS solution to the Einstein equations in 1 timelike and 2 spacelike dimensions. Its static variant is interpreted as a black hole whose mass is related to the period of the angular coordinate. This solution can be parametrically continued into one without horizons but with a conical deficit in the center. Such a solution is interpreted as a spacetime with a massive particle. It has been shown that this particle can be in static equilibrium with a cosmic string passing through the spacetime to infinity. In this work, we explore the interaction of point particles with strings, such as a bound system of two particles connected by a string of finite length. We identify additive local mass in the static spacetimes and apply it to the case of particles and strings. Finally, using the cut and glue method, we construct the system of two particles oscillating on the string, which goes out of the scope of static systems.
    \end{abstract}

    \maketitle

\section{Introduction}

Gravitational interaction between massive bodies in 3+1 spacetime dimensions has been successfully described by the Einstein equations (among a myriad of articles, let us name at least \cite{Swarzschild_orig,Einstein_GW,Kerr_BH}). However, for long, the theory has attracted only lesser interest in the 2+1 setting, where the Einstein equations determine the curvature fully in terms of the matter \cite{Fock_1962} and there are no additional degrees of freedom of the gravitational field, i.e., no gravitational waves and radiation.

The 2+1-dimensional gravity gained some attraction with the work of Staruszkiewicz \cite{Staruszkiewicz:1963} and subsequently of Clement, Deser, Jackiw, Witten, and t'Hooft \cite{Clement:1983nk,DESER1984220,DESER1984405,Deser:1985pk,DESER1989352,Witten:1988hc} with the discovery of distributional solutions representing massive particles. These particles alter the global properties of the spacetime. Spatially, their influence can be compared to cutting out an angular wedge from the spacetime and forming a conical defect by gluing the rest back together. This idea of spacetime with a conical defect was already investigated in \cite{Levi-Civita2011}  (in 3+1 dimensions), later in more modern formalism (but not yet identified in detail as a matter object) in \cite{MarderL.1959Fswg}.

Another important discovery in the field of 2+1 gravity was that of the black hole solution on AdS background by Ba\~{n}ados, Teitelboim, and Zanelli (BTZ) \cite{BTZ_original,Banados:1992gq}. This black hole exhibited mass, rotation, and charge, and it was widely discussed in the literature, e.g., \cite{Carlip:1995qv,Lemos:1995cm,Clement:1995zt,Martinez:1999qi,Cataldo:2000we,Kunz:2006eh,Akbar:2007zz,Gurtug:2010dr}, see also recent \cite{Mir:2016dio,Blazquez-Salcedo:2016rkj,Hennigar:2020drx,Maeda:2023oei,Hale:2024zvu} for a discussion of the charged case. The BTZ solution served as an interesting toy model for many novel theories of the time and is used, for example, in AdS/CFT discussion \cite{BTZ_CFT}. Also, recently, a new classification of 2+1 spacetimes based on the Cotton tensor was provided in \cite{Papajcik:2023zen}. The uncharged version of the BTZ solution is locally AdS everywhere, which means that the BTZ geometry could be obtained by a suitable regluing of the cut empty AdS spacetime.

In this work, we study spacetimes containing point particles interacting with cosmic strings and struts. In four dimensions, the cosmic strings have been discussed a long time ago \cite{Vilenkin1981,Zeldovich1980,Gott:1984ef}, with an overall overview of the topic in \cite{Vilenkin2000}. However, the situation relevant for this work, discussing accelerated particles potentially attached to the strings or struts in 2+1 dimensions, was only introduced in \cite{Astorino:2011} by studying the three-dimensional C-metric. Since then, the topic has attracted renewed interest, mainly its AdS flavor within the context of holography \cite{cmetric_ruth,Bunney:2025,BH_2+1_holography,Cong:2024pvs}, black hole thermodynamics \cite{Tian2024,GregoryKubiznak2016,Gregory2021,Hale2025,Appels2017,Gregory:2017ogk,Gregory:2019dtq,Anabalon:2018qfv}, and others \cite{Kubiznak:2024ijq,Cong:2021tnk,Chrusciel:2024vle}. In our paper, we want to emphasize that the three-dimensional C-metric does not describe only an accelerated black hole on the string but also the point-like source.

In three spacetime dimensions, both point particles and strings can be understood as lower-dimensional geometrical defects obtained by a suitable cut and glue method from the empty AdS spacetime. This method allows us to go beyond the situation of a single particle attached to the string described by the C-metric. Even when restricted to the static case, one can construct more complicated systems of particles interacting with each other through the strings, and one can also obtain spacetimes with nontrivial asymptotics, such as describing a wormhole connecting two AdS infinities. Using the cut and glue method, we also construct a dynamical situation describing oscillating particles connected by a string.

One of the purposes of this article is to inspect physically simple static matter distributions and focus the analysis on a deeper understanding of mass. It includes a careful discussion of the asymptotic regions containing an infinite amount of cosmological dark energy. It is achieved by choosing appropriate simple canonical spacetimes. Comparing with them allows us to understand more complicated spacetimes, including those with a nontrivial Euler characteristic.

The content of the paper is as follows: In \Cref{sec:ads_intro}, basic notions are reviewed and introduced. First, we discuss a static massive particle as an extension of the BTZ metric. Then, the system of an accelerated (but still static) particle on a static semi-infinite string or strut is introduced. The acceleration of the particle is caused by the string or strut, and together with the cosmological attraction, it keeps the particle static. We then introduce a novel static system of two particles at the ends of a finite string or strut.\footnote{See also a recent work \cite{Bunney:2025} in a slightly different context.}

In \Cref{sec:mass}, we investigate the notion of locally additive mass for static spacetimes. The important feature of this notion is that, although the mass in a domain is additive over subdomains, it can be evaluated solely on the boundary of the domain. It allows calculating the mass asymptotically. To handle the infinite contribution of dark energy in the asymptotic region, we introduce the notion of canonical spacetimes, and by comparing with them, we can focus on the nontrivial matter content. We find that the local mass characteristic of particles is not simply related to the global mass parameter of the BTZ metric.

In \Cref{sec:oscillator}, we introduce a new dynamical spacetime representing two particles oscillating on a string. This system does not belong to the class of discussed static spacetimes. It shows the necessity of an extension of the class of canonical spacetimes. This is somewhat surprising as the spacetime can be obtained by gluing together just two subregions of empty AdS in a symmetric way,\footnote{One can easily step out of the static context by cutting and regluing the empty AdS spacetime in a non-symmetric way, introducing a rotation. It is not the case of the system discussed here.} and one could expect a simple static asymptotic.

Finally, in \Cref{sec:spacetimes}, we give a glimpse of the possibilities of the cut and glue method, allowing us to construct various novel spacetimes combining features of the previously discussed simple cases. It is revealed that we can introduce geometries with different spatial topology (e.g., a traversable eternal wormhole or the M\"{o}bius strip) as well as spacetimes with multiple centers of gravity, strings, and particles.

\section{Particles and strings in AdS spacetime in 3~dimensions}
\label{sec:ads_intro}

The gravity in 3 spacetime dimensions is simple because the Riemann and Ricci tensors have the same number of independent components. Thus, all curvature is sourced by matter through Einstein's equations, and there are no propagating degrees of freedom (gravitational waves) as the Weyl tensor vanishes identically \cite{Staruszkiewicz:1963,Carlip:1998}, comprehensive review in \cite{Garcia-Diaz_2017}. 

A spacetime is said to be locally AdS if it is everywhere a solution of the $\Lambda$-vacuum Einstein equations with a negative cosmological constant. As a result, it is of constant negative curvature.

The simplest example of a locally AdS spacetime is the \emph{global AdS universe}. Other locally AdS spacetimes can be understood as various factorizations of the global AdS spacetime. An example of such a spacetime is the well-known BTZ black hole \cite{BTZ_original}. We will study a broader class of spacetimes that are locally AdS except for some lower-dimensional defects. Some of them can be interpreted as vacuum spacetimes with point-particle or string sources.


\subsection{Coordinate description and symmetries of AdS spacetime} \label{ssc:AdS_basics}

The AdS spacetime in 3 dimensions can be acquired by the pullback of a 4-dimensional flat metric with signature $(--++)$ onto a pseudo-sphere of negative square radius:
\begin{equation}
    \label{eq:4D}
    \begin{aligned}
        \ts{g}_{\mathsf{E}^{2,2}}   &= -\dif u^2 -\dif v^2 +\dif x^2 + \dif y^2\,, \\
        -\ell^2     &= -u^2 -v^2 +x^2 +y^2\,.
    \end{aligned}
\end{equation}
The global AdS is a cover of this pseudo-sphere with an unfolded time direction. The common coordinate description is
\begin{equation}
	\label{eq:AdS_R}
	\ts{g}_{\AdS} = -\Bigl(1+\tfrac{R^2}{\ell^2}\Bigr)\,\dif T^2 + \frac{1}{1+\frac{R^2}{\ell^2}}\, \dif R^2 + R^2\,\dif \ph^2\,,
\end{equation}
with $T\in(-\infty,\infty)$, $R\in(0,\infty)$ and $\ph\in(-\pi,\pi)$. Relations to coordinates of \eqref{eq:4D} are 
\begin{equation}
    \begin{aligned}
            u&=\sqrt{\ell^2{+}R^2}\, \cos\frac T\ell\,,& v&=\sqrt{\ell^2{+}R^2}\, \sin\frac T\ell\,,\quad\\
    x&=R \cos\ph\,,& y&=R \sin\ph\,.
    \end{aligned}
\end{equation}
These are called the global static coordinates. We call any static coordinate system with the same time foliation \emph{global static frame}.

Other coordinates suitable for the global static frame can be obtained by conveniently rescaling $R=\ell\tan\chi$ and $T=\ell\tau$. The metric takes the form
\begin{equation}
	\label{eq:AdStauchiph}
	\ts{g}_{\AdS} = \frac{\ell^2}{\cos^2\chi}\left(-\dif \tau^2 + \dif\chi^2 + \sin^2\chi\,\dif\ph^2\right)\,.
\end{equation}
The radial distance from the origin is $\ell\arcsech\cos\chi$, the length of the arc $\Delta\ph$ on the circle $\chi=\const$ is $\ell\Delta\ph\tan\chi$.

Metrics \eqref{eq:AdS_R} and \eqref{eq:AdStauchiph}, respectively, satisfy the vacuum Einstein equations with cosmological constant ${\Lambda<0}$, where the length scale $\ell$ is related to the cosmological constant as
\begin{equation}
    \Lambda=-\frac{1}{\ell^2}\,.
\end{equation}

The spacetime possesses a six-dimensional isometry group $SO(2,2)$ of continuous symmetries.
These are generated by Killing vector fields. Obviously, the coordinates of \eqref{eq:AdStauchiph} are adapted to two  Killing vectors ${\ts\partial}_\tau$ and ${\ts\partial}_\ph$. Four other independent  ``boost-like'' Killing vectors can be chosen to form a basis in the Lie algebra of isometries, cf.~\Cref{ssc:boostKV}. The Killing metric in the Lie algebra of isometries has signature ${(--++++)}$, with ${\ts{\partial}}_\tau$ and ${\ts\partial}_\ph$ having a negative square of the Killing norm and the boost-like Killing vector having a positive square of the Killing norm.

It can be shown that the static observers $R,\,\ph=\const$, with the worldline given by the orbit of the Killing vector ${\ts\partial}_\tau$, have an acceleration of a constant magnitude $a$,
\begin{equation}
    \label{eq:acceleration}
    a = \frac{R}{\ell\sqrt{\ell^2+R^2}} = \frac1\ell\sin\chi\,,
\end{equation}
directed outwards from the center $R=0$. Thus, these observers are not free; they need an external agent that would keep them on their worldlines. 

Notice that the central static observer at $R=0$ is geodesic, and all static observers in the global static frame have an acceleration of magnitude bounded by the cosmological scale, $a < \frac1\ell$.
We call them subcritical uniformly accelerated observers. In contrast, the observers adapted to the boost-like Killing vectors have supercritical acceleration $a>\frac1\ell$, cf.~\Cref{ssc:boostKV}.


\subsection{Coordinates centered on an observer with sub-critical acceleration} \label{ssc:AdSAccSmallAcc}

The metric \eqref{eq:AdStauchiph} is conformally related to half of the Einstein static universe. The time section $\tau=\const$ of the Einstein static universe is a full homogeneous sphere $S^2$ described by extended coordinate $\chi\in(0,\pi)$. The conformally related AdS spacetime corresponds to a hemisphere $\chi<\frac\pi2$, with $\chi=\frac{\pi}{2}$ being the AdS spatial infinity. 

\begin{figure}
    \centering
        \includegraphics[]{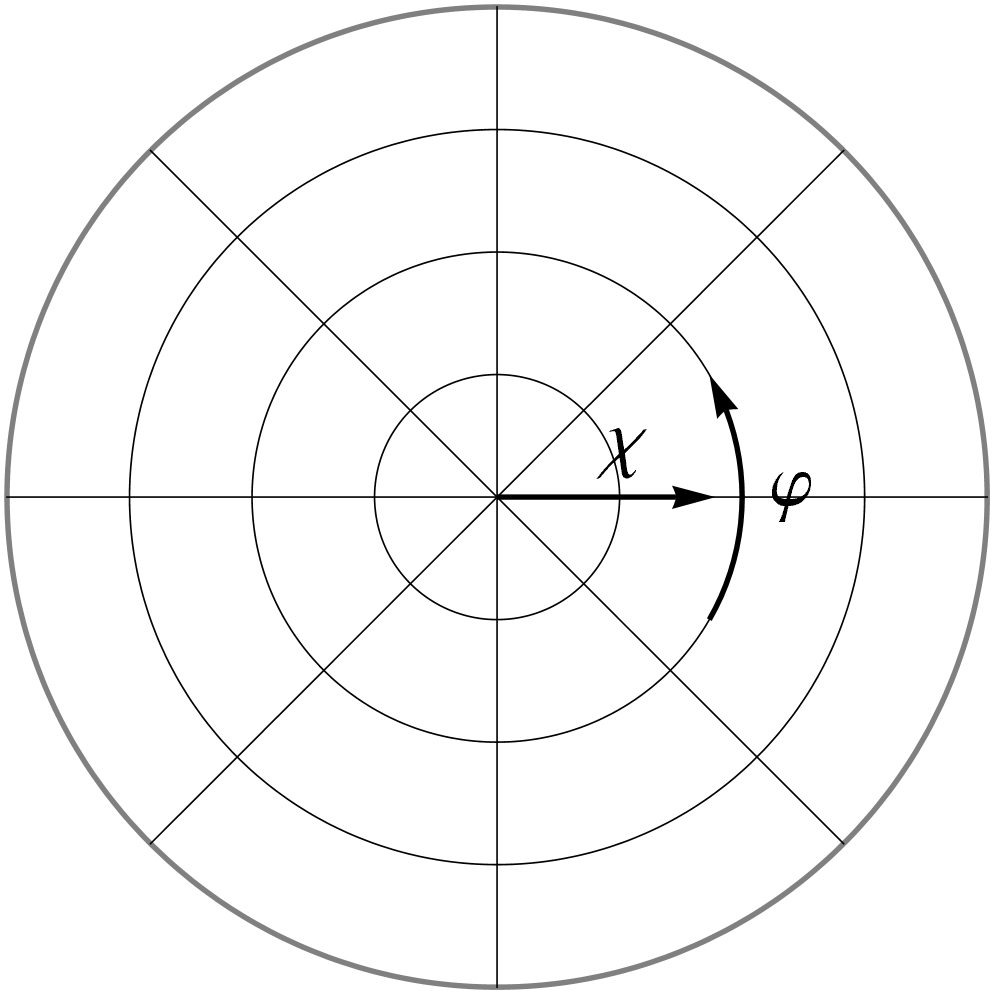}\hfill
        \includegraphics[]{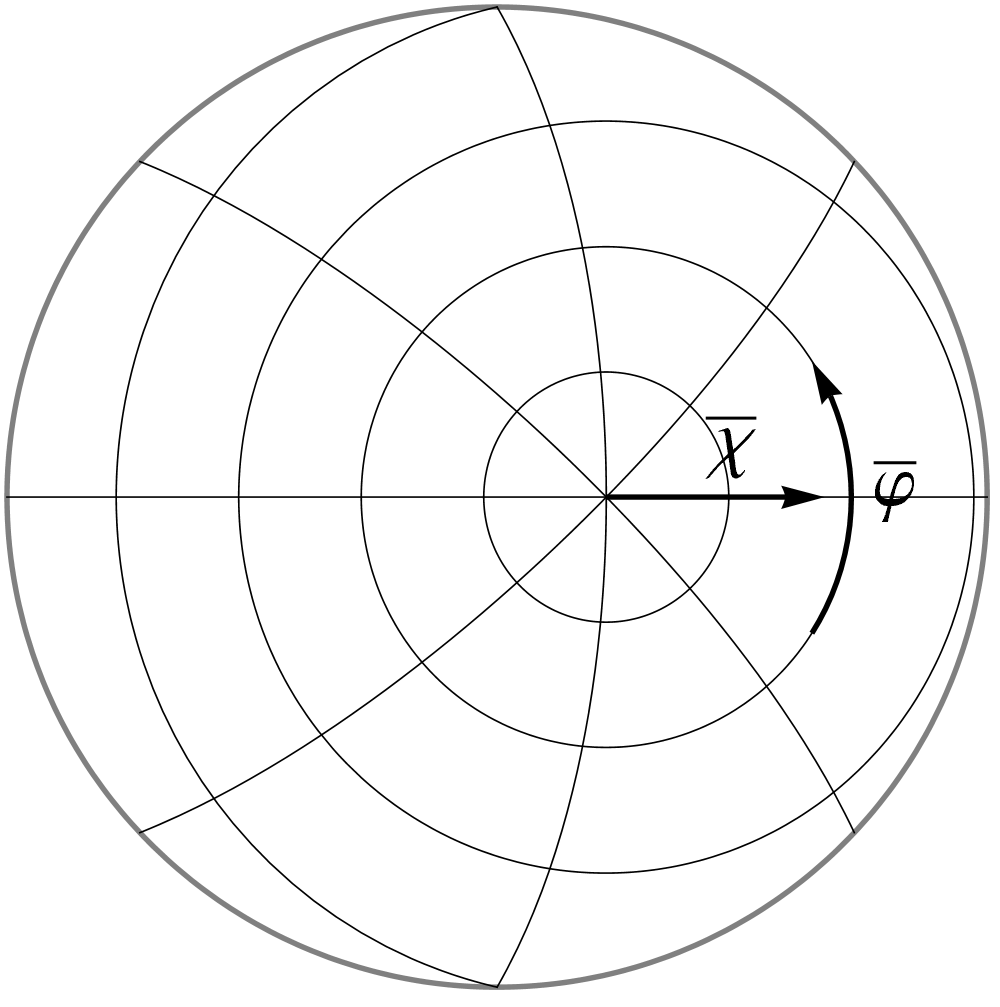}
    \caption{\label{fig:coords}%
    Left: coordinate lines of $\{\chi,\ph\}$ on a global time-slice with geometry of a hyperbolic plane. The time-slice is conformally related to a hemisphere with $\{\chi,\ph\}$ being the standard spherical coordinates. Right: coordinate lines of $\{\bchi,\bph\}$. These are spherical coordinates on a conformally related sphere centered on a pole rotated by $\chi_\oi$ with respect to the left diagram. The curves $\bph=\const$ are exocycles of the hyperbolic plane.
    }    
\end{figure}

The sphere can be rotated by an angle $\chi_\oi\in(-\frac{\pi}{2},\frac{\pi}{2})$ in the direction of axis $\ph=0$. Introducing rotated spherical coordinates $\bchi$, $\bph$ as
\begin{equation} \label{eq:S2rotation}
	\begin{aligned}
		\cos\chi         &= \cos\bchi\cos\chi_\oi - \sin\bchi\cos\bph\sin\chi_\oi\\
		\sin\chi\cos\ph  &= \cos\bchi\sin\chi_\oi + \sin\bchi\cos\bph\cos\chi_\oi\\
		\sin\chi\sin\ph  &= \sin\bchi\sin\bph\\
            \tau             &= \btau,
	\end{aligned}
\end{equation}
we have 
$\dif\chi^2 + \sin^2\!\chi\,\dif\ph^2 = \dif \bchi^2 + \sin^2\!\bchi\,\dif\bph^2$.
The metric \eqref{eq:AdStauchiph} thus transforms to 
\begin{equation}
    \ts{g}_{\AdS} = \frac{\ell^2 \left(-\dif \btau^2 + \dif \bchi^2 + \sin^2\bchi\,\dif\bph^2\right)}{(\cos\bchi\cos\chi_\oi{-}\sin\bchi\cos\bph\sin\chi_\oi)^2}
    \label{eq:AdSAcctauchiph}
\end{equation}
The infinity of AdS spacetime is determined by the zero of the conformal factor, $\cos\chi=0$. In the rotated coordinates, it reads
\begin{equation}
\label{eq:confinfAccchiph}
  \cot\bchi\cot\chi_\oi = \cos\bph\;.
\end{equation}
It determines the range of coordinates $\bchi$ and $\bph$. See \Cref{fig:coords} for diagrams of the coordinate systems $\{\chi,\ph\}$ and $\{\bchi,\bph\}$.

Next, let us introduce coordinates centered on an accelerated observer (shortly, accelerated coordinates)\footnote{The relation between $\bT$, $\bR$ and $\btau$, $\bchi$ is the same as between $T$, $R$ and $\tau$, $\chi$ with `modified' scale $\overline\ell=\frac\ell{\cos\chi_\oi}$.} 
$\bT,\ \bR,\ \bph$,
\begin{equation}
    \bT = \frac{\ell}{\cos\chi_\oi}\,\btau \;,\quad
	\bR = \frac{\ell}{\cos\chi_\oi}\,\tan\bchi\,,
\end{equation}
and parameter $a_\oi$ alternative to $\chi_\oi$,
\begin{equation}
	\label{eq:aodef}
    a_\oi = \frac{1}{\ell}\sin\chi_\oi\,.
\end{equation}
It transforms the AdS metric into
\begin{equation}
    \label{eq:AdSAccCoor}
    \begin{aligned}
        \ts{g}_{\AdS} &= \frac{1}{(1-a_\oi \bR \cos\bph)^2}
        \Bigl(-\Bigl(1+\tfrac{\bR^2}{\ell^2} \bigl(1{-}a_\oi^2\ell^{2}\bigr)\Bigr)\dif \bT^2\\
        &+ \frac{1}{1+\frac{\bR^2}{\ell^2} \bigl(1{-}a_\oi^2\ell^{2}\bigr)}\, \dif \bR^2 
        + \bR^2\dif \bph^2\Bigr)\;.
    \end{aligned}
\end{equation}
These coordinates are called accelerated because they play the role of spherical-like coordinates around the accelerated observer with worldline $\bR=0$. In coordinates of \eqref{eq:AdStauchiph}, $\bR=0$ corresponds to $\chi=\chi_\oi$, $\ph=0$, respectively, $R=R_\oi\equiv\ell\tan\chi_\oi$. This central observer has acceleration $a_\oi$ in the direction of the axis $\ph=0$.
\footnote{Formulae can also be used for negative values $\chi_\oi<0$ and $a_\oi<0$. Indeed, the change of sign of $\chi_\oi$ in \cref{eq:S2rotation} is equivalent to the change of angles $\ph$, $\bph$ by $\pi$. Rotation \cref{eq:S2rotation} represents the rotation of the conformal sphere such that the point $\bchi = 0$ corresponds to $\chi = \chi_\oi,\ \ph = 0$. If $\chi_\oi<0$, the origin $\bchi=0$ is thus equivalent to $\chi=\lvert{\chi_\oi}\rvert,\ \ph = \pm\pi$, without a need to expand the ranges of $\chi$ or $\bchi$ into negative values. Employing negative values of $\chi$ or $\bchi$ can be safely done only on the axis $\ph=0,\pm\pi$. In such a case, the positive direction on this axis is understood as the direction of growing $\chi\in\realn$, and $a_\oi$ is understood as a component of the acceleration in this direction. Negative $a_\oi$ at the position of negative $\chi_\oi$ thus means the acceleration pointing away from the center $\chi=0$.}

The ranges of the are $\bT\in\realn$, $\bph\in(-\pi,\pi)$, and the range of the radial coordinate $\bR$ is restricted by the condition \eqref{eq:confinfAccchiph} for the conformal infinity,
\begin{equation}\label{eq:Rbarrange}
    a_\oi\bR\cos\bph < 1\;.
\end{equation}
For $a_\oi\cos\bph<0$, one has to allow also $\bR<0$ satisfying the inequality \eqref{eq:Rbarrange} in order to cover the whole spacetime. Regularity across $\bR\to\pm\infty$ can be achieved using, e.g., coordinate $\bchi$. $\bR<0$ corresponds to $\bchi\in(\frac\pi2,\pi)$.

Both in four and three spacetime dimensions, the solution called the C-metric describes uniformly accelerated sources (typically black holes) \cite{KinnersleyWalker:1970,Bonnor:1982,AshtekarDray:1981,GriffithsPodolsky:2005,GriffithsKrtousPodolsky:2006,DiasLemos:2003a,PodolskyGriffiths:2006,Krtous:2005}. In three dimensions, metric \eqref{eq:AdSAccCoor} is one of the possible forms of the C-metric. However, to obtain a nontrivial spacetime describing accelerated sources, one has to choose a different range of coordinate $\bph$. With $\bph\in(-\pi,\pi)$, the metric \eqref{eq:AdSAccCoor} describes just the empty global AdS spacetime written in the accelerated coordinates.\footnote{In four dimensions, the C-metric is not locally isomorphic to empty AdS spacetime. It differs not only by a choice of coordinate ranges but also by a suitable modification of metric functions.}  We will discuss the case of a nontrivial C-metric in \Cref{ssc:Cmetric}.


\subsection{Spatial description in the global static frame}
\label{ssc:spcdescr}

We have been describing AdS spacetime in full three-dimensional formalism until now. It is often useful to employ a description of splitting space and time, the so-called 2+1 formalism. The global static frame gives us a natural choice of time splitting. The time-flow vector is given by the Killing vector ${\ts\partial}_T$, and the time foliation is given by the time coordinate $T$. Inspecting the metric \eqref{eq:AdS_R}, hypersurfaces $T=\const$ have a geometry of the hyperbolic plane,
\begin{equation}
  \label{eq:mtrTconst}
	\ts{q}_{\mathsf{H}} = \frac{1}{1+\frac{R^2}{\ell^2}}\, \dif R^2 + R^2\,\dif \ph^2\;,
\end{equation}
and the static time function $T$ is related to the proper time of static observers by a lapse function
\begin{equation}
  \label{eq:lapse}
	N \equiv \exp\Phi = \sqrt{1+\tfrac{R^2}{\ell^2}} = \frac{1}{\cos\chi}\;.
\end{equation}
Here, $\Phi$ is usually understood as a gravitational potential in the static frame. The acceleration of static observers is given by 
\begin{equation}
  \label{eq:obsacc}
	\frac1N\dif N = \dif \Phi = \frac1\ell\frac{R}{\sqrt{\ell^2+R^2}}\,\frac1N\dif R\;,
\end{equation}
with magnitude given by \eqref{eq:acceleration}.

When working in a frame associated with a family of preferred observers (for example, observers following orbits of a Killing vector), one wishes to interpret them as the observers at rest. If the observers are not free, not moving along geodesics in the spacetime sense, they must be held at their position by an external agent. Then, the geodesic observers move with respect to the `rest' observers of the frame. Therefore, in spatial description, one has to introduce a ``fictitious'' gravitational force that is responsible for the spatial motion of free observers and which is in equilibrium with the external agent, keeping the frame observers at their position. For the static frame with preferred static observers, the gravitational force is thus exactly opposite to the spacetime acceleration of the observer. The gravitational force per unit mass is hence given by $-\dif\Phi$.

In the static splitting of the AdS universe, the gravitational force is pointing toward the origin of the static frame, vanishing at the origin, growing in the radial direction, with magnitude approaching $\frac1\ell$ at infinity. This ``apparent'' force is also called the cosmological attraction of the AdS universe. Exploring the AdS universe is thus doing physics in the hyperbolic space with the central gravitational force given by the potential $\Phi$ \eqref{eq:lapse}.

Let us remark that the global static frame is not unique. Such a frame can be built around any spacetime geodesic, with this geodesic in the center. Even with a fixed central geodesic, one still has rotational symmetry around the origin. In general, any two globally static frames are related by an isometry. It is thus usually sufficient to discuss systems from the point of view of one such frame. However, in \Cref{sec:oscillator}, we will encounter a case in which two global-like frames will be relevant.


\subsection{Point particle}
\label{ssc:pointparticle}

As mentioned above, any solution of the $\Lambda$-vacuum Einstein equations with $\Lambda<0$ is locally isomorphic to a region in global AdS spacetime. However, it can differ globally and can be obtained from the global AdS universe by appropriately identifying its subregions (idea first conceived in \cite{DESER1984220,DESER1984405}). These identifications can lead to the formation of lower-dimensional defects interpreted as massive sources.

The simplest identification of subregions can be achieved by a restriction to the subrange of a suitable (e.g., Killing) coordinate and making it periodic. As the archetypal example, consider a \mbox{2-dimensional} flat plane. Restricting its angular coordinate to $\ph\in(-\ph_\oi,\ph_\oi)$, with $\ph_\oi\in(0,\pi)$, can be imagined as cutting an angle from the plane, and the periodization of $\ph$ at $\ph=\pm\ph_\oi$ as gluing the cut edges, thus forming a cone. An analogous procedure will be the key in the following constructions. 

Let us repeat the described procedure in the metric \eqref{eq:AdS_R}. Identifying points ${\ph = -\ph_\oi}$ and ${\ph = \ph_\oi}$ with the same $R$ and $T$, we obtain the same form of the metric \eqref{eq:AdS_R}, only with a periodic coordinate ${\ph\in(-\ph_\oi,\ph_\oi)}$.

\begin{figure}\centering
    \includegraphics[]{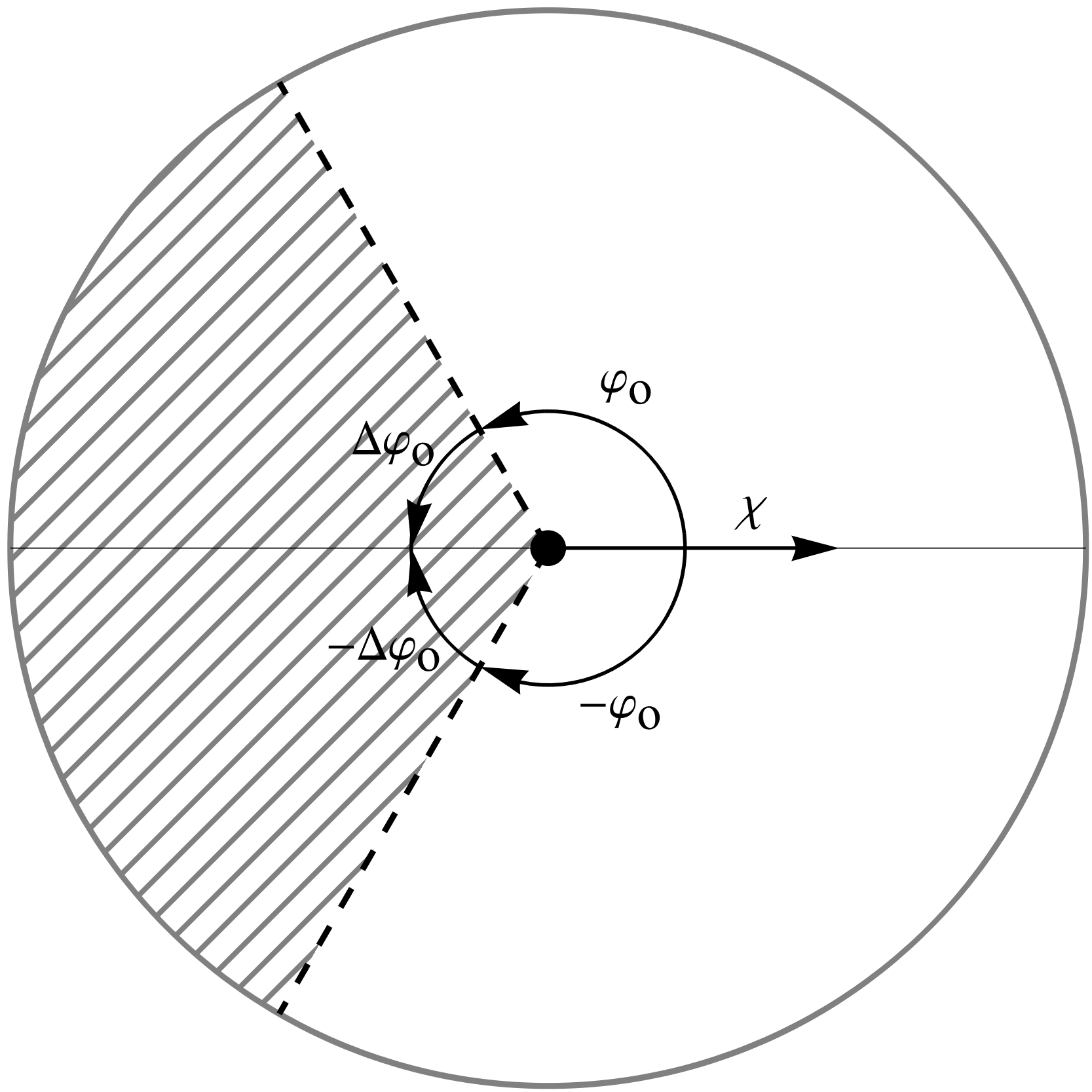}
    \caption{\label{fig:one_particle}%
    Diagram of a spatial section ${\tau=\const}$ of the spacetime with a conical deficit. It is drawn using the coordinates of the global AdS universe: radial coordinate ${\chi\in\left(0,\frac{\pi}{2}\right)}$ and angular coordinate ${\ph\in(-\pi,\pi)}$. The boundary circle ${\chi=\frac{\pi}{2}}$ represents the conformal infinity. The angular coordinate is restricted to the interval ${\ph\in(-\ph_\oi,\ph_\oi)}$ and radial edges ${\ph=\pm\ph_\oi}$ represent the identified hyperplanes. The hatched angle parameterized by $\Delta\ph$ is excluded. There is a conical deficit at the origin $\chi=0$ which represents a static point particle (represented by the black disk). Although the metric is not smooth at the conical singularity, one can estimate the acceleration of close-by worldlines and conclude that it vanishes. The point particle is thus in a free motion.}
\end{figure}

Since the hypersurfaces $\ph=\pm\ph_\oi$ are isomorphic, without extrinsic curvature, the result of the gluing procedure is transparent. The metric is smooth everywhere except the origin $R=0$. At the origin, the geometry is singular, with a conical singularity.\footnote{To demonstrate the smoothness of the gluing, one has to introduce another coordinate patch covering smoothly the axis $\ph=\pm\ph_\oi$, which is straightforward and common for polar-like coordinates. However, for $\ph_\oi\neq\pi$, one cannot find coordinates covering also the origin $R=0$ in such a way that the metric would be smooth there.} In three spacetime dimensions, it can be demonstrated that the singularity has a delta-like character in the curvature. Namely, the singular geometry can be obtained by a limit of regular geometries with a mass distribution localized around the origin, squeezing the matter to that point. The deficit angle gives the local mass $m_\oi$ corresponding to the conical singularity,
\begin{equation}
    \label{eq:mascondef}
	\frac{\kap m_\oi}{2\pi} = \frac{\Delta\ph_\oi}{\pi} = \Bigl(1-\frac{\ph_\oi}{\pi}\Bigr)\;,
\end{equation}
where ${\Delta\ph_\oi = \pi - \ph_\oi}$ is the angular deficit,\footnote{We define $\Delta\ph_\oi$ as half of the total deficit angle to simplify notation in our figures. It thus differs by a factor of 2 from the common usage.} see \Cref{ssc:PointParticleAdS}. We can observe that for $\ph_\oi>\pi$ (excess angle, ${\Delta\ph_\oi<0}$) the mass $m_\oi$ is negative; for ${\ph_\oi=\pi}$ (${\Delta\ph_\oi=0}$) mass is vanishing and there is no conical singularity; and for ${\ph_\oi\in(0,\pi)}$ (deficit angle, ${\Delta\ph_\oi>0}$) the conical singularity describes a point-like particle of a positive mass, such that ${\frac{\kap m_\oi}{2\pi}\in(0,1)}$. For ${\ph_\oi\le 0}$, the described construction does not make sense, but one can find its generalization leading to the spacetime containing a BTZ black hole.

We will use the deficit-angle parameter $\Delta\ph_\oi$ as a geometric characterization of the conical singularity and the mass $m_\oi$, related by \eqref{eq:mascondef}, as a physical characterization of the corresponding point particle.

\subsection{Subcritical BTZ metric}
\label{ssc:subBTZ}

We can rescale the coordinates of \eqref{eq:AdS_R} as
\begin{equation}
    \label{eq:rescaling}
	T = \frac{\ph_\oi}{\pi}\,T_\BTZ\;,\quad R = \frac{\pi}{\ph_\oi}\,R_\BTZ\;,\quad \ph = \frac{\ph_\oi}{\pi}\,\phi_\BTZ\;,
\end{equation}
with $T_\BTZ\in\realn$, $R_\BTZ\in(0,\infty)$, and $\phi_\BTZ\in(-\pi,\pi)$. Skipping, for simplicity, the subscript $\BTZ$ of the coordinates, the metric reads
\begin{equation}
	\label{eq:BTZ_metric}
	\ts{g}_\BTZ = -\left(1{-}\mBTZ{+}\tfrac{R^2}{\ell^2}\right)\dif T^2 
        + \frac{1}{1{-}\mBTZ{+}\tfrac{R^2}{\ell^2}}\, \dif R^2 
        + R^2\,\dif \phi^2 \;,
\end{equation}
with 
\begin{equation}
	\label{eq:ourBTZ_mass}
	\mBTZ = 1-\left(\frac{\ph_\oi}{\pi}\right)^2 
     = \frac{\Delta\ph_\oi}{\pi}\Bigl(2-\frac{\Delta\ph_\oi}{\pi}\Bigr)\;.
\end{equation}
Metric \eqref{eq:BTZ_metric} is of the same form as the famous BTZ black hole metric \cite{BTZ_original}, only with a different mass parametrization. Our mass parameter $\mBTZ$ differs from the parameter $\MBTZ$ of \cite{BTZ_original} as ${\mBTZ=1+\MBTZ}$. The point-like particle described here thus corresponds to ${\mBTZ<1}$, i.e., ${\MBTZ<0}$, and we call it the subcritical BTZ case. Black hole spacetimes (the supercritical case) are parametrized by ${\mBTZ>1}$ (${\MBTZ>0}$), and we will discuss them elsewhere \cite{BHs_particles}. Parametrization using $\mBTZ$ seems more natural since the case $\mBTZ=0$ corresponds to the empty global AdS universe, which is not true for $\MBTZ=0$. However, it suggests the mass gap $\mBTZ=1$ for the black hole solutions. It is a delicate question related to the renormalization of the global mass of spacetime. Note that $\mBTZ$ is a dimensionless version of the full mass parameter $\frac{2\pi}{\kap}\mBTZ$.
To avoid cumbersome expressions, we use just geometric $\mBTZ$ to parametrize the BTZ solution.

We conclude by comparing two mass characteristics of a point particle: The local mass $m_\oi$ given by~\eqref{eq:mascondef} is related to $\mBTZ$ given by~\eqref{eq:ourBTZ_mass} as
\begin{equation}
	\label{eq:massrel}
    \mBTZ = \frac{\kap m_\oi}{2\pi} \Bigl(2-\frac{\kap m_\oi}{2\pi}\Bigr)\;,\quad
    \frac{\kap m_\oi}{2\pi} 
    = 1- \sqrt{1-\mBTZ}\;.
\end{equation}
For a deficit angle ${\ph_\oi\in(0,\pi)}$, both $\mBTZ$ and $\frac{\kap m_\oi}{2\pi}$ belong to the interval ${(0,1)}$. For an excess angle, ${\ph_\oi>\pi}$, both masses are negative.


\subsection{A cosmic string or strut}
\label{ssc:string}

Let us now introduce a conical deficit to the accelerated observer at the origin of the accelerated coordinates of the metric \eqref{eq:AdSAccCoor}. Similarly to the previous case, this is achieved by restricting ${\bph\in(-\ph_\oi,\ph_\oi)}$ and rescaling the coordinate to $2\pi$ periodicity. In other words, we cut the global AdS spacetime along hypersurfaces ${\bph=\pm\ph_\oi}$ and identify these hypersurfaces. 

Near the origin ${\bchi=0}$, the geometry has again a conical deficit, which can be interpreted again as a point particle of a local mass given by \eqref{eq:mascondef}. However, we now have to pay attention to the identified hypersurfaces.

Since the hypersurfaces ${\bph=\pm \ph_\oi}$ were chosen symmetric (a mirror symmetry $\bph\to-\bph$) they have the same intrinsic metric $\ts{h}$,
\begin{equation}
	\ts{h} = \frac{\ell^2}{(\cos\bchi\cos\chi_\oi{-}\sin\bchi\cos\ph_\oi\sin\chi_\oi)^2}\left(-\dif \btau^2 + \dif \bchi^2 \right)\;,
	\label{eq:AdSAcctauchiph_surface}
\end{equation}
and they can be glued together. However, they have a nontrivial extrinsic curvature. It is shown in \Cref{apx:string_curvature} that hypersurfaces $\bph=\const$ are umbilical, i.e., their extrinsic curvature $\ts{K}$ is proportional to the induced metric $\ts{h}$, 
\begin{equation}
    \ts{K} = \scur\,  \ts{h}\;.
    \label{eq:extrcurv}
\end{equation}
Here, $\scur$ is a mean curvature $\scur=\frac12 \Tr \ts{K}$ and it fully characterizes the curvatures of the umbilical surface.

Another interesting property of the surface $\bph=\const$ is that its spatial representation on the hypersurface $T=\const$ is an exocycle, i.e., a curve of constant curvature\footnote{Here, we choose the normal to the surface $\ts{n}$ oriented from the umbilical surface into the domain of the spacetime, which is not removed in the identification. The surface normal $\ts{n}$ is equal to the normal to the corresponding curve in the static time slice. For such a choice, ${{\ts K}\equiv {\ts h}\cdot\ts\nabla {\ts n} = \scur\, {\ts h}}$, and ${{\ts e}\cdot\ts\nabla {\ts e} = -\scur\, {\ts n}}$ with $\ts{e}$ being unit a tangent vector to the curve.} $\scur$. In particular, it is equidistant to a straight line through the gravity center, understood in the hyperbolic geometry \eqref{eq:mtrTconst}. 

\begin{figure}
    \centering
    \includegraphics[]{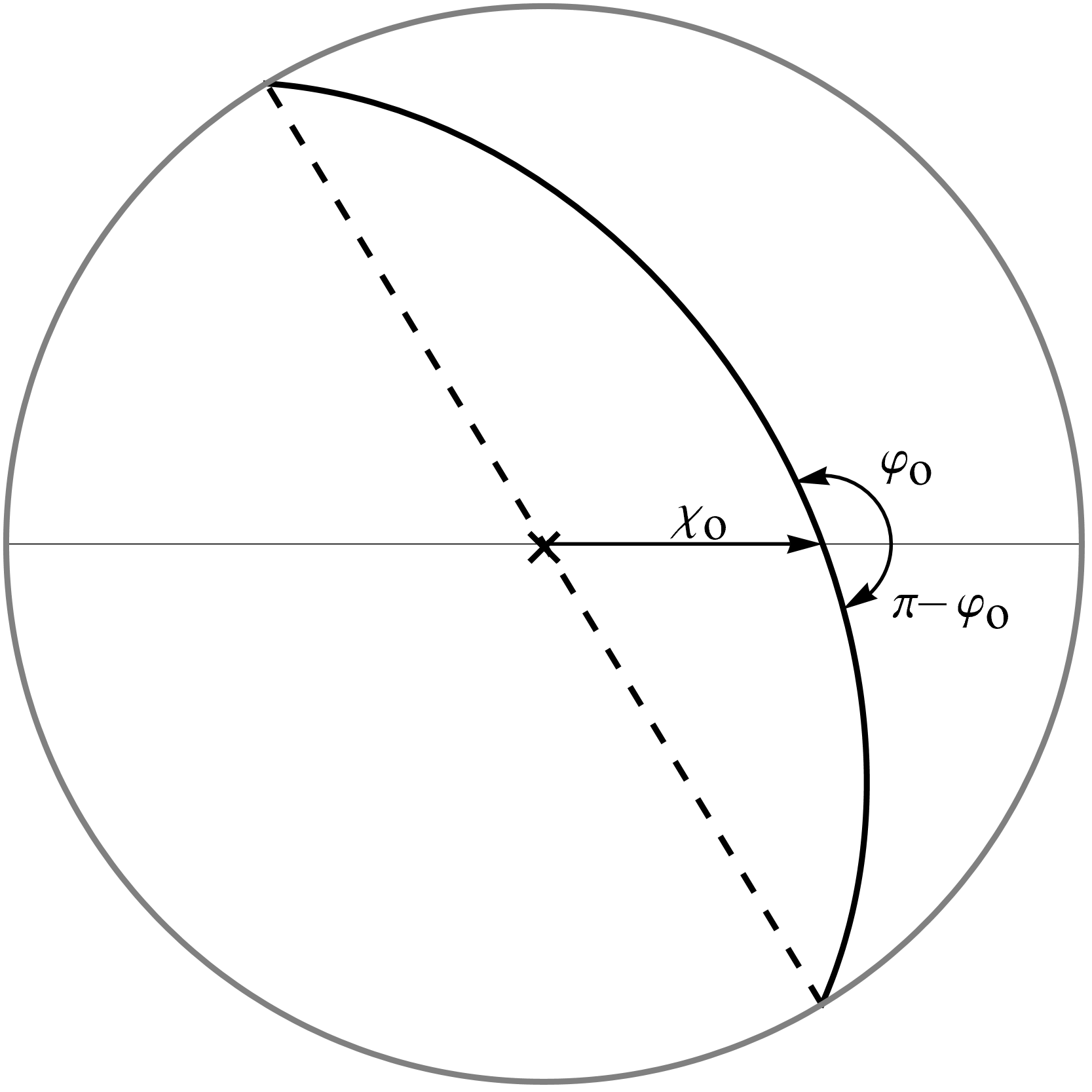}
    \caption{\label{fig:equidistant}%
    Diagram of a global time-slice $T=\text{const}$. The origin of the accelerated frame and its shift by $\chi_\oi$ from the origin of the unaccelerated frame is indicated. Two curves, $\bph=\ph_\oi$ and $\bph=\pi-\ph_\oi$, starting at this point are shown. These curves together form an exocycle, i.e., the equidistant line to the axis (dashed line) going through the center of gravity.}
\end{figure}

It is actually the characteristic property of the static umbilical surface: Any surface static with respect to Killing vector ${\ts\partial}_T$ is umbilical if, and only if, the corresponding curve at ${T=\text{const}}$ is equidistant to a straight line going through the gravity center $R=0$ of the slice, cf.~\Cref{fig:equidistant}. See \Cref{apx:string_curvature} for details.

Gluing two symmetric umbilical hypersurfaces, we have to employ the Israel junction conditions, cf.~\Cref{apx:IJC}. The identified hypersurface $\Sigma$ carries a matter source with a stress-energy tensor ${\Terg = \ts{S}\, \delta_\Sigma}$ related to a jump in the extrinsic curvature as ${-[\ts{K}]+\Tr [\ts{K}]\,\ts{h}=\kap \ts{S}}$, namely,
\begin{equation}
    \label{eq:surface_stress_energy}
    \ts{S} = \frac{2\scur}{\kap} \ts{h} = -\mu\,\ts{h}\;,\quad 
    \mu \equiv -\frac{2\,\scur}{\kap}\;.
\end{equation}
In three dimensions, the hypersurface $\Sigma$ can be understood as a two-dimensional worldsheet of a one-dimensional spatial object -- of a string or strut. The stress-energy tensor \eqref{eq:surface_stress_energy} proportional to the metric is usually interpreted as that of a cosmic string ($\mu>0$) or strut ($\mu<0$) (first conceptualized already in \cite{Bach1922} by Weyl, but only in \cite{DESER1989352} for the case of three dimensions). It has a linear energy density $\mu$ and a linear tension $-\mu$. For simplicity, in further text we will call any two-dimensional defect with $\ts{S}\propto\ts{h}$ a string, although for negative $\mu$, it should be understood automatically as the strut. Typically, the discussion of both strings and struts is similar; it differs just by the actual sign of $\mu$.

\begin{figure*}\centering         
    \vspace*{-6ex}
    \includegraphics[]{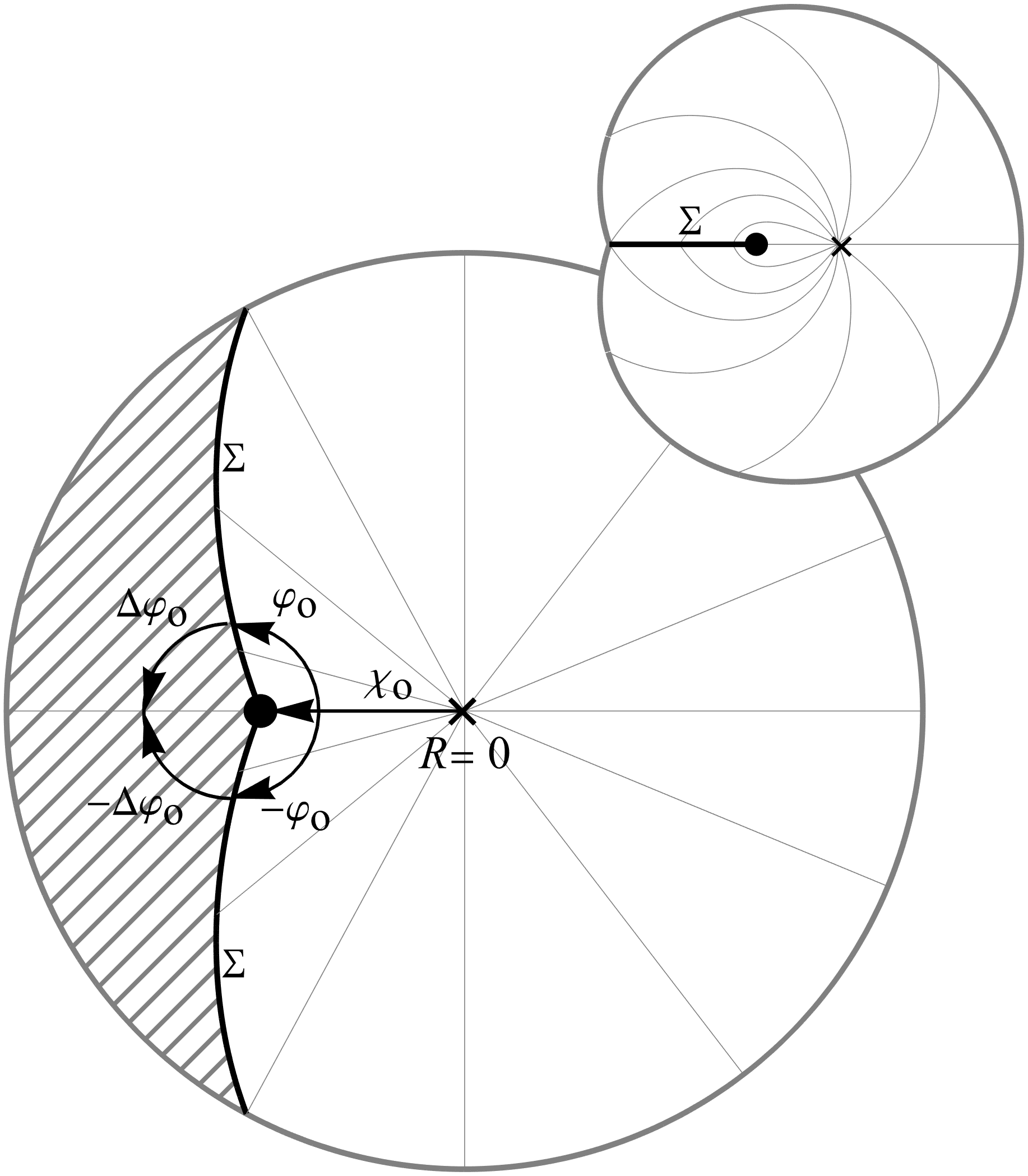}\qquad\qquad
    \includegraphics[]{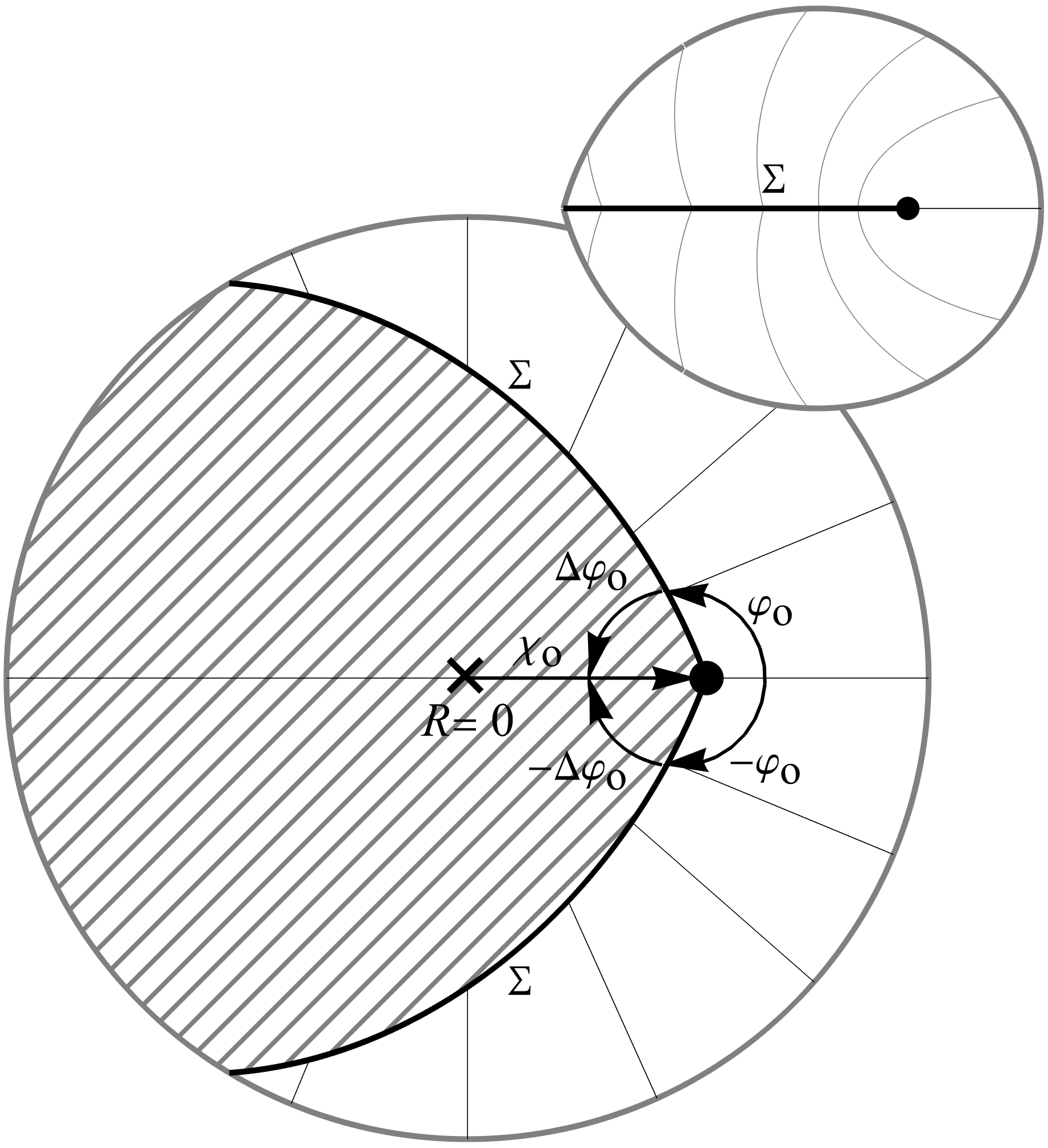}
    \caption{\label{fig:string_strut}%
    Left: A particle at $\bchi=0$, positioned at $\chi=\chi_\oi<0$, is in equilibrium between a pull of the semi-infinite string and gravitational attraction toward the origin. The excluded region is hatched, and the thick curves denote the hypersurfaces identified into a string $\Sigma$. The diagram is drawn in standard polar coordinates $\chi,\,\ph$. Radial lines show the direction of the apparent gravitational force in the global static frame. Relation between the particle's acceleration, $a_\oi<0$, and parameter $\chi_\oi<0$ is given by~\cref{eq:aodef}.
    The small diagram suggests how an observer inside this spacetime might perceive the situation. The line $\Sigma$ represents the semi-infinite string depicted as a spatially linear defect. The same `radial' lines as in the main diagram are drawn, showing the direction of the gravitational force.\\
    Right: Setting the particle's position at $\chi=\chi_\oi>0$ leads to the particle being pushed by a strut against the gravitational pull. Again, radial lines show the gravity direction. The acceleration parameter $a_\oi$ is positive. Notice that the center of gravity is excluded from the spacetime in this case.
    The small diagram is again deformed in such a way that the strut $\Sigma$ is depicted as a one-dimensional object. In this case, gravity points everywhere in the spacetime towards the strut, which acts as its source. This is because the original center of gravity is in the removed region.\\[-4ex]}
\end{figure*}

In our case, we are gluing two symmetric surfaces $\bph=\pm\ph_\oi$ (removing domains $\bph>\ph_\oi$ and $\bph<-\ph_\oi$) which, however, intersect at $\bR=0$. The identified surfaces end in the intersection, forming the conical deficit $2\Delta\ph_\oi$. The constructed spacetime thus describes a point particle (characterized geometrically by deficit angle $\Delta\ph_\oi$ or physically by mass $m_\oi$) which is joined to a string $\Sigma$ (characterized geometrically by its mean curvature $\scur_\oi$ or physically by its linear density $\mu_\oi$). One end of the string stretches to infinity while the other is attached to the particle onto which it exerts a pull or push, which compensates for gravitational attraction $a_\oi$ acting towards the gravity center, keeping the particle at rest in the static frame. 

In particular, the mean curvature $\scur_\oi$ of the hypersurface $\bph=\ph_\oi$ is related to the parameters $a_\oi$ of \cref{eq:aodef} and $\Delta\ph_\oi=\pi-\ph_\oi$ as
\begin{equation}
    \scur_\oi = a_\oi\, \sin\Delta\ph_\oi = \frac{\sin\chi_\oi\,\sin\Delta\ph_\oi}{\ell}\;.
    \label{eq:meancurvaopho}
\end{equation}
For more discussion, see \Cref{apx:string_curvature}.

Notice that for $\ph_\oi\in(0,\pi)$ (i.e., $m_\oi>0$), the sign of the string linear energy density $\mu_\oi=-\tfrac{2\scur_\oi}{\kap}$ is opposite to the sign of the acceleration parameter $a_\oi$. Recall also that $a_\oi$ is related to the particle's position by \eqref{eq:aodef}. The string extends on the opposite side of the particle than the infinity of the semi-axis $\ph=0$ (aiming to the left in \Cref{fig:string_strut}). Clearly, if the particle is localized at $\chi_\oi<0$ (left to the origin), the gravity strength $-a_\oi$ is positive, the gravity is pointing to the right (towards the center) which is compensated by the pull of the string (to the left) with the energy density $\mu_\oi>0$, see \Cref{fig:string_strut} left. If the particle is localized at $\chi_\oi>0$, the gravity strength $-a_\oi$ is negative, and the gravity is pointing to the left (also towards the center), which reflects the push of the strut (to the right) with the energy density $\mu_\oi<0$, see \Cref{fig:string_strut} right.

We could deform the visualization of the spatial section of our spacetime in such a way that the identified surface $\Sigma$ is depicted as a linear object localized on the semi-axis pointing to the left from the particle, see the small diagrams in \Cref{fig:string_strut}. These diagrams supply a natural intuition to our understanding of the system: the linear object supports the point particle at its static position. However, these diagrams employ an arbitrary transformation particular to the given cases. There does not seem to be a similar preferred way to deform the diagrams in more general situations.

\begin{figure*}\centering
    \vspace*{-4ex}
    \includegraphics[]{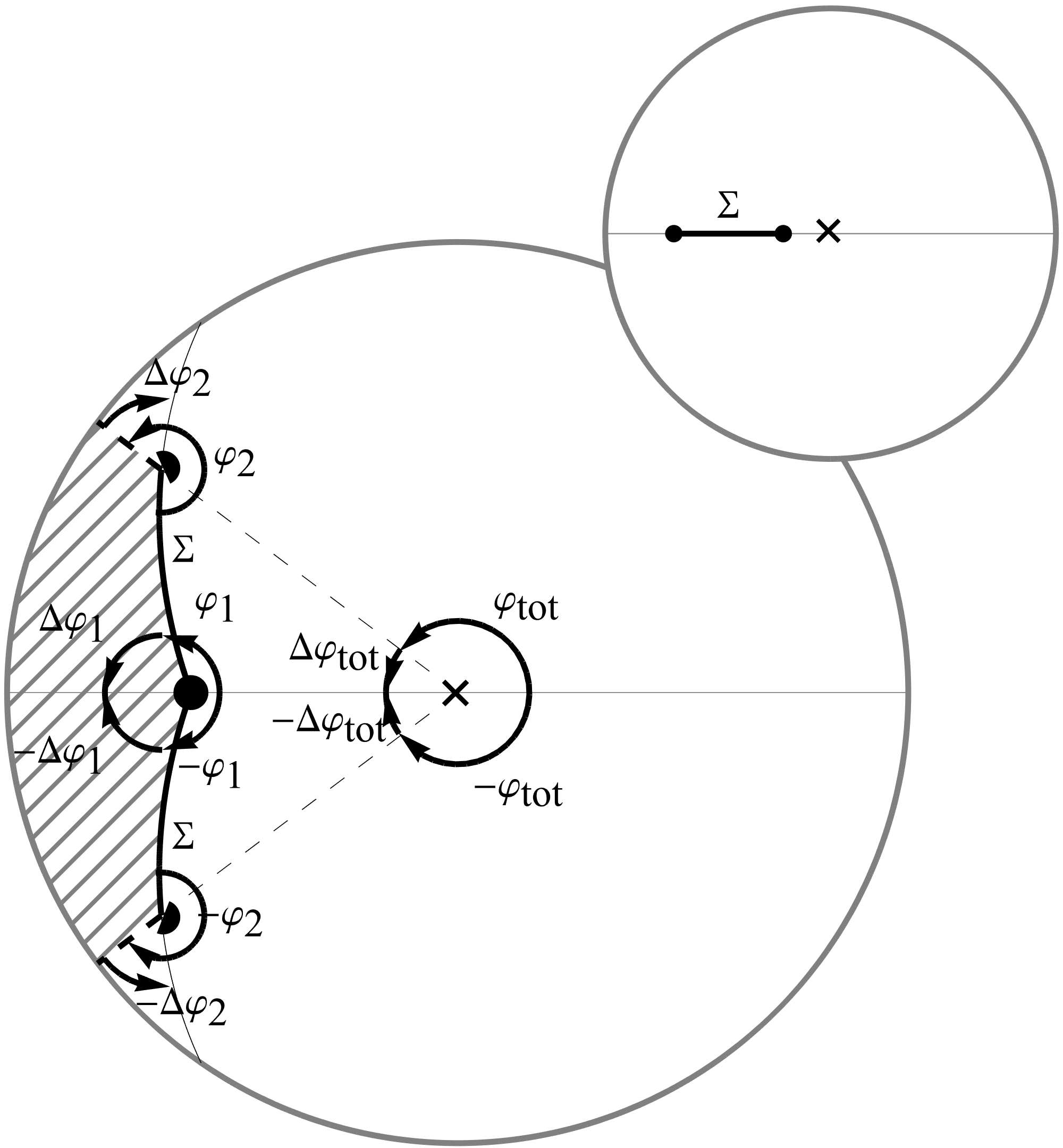}\qquad
    \includegraphics[]{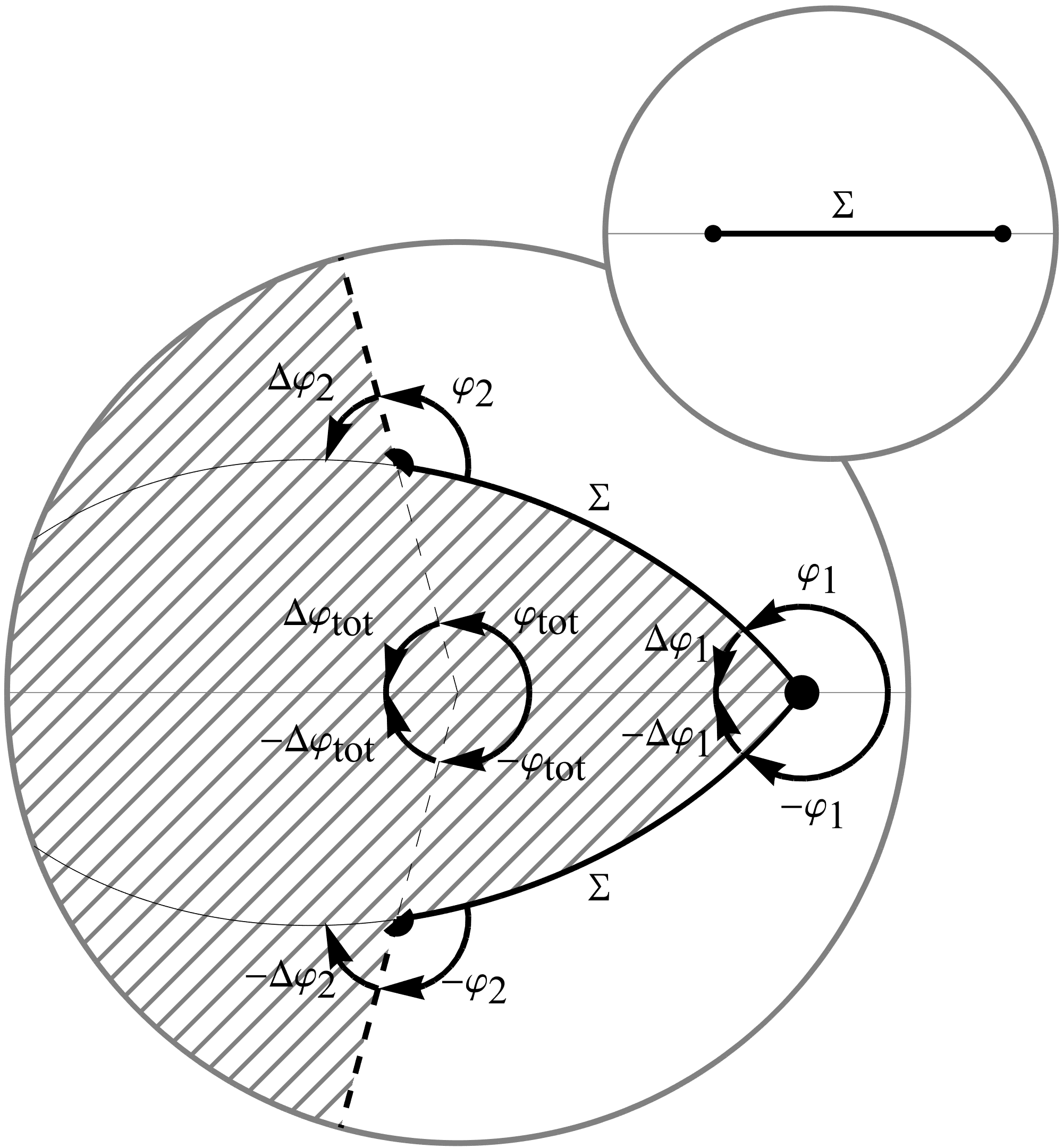}
    \caption{\label{fig:two_particles_string/strut}%
    Left: Two particles on a string. The thick solid curves suggest the two hypersurfaces forming the history of the finite string. The thick dashed line segments represent the flat hypersurfaces reaching to infinity. The hatched area again represents the excluded region. $\Delta\ph$'s parameterize the conical defects at respective points. $\Delta\ph_{\tot}$ parameterizes the mass of the whole system as observed in the far region. Notice that $\Delta\ph_2 < 0$ and the corresponding particle thus has a negative mass. We thus have a self-accelerating system of negative and positive mass particles joined by the string. The diagram in the top right corner shows a qualitative suggestion of how an observer inside the spacetime might perceive it. The thick curve represents the string, the small discs represent particles (semi-discs are used when identification of multiple points forms a single particle), the thin and dashed semi-lines indicate flat surfaces, and the small cross denotes a location of the center of gravity. \\ 
    Right: The situation represents two particles separated by a finite strut. It keeps both particles in equilibrium and compensates for the AdS gravitational pull. Note that this time $\Delta\ph_2 > 0$ and thus both particles have positive mass. Again, a qualitative representation of what an observer inside might perceive is presented in the top right corner. Notably, this time, the center of gravity is not part of the spacetime.\\[-3ex]}
\end{figure*}

The metric of the spacetime can be acquired from \eqref{eq:AdSAccCoor} by introducing angular restriction ${\ph\in(-\ph_\oi,\ph_\oi)}$, rescaling similarly to \eqref{eq:rescaling}, substituting \eqref{eq:ourBTZ_mass}, and introducing 
${\ell' = \tfrac{\ell}{\sqrt{1-a_\oi^2\ell^2}}}$. We obtain
\begin{equation}
    \begin{split}
        \ts{g}_{\AdS} &= \frac{1}{\Bigl(1-\frac{a_\oi \bR \cos(\sqrt{1-M}\,\bphi)}{\sqrt{1-M}}\Bigr)^2}
        \biggl(-\Bigl(1-M+\tfrac{\bR^2}{\ell'^2}\Bigr)\dif \bT^2\\
        &\quad+ \frac{1}{1-M+\frac{\bR^2}{\ell'^2}}\, \dif \bR^2 
        + \bR^2\dif \bphi^2\biggr)\;,
    \end{split}\raisetag{5ex}
\end{equation}
with ${\bphi\in(-\pi,\pi)}$. The conical defect is located at ${\bR = 0}$ around which the conformal prefactor behaves as $1$. Therefore, near the origin, the spacetime behaves as a conical deficit in the same way as in the unaccelerated case. Moreover, the acceleration does not interfere with the parametrization of mass: it is still given by \cref{eq:ourBTZ_mass} even in the accelerated case.

Although we refer to the spatially linear object as a string, in three spacetime directions, such an object is also equivalent to a domain wall: an object of codimension one, whose worldsheet separates two domains of the spacetime by a massive membrane. We keep the naming analogous to a four-dimensional situation, where the agents pushing or pulling black holes in the C-metric spacetime are cosmic strings. However, some of the features of the cosmic strings in three dimensions can be more understandable from the point of view of the domain wall interpretation.

\subsection{C-metric}
\label{ssc:Cmetric}

As we already mentioned, the metric \eqref{eq:AdSAccCoor} with specification $\bph\in(-\ph_\oi,\ph_\oi)$ is one of the possible forms of the three-dimensional C-metric. In four dimensions, the \mbox{C-metric} was first conceived already in 1917 in \cite{Weyl:1917}, coined in \cite{EhlersKundt:1962}, but studied and understood much later \cite{KinnersleyWalker:1970,Bonnor:1982,AshtekarDray:1981,HongTeo:2003,GriffithsPodolsky:2005}, for review see \cite{GriffithsKrtousPodolsky:2006}, and in the AdS context cf.~e.g.~\cite{DiasLemos:2003a,Krtous:2005,PodolskyGriffiths:2006}. The discussion in three dimensions opened in \cite{Astorino:2011}, among recent articles are \cite{cmetric_ruth,Bunney:2025,BH_2+1_holography}). 

We have justified that \eqref{eq:AdSAccCoor} with the restriction on $\bph$ describes an accelerated point particle connected to the string. Let us introduce another coordinate system, which brings the metric to another customary form analogous to the standard \mbox{C-metric} form in four dimensions. Namely, we define 
\begin{equation}
  \label{eq:CMcoors}
    t = a_\oi\bT \;,\quad
	y = \frac{1}{a_\oi\bR}\;,\quad
    x = \cos\bph
\end{equation}
which brings \eqref{eq:AdSAccCoor} into form
\begin{equation}
	\label{eq:Cmetric}
	\ts{g}_\Cmt = \frac{1}{a_\oi^2(y-x)^2}
        \left(-P(y)\dif t^2 
        + \frac{1}{P(y)} \, \dif y^2 
        + \frac{1}{Q(x)} \, \dif x^2\right)\;,
\end{equation}
with metric functions 
\begin{equation}
  \label{eq:PQdef}
	P(y) = \frac{1}{a_\oi^2\ell^2}-1+y^2\;,\quad
    Q(x) = 1- x^2\;,
\end{equation}
and restriction $x\in(0,x_\oi)$, where $x_\oi=\cos\ph_\oi<1$.

\subsection{Two static particles on a finite string or strut}
\label{ssc:two_particles}

The just-discussed C-metric describes the point particle attached to the string that reaches up to infinity. Now, we will construct the finite system of two particles connected by a string, cf.~\Cref{fig:two_particles_string/strut}.

We start with the particle attached to a semi-infinite string depicted in \Cref{ssc:string}. We rename the angle under which the string leaves ${\bR = 0}$ as $\ph_1$. To construct another particle on the same string, let us choose an angle $\varphi_\tot$ such that the radial surface ${\ph=\ph_\tot}$ starting from ${R=0}$ intersects the surface $\bph = \ph_1$. We denote $\ph_2$ the angle at the intersection. Similarly, the symmetric surface ${\ph=-\ph_\tot}$ intersects the surface ${\bph=-\ph_1}$ under angle ${-\ph_2}$.

As before, we identify the curved surfaces $\bph=\pm\ph_1$ from $\bR=0$ until the intersection with the flat surfaces $\ph=\pm\ph_\text{tot}$. Next, we identify the flat surfaces $\ph=\pm\ph_\text{tot}$, starting at the intersection up to infinity. We thus remove the hatched region in \Cref{fig:two_particles_string/strut}. Since near infinity only hypersurfaces with no extrinsic curvature are identified, there is no mass-energy distribution along the gluing and therefore neither in the whole asymptotic region. The parts of the identified curved surfaces $\bph=\pm\ph_1$ are now spatially finite and thus represent a string of finite length.

There are non-trivial conical deficits $\Delta\ph_{1,2} = \pi - \ph_{1,2}$ at both ends of the string. They represent point particles of local masses $m_{1,2}$ given by~\eqref{eq:mascondef}. The string pulls or the strut pushes the particle at either of its ends so that they are in static equilibrium with the AdS gravitational attraction. Non-zero mass-energy density of static objects is thus localized only within a finite region.

In \Cref{fig:two_particles_string/strut}, we have chosen $\Delta\ph_1>0$, corresponding to the positive subcritical mass $m_1$. On the left, the conical deficit $\Delta\ph_2$ is negative, which corresponds to the angular excess and negative mass $m_2<0$. Negative and positive mass particles connected by a string accelerate each other and keep themselves hanging above the gravitational center. On the right, both conical deficits are positive, and both particles have positive masses. The strut between them keeps them at rest around the gravitational center, which, however, is not present in the spacetime -- it is replaced by the strut between the particles.

\section{Finite static systems and their mass}
\label{sec:mass}

Let us now study systems localized in a finite domain of spacetime, i.e., no matter distribution reaches the spatial infinity (of course, ignoring the dark energy/cosmological constant contribution). We formulate two simple definitions of total mass for such systems and discuss them in the simplest cases.

When discussing a single point particle in \Cref{sec:ads_intro}, we have introduced the mass parameter $m_\oi$ proportional to the angle deficit $\Delta\ph_\oi$, cf.~\eqref{eq:mascondef}. In the context of the BTZ metric, we defined the BTZ mass $\mBTZ$, which is related to the local particle mass by \eqref{eq:massrel}. Now, we relate these quantities to more general characteristics of finite static systems.

For the rest of this section, let us consider a static system with arbitrary matter content. The metric thus reads
\begin{equation}
    \ts{g} = - N^2 \dif T^2 + \ts{q}\;,
\end{equation}
where $T$ is the static time coordinate, $N$ is lapse, and $\ts{q}$ spatial metric on the slice $T=\const$. We assume the Einstein equation in the form
\begin{equation}
    \Ein = \kap\, \Terg\;,
\end{equation}
i.e., we include the cosmological term in the stress-energy tensor $\Terg$ as a dark energy term.


\subsection{Local mass}
\label{ssc:localmass}

The total rest energy density $\eps$ is given by the projection of $\Terg$ on the time normal $\tnrm=\frac{1}{N}\ts{\partial}_T$,
\begin{equation}
    \eps 
       = \tnrm\cdot\Terg\cdot \tnrm\;.
\end{equation}
A natural definition of the mass localized in a spatial domain $\dom{D}$ is a plain integral of the rest energy,
\begin{equation}
    \label{eq:locmass}
    \locmass[\dom{D}] = \int_{\dom{D}} \eps\, dS\;.
\end{equation}
Here, $dS$ is the area element associated with the spatial metric $\ts{q}$. We call this quantity the \emph{local mass}.

In 3 dimensions, the local mass has a very close relation to geometric quantities. Recall that in the 3-dimensional context, the normal-normal projection of the static Einstein tensor is given by the Gauss curvature\footnote{The Gauss curvature $\spgc$ is just the rescaled scalar curvature $\spsc$ of the spatial metric $\ts{q}$, $\spgc=\frac12\,\spsc$.} of the spatial geometry \cite{DESER1984220}
\begin{equation}
    \tnrm\cdot\Ein\cdot\tnrm = \spgc\;.
\end{equation}
Thus, the local mass is given by
\begin{equation}
    \label{eq:locmassspsc}
    \locmass[\dom{D}] = \frac{1}{\kap}\int_{\dom{D}} \spgc\, dS\;.
\end{equation}
The integral of the Gauss curvature over a domain $\dom{D}$ is related by the Gauss-Bonnet theorem to the integral over its boundary $\partial \dom{D}$. For a compact domain $\dom{D}$ with a piecewise smooth boundary $\partial \dom{D}$, the Gauss-Bonnet theorem states \cite{Tapp:2016}
\begin{equation}
    \label{eq:GaussBonnet}
    \int_{\dom{D}} \spgc\, dS = 2\pi\,\chi[\dom{D}] 
    + \int_{\partial \dom{D}} \alpha\, ds 
    -\sum_j \psi_j\;,
\end{equation}
where $\chi[\dom{D}]$ is the Euler characteristic of the domain $\dom{D}$, $\alpha$ is the boundary curvature defined with respect to the outer normal\footnote{Let $\ts{e}$ be a unit vector tangent to the boundary and $\ts{n}$ the outer normalized normal of the boundary. The curvature vector $\ts{a}=\ts{e}\cdot\spcd\ts{e}$ ($\spcd$ being the spatial covariant derivative) has the normal direction, $\ts{a}=\alpha\,\ts{n}$. I.e., $\alpha=\ts{n}\cdot\ts{a}$. In \cite{Tapp:2016}, they effectively assume $\ts{a}=-\alpha\,\ts{n}$ compared to \cref{eq:GaussBonnet} here, so they acquire opposite sign in the geodesic curvature term.},
$dS$ is the area element of the spatial geometry, and $ds$ is the linear element on the boundary induced by the spatial geometry. The last term enters when the boundary $\partial \dom{D}$ contains corners. $\psi_j$ are angular jumps in the change of the boundary tangent vector at the corners. The Euler characteristic for a compact surface of genus $g$ with the boundary composed of $b$ parts of topology $S^1$ is\footnote{Well-known classification result for the topology of two-dimensional compact surfaces tells that the topology of any surface is given by its genus $g$, and by the number $b$ of $S^1$-boundary parts (i.e., the number of removed discs). For an orientable surface, the genus is the number of `handles' attached to $S^2$ (the number of connected sums with tori). For a non-orientable surface, the genus is half of the attached `crosscups' to $S^2$ (the number of connected sums with projective planes). The Euler characteristic is the topological invariant of the surface completely determined by $g$ and $b$ through \eqref{eq:EulerCh}.}
\begin{equation}
    \label{eq:EulerCh}
    \chi[\dom{D}] = 2- 2g - b \;.
\end{equation}
For our applications, typically $g=0$, and in this section, we assume a smooth boundary without corners.

The local mass in a domain $\dom{D}$ thus can be determined by measuring the boundary curvature $\alpha$ (which is closely related to the parallel transport along the boundary), without exploring the interior of the domain explicitly,
\begin{equation}
    \label{eq:locmaspardef}
    \locmass[\dom{D}] = \frac{2\pi}{\kap}\chi[\dom{D}] 
       + \frac1{\kap}\int_{\partial \dom{D}} \alpha\, ds\;.
\end{equation}


\subsection{Killing mass}
\label{ssc:Killingmass}

Another natural definition of mass is related to the static Killing vector $\ts{\partial}_T$. The relevant energy density is given by 
\begin{equation}
    \eps_T 
      = \tnrm\cdot\Terg\cdot \ts{\partial}_T=N\eps\;,
\end{equation}
and the energy localized in the domain $\dom{D}$ is
\begin{equation}
    \label{eq:KVmass}
    \MKV[\dom{D}] = \int_{\dom{D}} \eps_T\, dS = \int_{\dom{D}} N \eps\, dS= \frac{1}{\kap}\int_{\dom{D}} N\spgc\, dS\;.
\end{equation}
We call this quantity the Killing mass.

Conservation of the Killing mass for a static matter distribution is trivial. But thanks to the local conservation of the stress-energy tensor $\ts\nabla\cdot\Terg=0$ and the Killing equation, this quantity is conserved even for the dynamical evolution of test matter on the static background.


\subsection{Dark-energy contribution}
\label{ssc:DarkEnergt}

Let us recall that we included the cosmological term into the stress-energy tensor $\Terg_{\!\Lambda} = -\frac{\Lambda}{\kap}\,\ts{g}$. In the AdS context, it means that spacetime is filled by the dark energy of density $\eps_\Lambda = \frac{\Lambda}{\kap}$. We thus cannot expect that the defined mass will be finite for a whole spatial section. However, we can evaluate a difference between the mass content of the studied spacetime and a reference `empty' spacetime containing only the dark-energy contribution. 

The typical approach is subtracting the contribution of a corresponding domain of the empty AdS spacetime $\dom{M}_\AdS$. Of course, one has to specify what the corresponding domains are. For spherically symmetric systems, one can consider spherical domains $\dom{B}_R$ and ${}^{\AdS}\!\dom{B}_R$ of the same circumference $2\pi R$. For example, the renormalized global mass of the whole spacetime then reads
\begin{equation}
 \label{eq:renKVmass}
 \MKV_\ren 
  = \lim_{R\to\infty} \bigl(\MKV[\dom{B}_R]-\MKV_{\AdS}[{}^{\AdS}\!\dom{B}_R]\bigr)
  \;.
\end{equation}

However, if our aim is to subtract exactly the asymptotic contribution of the dark energy, it is more natural to compare the studied spacetime $\dom{M}$ with another `canonical' spacetime $\dom{M}_\can$ which is asymptotically identical, but `empty' as much as possible. In such an approach, one has to identify the canonical spacetimes given by their asymptotic behavior and contain only dark energy content, except for a sufficiently simple matter distribution necessitated by the given asymptotic. In a sense, we need to find a `canonical monopole' spacetime with a given asymptotic and subtract its contribution,
\begin{equation}
    \label{eq:difKVmass}
    \Delta\MKV =  \MKV[\dom{B}_{R_\ai}]-\MKV_\can[{}^\can\!\dom{B}_{R_\ai}]\;.
\end{equation}
Here, $R_\ai$ is the circumference radius in the asymptotic region, where the studied spacetime can be identified with the canonical one. Clearly, the difference $\Delta\MKV$ does not depend on a particular choice of $R_\ai$ in the asymptotic region, since the spacetimes are identical in the asymptotic region. 

The quantity $\Delta\MKV$ estimates how much the matter content of the studied spacetime differs from the canonical one and how the inner parameters of the matter distribution are related to the asymptotic characteristics.

We have mentioned in \eqref{eq:locmaspardef} that the local mass $\locmass[\dom{D}]$ is fully given by the properties of the spacetime on the boundary $\partial \dom{D}$. It implies that when comparing the local mass inside the domain $\dom{D}$, which reaches the asymptotic of the studied spacetime, with the local mass of the corresponding\footnote{The correspondence of the domains $\dom{D}$ and ${}^\can\!\dom{D}$ means that they have identical boundaries in the asymptotic region, where both spacetimes can be identified. It is a generalization of spherical domains $\dom{B}_{R_\ai}$ and ${}^\can\!\dom{B}_{R_\ai}$ discussed above.}
domain ${}^\can\!\dom{D}$ in the canonical spacetime, the difference must vanish,
\begin{equation}
    \label{eq:diflocmass}
    \Delta\locmass = \locmass[\dom{D}]-\locmass_\can[{}^\can\!\dom{D}] = 0\;.
\end{equation}
We thus obtain just a relation $\locmass[\dom{D}]=\locmass_\can[{}^\can\!\dom{D}]$ between the inner characteristics of the matter distribution in $\dom{D}$ and its asymptotic behavior given by the canonical spacetime. Notice that we silently assumed that the Euler characteristics $\chi[\dom{D}]$ and $\chi[{}^\can\!\dom{D}]$ of both domains are the same. It does not have to be true when analyzing, for example, the black hole situation. We will return to this point below and in \cite{2+1_mass}.


\subsection{Inertial mass}
\label{ssc:Inertialmass}

Finally, we can also estimate the mass of the point particle by its response to the force. In \Cref{ssc:string}, we discussed the static particle attached to the string. In the spacetime description, the particle is moving along a non-geodesic Killing orbit under the influence of the string's tension. The relation \eqref{eq:meancurvaopho} corresponds to the Newton law $\text{mass}\times\text{acceleration} = \text{tension}$, namely,
\begin{equation}
    m_\oi a_\oi\, \Bigl(\frac2{\kap m_\oi}\,\sin\frac{\kap m_\oi}{2}\Bigr) = -\mu_\oi\;.
    \label{eq:stringparticleNewto}
\end{equation}
Alternatively, in spatial description, we have to introduce the fictitious gravity force. The static particle is then understood in the equilibrium of gravity and the string force. The relation \eqref{eq:meancurvaopho} can be understood as the equilibrium condition $\text{tension}+\text{gravity}=0$, namely,
\begin{equation}
    -\mu_\oi - m_\oi a_\oi\, \Bigl(\frac2{\kap m_\oi}\,\sin\frac{\kap m_\oi}{2}\Bigr) = 0\;.
    \label{eq:stringparticleeqcond}
\end{equation}
We see that the effective mass estimating the response to the force (entering into the Newton law), and at the same time playing the role of gravitational charge, is $m_\oi\bigl(\frac2{\kap m_\oi}\,\sin\frac{\kap m_\oi}{2}\bigr)$. For small $m_\oi$, it is exactly $m_\oi$. For a general value of $m_\oi$, we have an additional factor $\frac{2}{\kap m_\oi}\sin\frac{\kap m_\oi}{2}$. Its exact meaning is not satisfactorily understood. It is related to a singular character of the attachment of the point particle to the string.


\subsection{Static rotational symmetric space}
\label{ssc:StatRotSym}

Next, we want to study simple static systems with rotational symmetry. We start by summarizing an elementary two-dimensional geometry with rotational symmetry \cite{2+1_mass}. In such a case, the spatial metric can be written
\begin{equation}
    \label{eq:statrotsymmtrc}
    \ts{q} = \dif r^2 + R^2 \dif\phi= \frac{1}{{R'}^2}\,\dif R^2 + R^2 \dif\phi\;,
\end{equation}
with $\phi\in(-\pi,\pi)$. $r$ measures the radial distance, ${R=R(r)}$ is the circumference radius and ${'\equiv\frac{d}{dr}}$. In the second form of the metric, we usually understand $R'=R'\bigl(r(R)\bigr)$ as a function of $R$ and ${' = R'\frac{d}{dR}}$. 

The Gauss curvature $\spgc$ of the spatial geometry is 
\begin{equation}
    \label{eq:GaussCsrsym}
    \spgc = - \frac{R''}{R}\;.
\end{equation}
The boundary curvature $\alpha$ of the circle $r=\const$ with respect to the normal $\ts{\partial}_r$ is 
\begin{equation}
    \label{eq:boundcurvsrsym}
    \alpha = - \frac{R'}{R}\;.
\end{equation}
Area and line elements are 
\begin{equation}
    \label{eq:intdens}
    dS= R\, dr d\phi\,,\qquad ds = R\, d\phi\;.
\end{equation}

\subsection{Point particle in the Minkowski spacetime}
\label{ssc:PointParticleMink}

Let us analyze shortly the case $\Lambda=0$. We do not have to worry about the dark energy contribution for a vanishing cosmological constant. The local mass can be evaluated directly, and for finite systems, it leads to a finite result when evaluated on the whole spatial section. Moreover, thanks to \eqref{eq:locmaspardef}, it is entirely determined by the asymptotic geometry.

The asymptotic geometry is locally flat, but it can be some non-trivial composition of the Minkowski regions globally. If we assume (2-dimensional) spherical symmetry and simple topology $S^1\times \realn$ of the asymptotic region, the metric in the far region is conical:
\begin{equation}
    \label{eq:flatsph}
    \ts{g}_\con 
        = - \dif t^2 + \dif r^2 + r^2 \dif\ph^2 \;,\quad
    \text{where}\ \ph\in(-\ph_\oi,\ph_\oi)\;.
\end{equation}
Here, $t$ is the static time and $r$ the radial distance. The asymptotic geometry is fully characterized by the constant $\ph_\oi$. Rescaling all coordinates as
\begin{gather}
        t=\sqrt{1{-}\mBTZ}\,T\,,\quad r=\frac{1}{\sqrt{1{-}\mBTZ}}R\,,\quad \ph=\sqrt{1{-}\mBTZ}\,\phi\,,\notag\\ \text{with}\quad  \mBTZ=1-\Bigl(\frac{\ph_\oi}{\pi}\Bigr)^2 \;,
    \label{eq:flatcoorresc}
\end{gather}
leads to the BTZ-like form of the metric
\begin{equation}
    \label{eq:flatcon}
    \ts{g}_\con = - (1-\mBTZ)\,\dif T^2 
      + \frac{1}{1-\mBTZ}\, \dif R^2 + R^2\, \dif\phi^2 \;.
\end{equation}
Now $\phi\in(-\pi,\pi)$, and $R=\sqrt{1-\mBTZ}\,r$ is the circumference radius. 

For a circle $r=r_\ai$, the boundary curvature is ${\alpha=-\frac{1}{r_\ai}=-\frac{\sqrt{1-\mBTZ}}{R_\ai}}$. Since it is constant, the boundary contribution to the local mass \eqref{eq:locmaspardef} is ${2\pi R_\ai \alpha = - 2\pi\sqrt{1-\mBTZ}=-2\ph_\oi}$. As expected, it does not depend on the choice of the radius $r_\ai$ -- the conical space is empty everywhere except the origin. For a domain of topology of a disc with one boundary, the Euler characteristic is 1; therefore 
\begin{equation}
    \label{eq:locmassdefangrel}
    \locmass[{}^\con\!\dom{B}_{R_\ai}] 
      =\frac{2\pi}{\kap} \bigl(1 - \frac{\ph_\oi}{\pi}\bigr) 
      =\frac{2\Delta\ph_\oi}{\kap} 
      = m_\oi\;.
\end{equation}
The deficit angle $\Delta\ph_\oi=\pi-\ph_\oi$ corresponds to (half of) the vertex deficit angle of the conical space. We thus justified the definition of the mass parameter $m_\oi$ of a point particle \eqref{eq:mascondef} by demonstrating that it is equal to the local mass.  
From \eqref{eq:flatcoorresc} we also see that the BTZ mass parameter $\mBTZ$ is related to $m_\oi$ by \eqref{eq:massrel}, cf.~also \eqref{eq:ourBTZ_mass}.

Before concluding this simplest case, we should mention that we used relation \eqref{eq:locmaspardef} for evaluating the local mass in a domain encircled by the loop. This relation is valid for a sufficiently smooth energy distribution inside the domain. Clearly, the point particle does not satisfy this requirement. Indeed, the energy distribution in the conical space is trivially zero everywhere except the origin, where it is not well-defined. Generalization of \eqref{eq:locmaspardef} to this case suggests that the energy has $\delta$-function character at the origin. We intuitively agree with this interpretation. However, it is known that the conical singularity cannot be fully grasped in the traditional distribution theory mainly because of the nonlinearity of the curvature and the problem with differentiability of the spacetime at the point of the conical singularity \cite{Steinbauer_2006,Geroch_Trashen_1987}. The Gauss-Bonnet theorem, which allows shifting the evaluation of a domain integral to its boundary, is an alternative way to approach the distributional character of the conical singularity.


\subsection{Canonical static rotationally symmetric spacetimes}
\label{ssc:CanSTAdS}

In the AdS context, the rotationally symmetric static solution has the asymptotic form of the BTZ metric. 
\begin{equation}
	\label{eq:BTZ_metric2}
	\ts{g} = -\left(1{-}\mBTZ{+}\tfrac{R^2}{\ell^2}\right)\dif T^2 
        + \frac{1}{1{-}\mBTZ{+}\tfrac{R^2}{\ell^2}}\, \dif R^2 
        + R^2\,\dif \phi^2 \;,
\end{equation}
with ${\phi\in(-\pi,\pi)}$ and ${N=R'=\sqrt{1-\mBTZ+\frac{R^2}{\ell^2}}}$. The mass parameter $\mBTZ$ uniquely determines the asymptotic behavior. When extending the asymptotic geometry as $\Lambda$-vacuum for small $R$, one finds two qualitatively different cases: For ${\mBTZ>1}$ it leads to the BTZ black hole; for ${\mBTZ<1}$ we have seen in \Cref{ssc:subBTZ} that it is equivalent to the conical space representing the point particle. The spacetimes with this geometry play the role of canonical spacetimes to which we asymptotically compare a generic static solution. 

This is in contrast with perhaps a more common approach in which one assumes the asymptotic region to be a part of the global empty AdS universe, leading one to the conclusion that the empty AdS should be taken as a baseline for dark energy subtraction. Comparing with the canonical spacetime instead gives the exact cancellation of the dark-energy contribution in the asymptotic region.

It is straightforward to evaluate the Killing mass \eqref{eq:KVmass} for the domain $\dom{B}_{R_\iix}^{R_\oix}$ between two radii $R_\iix$ and $R_\oix$ for the canonical spacetime \eqref{eq:BTZ_metric2},
\begin{equation}
	\label{eq:BTZ_KmassRR}
	\MKV[\dom{B}_{R_\iix}^{R_\oix}] = \frac{2\pi}{\kap}\frac{R_\iix^2-R_\oix^2}{2\ell^2}\;.
\end{equation}
It describes the dark-energy contribution. Unfortunately, including matter would require a more detailed knowledge of the matter model. We leave the further discussion of the Killing mass to \cite{2+1_mass}.

However, thanks to formula \eqref{eq:locmaspardef}, one can make more conclusions about the local mass, even without knowing details of a particular matter model.


\subsection{Point particle in AdS}
\label{ssc:PointParticleAdS}

First, we focus on the case ${\mBTZ<1}$. Similarly to the $\Lambda=0$ case, we can evaluate the local mass inside the ball $\dom{B}_{R_\ai}$, given by ${R<R_\ai}$, using formula \eqref{eq:locmaspardef}. Taking $\chi[\dom{B}_{R_\ai}]=1$ and $\alpha$ given by \eqref{eq:boundcurvsrsym}, we get
\begin{equation}
	\label{eq:BTZ_locmass}
	\locmass[\dom{B}_{R_\ai}] = \frac{2\pi}{\kap}\Bigl(1-\sqrt{1{-}\mBTZ{+}\tfrac{R_\ai^2}{\ell^2}}\Bigr)\;.
\end{equation}
Of course, it decreases with growing $R_\ai$ because of the dark energy contribution and diverges for $R_\ai\to\infty$.

We suggest using this value as the local mass of the canonical space, which we compare with the local mass of generic systems with the asymptotic \eqref{eq:BTZ_metric2}, cf.~\eqref{eq:diflocmass}. 

Notice that for $\mBTZ<1$, when the spacetime represents conical geometry (cf.~Sec.~\ref{ssc:subBTZ}), the value of the local mass for $R_\ai\to0$ is $\locmass[\dom{B}_{0+}] =  \frac{2\pi}{\kap}\bigl(1-\sqrt{1{-}\mBTZ}\bigr)$, which, together with \eqref{eq:ourBTZ_mass}, gives 
\begin{equation}
	\label{eq:BTZ_locmass_R0}
	\locmass[\dom{B}_{0+}] =  \frac{2\Delta\ph_\oi}{\kap} = m_\oi\;.
\end{equation}
The local mass \eqref{eq:BTZ_locmass} thus takes into account not only the dark energy contribution but also the contribution of the point particle at the origin. The particle mass parameter $m_\oi$ is equal to the local mass localized just around the singular origin. 

We can also evaluate the local mass in the domain $\dom{B}_{R_\iix}^{R_\oix}$ between two circumference radii. Now ${\chi[\dom{B}_{R_\iix}^{R_\oix}]=0}$, and, at the inner radius $R_\iix$, we have to choose the boundary normal $\ts{n}=-\ts{\partial}_r$. We obtain
\begin{equation}
	\label{eq:BTZ_locmassRR}
	\locmass[\dom{B}_{R_\iix}^{R_\oix}] = \frac{2\pi}{\kap}\Bigl(\sqrt{1{-}\mBTZ{+}\tfrac{R_\iix^2}{\ell^2}}-\sqrt{1{-}\mBTZ{+}\tfrac{R_\oix^2}{\ell^2}}\Bigr)\;.
\end{equation}
We can write
\begin{equation}
    \label{eq:BTZ_locmassPPsplit}
    \begin{split}
        \locmass[\dom{B}_{R_\ai}] &= \locmass[\dom{B}_{0+}] + \locmass[\dom{B}_{0+}^{R_\ai}]\\
        &= m_\oi + \frac{2\pi}{\kap}\Bigl(\sqrt{1{-}\mBTZ{+}\tfrac{R_\iix^2}{\ell^2}}-\sqrt{1{-}\mBTZ{+}\tfrac{R_\oix^2}{\ell^2}}\Bigr)\;.
    \end{split}\raisetag{9ex}
\end{equation}
The first term is the local mass of the point particle, and the second term is the dark-energy contribution in the surrounding space. As discussed above, the contribution from the singular point particle is mathematically slightly problematic. But here we can understand it as a limit of an arbitrary smooth rotation-symmetric matter distribution squeezed closely around the origin. The formula \eqref{eq:locmaspardef} allowed us to show that such a contribution does not depend on a particular matter model for the particle.

\begin{figure*}\centering
    \vspace*{-3ex}
    \includegraphics[]{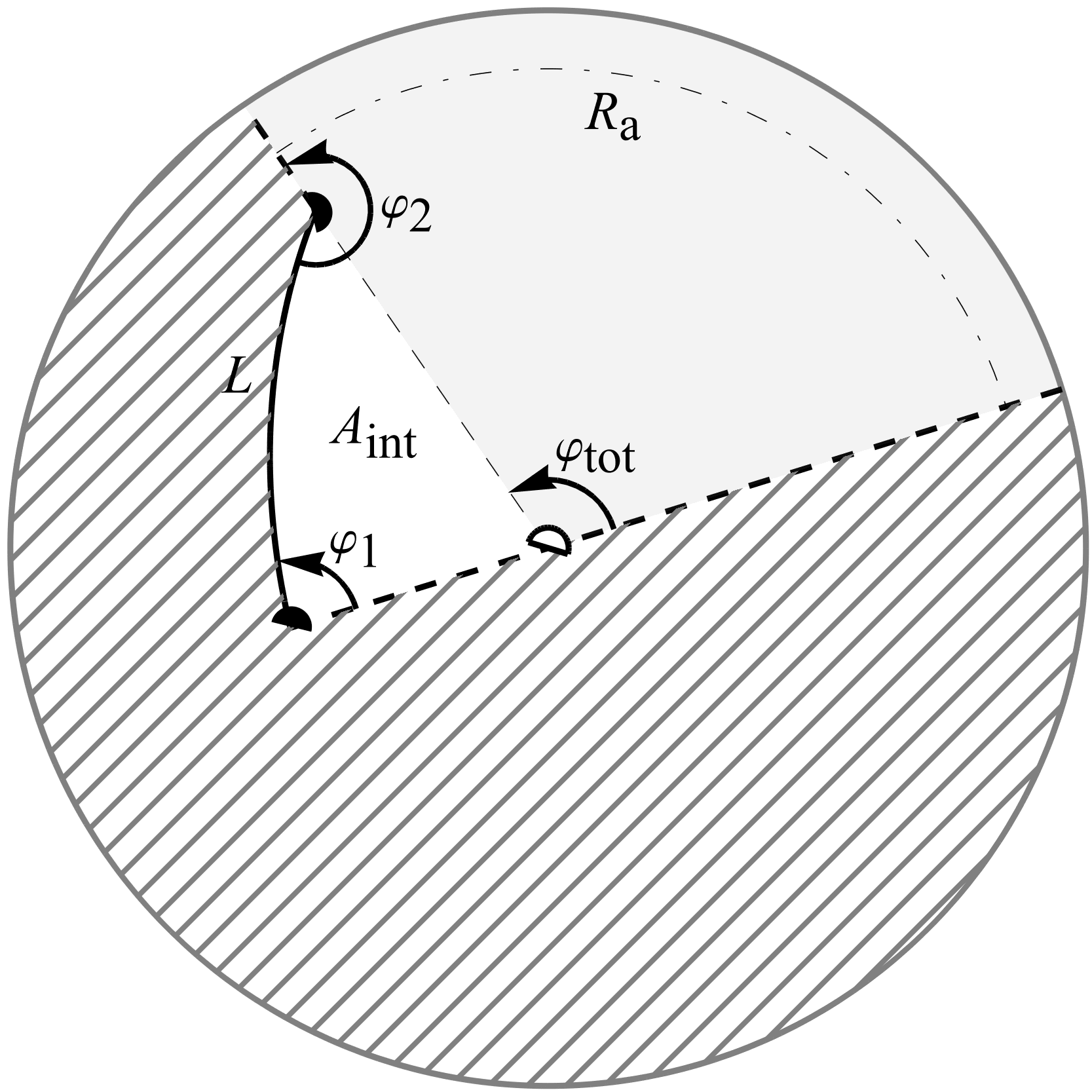}\qquad\qquad\qquad\qquad
    \includegraphics[]{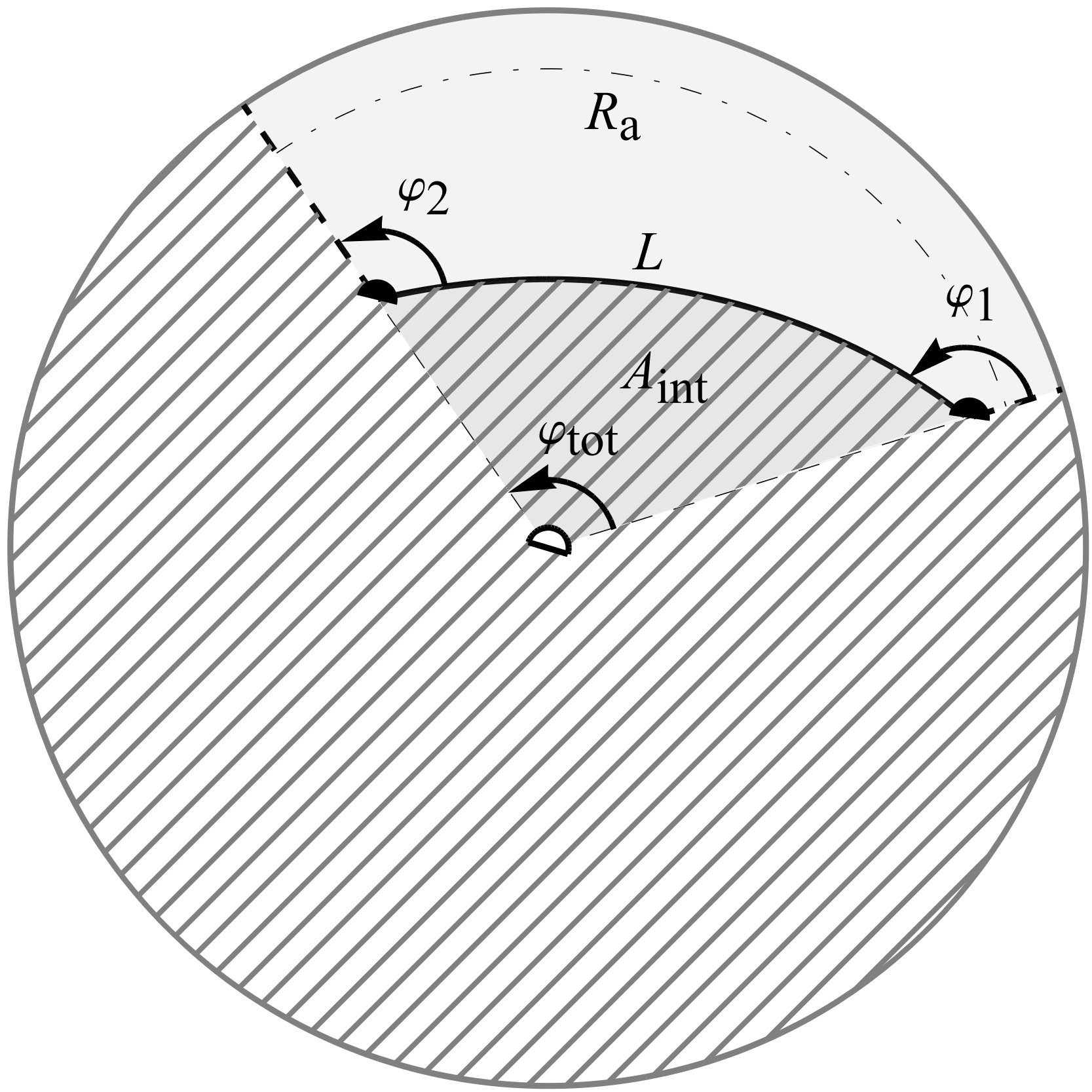}
    \caption{\label{fig:pptot}%
    Space with two point particles attached to a string (left) or a strut (right). The diagrams show only half of the space: the hatched area should be removed, and the cut should be glued with another copy of the diagram. The diagrams thus correspond to the upper half of the diagrams in \Cref{fig:two_particles_string/strut}. The gravity center is in the center of the diagrams. The string or strut, respectively, is represented by thick lines, and its linear energy density is given by its curvature. Their length is $L$. Masses of point particles are given by conical angles $\ph_1$, $\ph_2$. The spacetime is compared with the canonical conical space of the same asymptotic. The canonical space is depicted in gray, and its central particle is represented by the empty semi-disc. The local mass can be calculated and compared for the domain below the asymptotic circumference radius $R_\ai$. The domain of the physical space differs from the domain of the canonical conical space by $2A_\itr$, the factor $2$ referring to the other half, which is glued to form the whole space. For the string (left), the physical space is bigger than the canonical; for the strut, it is smaller.
    }
\end{figure*}


\subsection{BTZ black hole}
\label{ssc:BTZ}

For completeness, we briefly mention the black hole case. For $\mBTZ>1$, the minimal value of the circumference radius~$R$ in the static region of the geometry \eqref{eq:BTZ_metric2} is at the horizon $R_\hor=\ell\sqrt{\mBTZ{-}1}$. For a pure $\Lambda$-vacuum case, no rotation-symmetric ball with a disc topology exists. But we can still evaluate the local mass between two circumference radii, leading to \eqref{eq:BTZ_locmassRR}.

However, one can study a spacetime with a black hole asymptotic \eqref{eq:BTZ_metric2}, ${\mBTZ>1}$, but with a static rotation-symmetric matter model in the central region. In such a case, we can understand formula \eqref{eq:BTZ_locmass} as a local mass of a matter object surrounded by the dark energy. Splitting the ball $\dom{B}_{R_\ai}$ into part $\dom{B}_{R_*}$ containing the matter and part $\dom{B}_{R_*}^{R_\ai}$ containing just the dark energy, we get
\begin{equation}
\begin{aligned}
    \locmass[\dom{B}_{R_\ai}] &= \locmass[\dom{B}_{R_*}] + \locmass[\dom{B}_{R_*}^{R_\ai}]\\
    &= \frac{2\pi}{\kap}\Bigl(1-\sqrt{1{-}\mBTZ{+}\tfrac{R_*^2}{\ell^2}}\Bigr)\\
    &+ \frac{2\pi}{\kap}\Bigl(\sqrt{1{-}\mBTZ{+}\tfrac{R_*^2}{\ell^2}}-\sqrt{1{-}\mBTZ{+}\tfrac{R_\ai^2}{\ell^2}}\Bigr)\;.
\end{aligned}
    \label{eq:BTZ_locmassStarsplit}
\end{equation}
Squeezing the matter object as close to the horizon as possible, $R_*\to R_\hor$, one gets a rather unintuitive result
\begin{equation}
	\label{eq:BTZ_locmassBHsplit}
	\locmass[\dom{B}_{R_\ai}] = m_\hor  
    - \frac{2\pi}{\kap}\sqrt{1{-}\mBTZ{+}\tfrac{R_\ai^2}{\ell^2}}\;.
\end{equation}
The second term clearly describes the dark-energy contribution above the horizon. The first term is the local mass of the black hole horizon obtained as a limit of any matter distribution consistent with the asymptotics \eqref{eq:BTZ_metric2}, ${\mBTZ>1}$. It reads
\begin{equation}
	\label{eq:BTZ_locmasshor}
	m_\hor = \frac{2\pi}{\kap}\;.
\end{equation}
It is, surprisingly, independent of the black hole parameter $\mBTZ$. It suggests that $m_\hor$ represents the mass gap needed for the black hole solution (remember the mentioned mass gap $\mBTZ=1$ in terms of BTZ mass parameter), and further increase of the local mass of the black hole is related to the dark energy stored around the black hole.


\subsection{Two particles on the string}
\label{ssc:TwoParticlesMass}

Next, we will discuss the simplest case without rotational symmetry. In \Cref{ssc:two_particles} we studied two particles connected by a string or strut with a linear density $\mu$. In the following, we have to distinguish the cases of the string ($\mu>0$) and the strut ($\mu<0$).

First, let us consider two particles of positive mass separated by a strut. Half of the spatial section is depicted as a non-hatched region\footnote{As described in \Cref{ssc:two_particles}, the non-hatched region has to be glued on its boundary to a mirrored copy of the same region. \Cref{fig:pptot} corresponds to the upper half of \Cref{fig:two_particles_string/strut}. Remember that angles $\ph_j$ are thus only halves of the vertex angles of the conical singularities, cf.~\eqref{eq:mascondef}, and the boundary curvature $\alpha$ of the strut is only half of the curvature jump which relates to the linear energy density \eqref{eq:surface_stress_energy}. Similarly, any area in diagrams \ref{fig:pptot} enters into relations concerning the full space twice.} in \Cref{fig:pptot} right. This region is aligned with half of the canonical conical space of the same asymptotic -- depicted in gray. Particles of masses $m_1$, $m_2$ correspond to conical singularities characterized by angles $\ph_1$ and $\ph_2$. They are connected by the strut of the linear energy density $\mu$ and the length $L$. The canonical space is characterized by the central particle $m_\tot$ corresponding to the angle $\ph_\tot$. As explained in \eqref{eq:diflocmass}, local masses in both spaces up to an asymptotic radius $R_\ai$ must be the same, $\locmass[\dom{B}_{R_\ai}]=\locmass_\con[{}^\con\!\dom{B}_{R_\ai}]$. Since the local mass is just an integral of the energy density \eqref{eq:locmass}, these masses can be split into individual contributions
\begin{equation}
\label{eq:locmPPstrut}
    \begin{aligned}
        \locmass[\dom{B}_{R_\ai}] &= \locmass_{\text{particle 1}} + \locmass_{\text{particle 2}}\\
        &+ \locmass_{\text{strut}} + \locmass_{\text{dark energy}}\\
        &= m_1 + m_2 + \mu L + \eps_\Lambda A\;,\\
        \locmass_\con[{}^\con\!\dom{B}_{R_\ai}] &= \locmass_{\text{central particle}} +  \locmass_{\text{dark energy}}\\
        &= m_\tot + \eps_\Lambda A_\tot\;.
    \end{aligned}
\end{equation}
Here, we have used that the linear energy density of the strut and the dark energy density $\eps_\Lambda$ are constant, and the local mass is obtained just by multiplication by the length of the strut or the relevant area, respectively.
Comparing, we receive the relation for the central mass of the canonical conical spacetime,
\begin{equation}
\label{eq:locmPPstrut_final}
	m_\tot =  m_1 + m_2 + \mu L - \eps_\Lambda 2A_\itr\;,
\end{equation}
where $2A_\itr=A_\tot-A$ is the interaction area, which has to be removed from the canonical space to match the studied space with particles and the strut. We understand $A_\itr$ as a subdomain of the half of the spatial section, which is depicted in \Cref{fig:pptot} right; therefore, it enters the full spatial balance twice.

A geometric proof of \eqref{eq:locmPPstrut_final} is based on the Gauss-Bonnet theorem. Geometric quantities corresponding to physical ones in \eqref{eq:locmPPstrut_final} are $\ph_\tot$, $\ph_1$, $\ph_2$, $L$, $\alpha$ (the strut curvature), $A_\itr$, and $\spgc_\Lambda$ (the spatial Gauss curvature corresponding to the dark energy). If we apply the Gauss-Bonnet theorem \eqref{eq:GaussBonnet} just to one copy of the domain $A_\itr$, we use $\psi_1=\ph_1$, $\psi_2=\ph_2$, $\psi_\tot=\pi-\ph_\tot$, $\chi=1$, and employ relations \eqref{eq:mascondef}, \eqref{eq:surface_stress_energy} and $\kap\eps_\Lambda = \spgc_\Lambda$, we prove \eqref{eq:locmPPstrut_final}. 

In complete analogy, one can write down the central mass of the canonical conical spacetime for two particles connected by the string, cf.~\Cref{fig:pptot} left. Merely, in this case, one has to \emph{add} the interaction area $A_\itr$ to the canonical space to match the full space with the string.
\begin{equation}
\label{eq:locmPPstring}
	m_\tot =  m_1 + m_2 + \mu L + \eps_\Lambda 2A_\itr\;.
\end{equation}

\section{Two oscillating particles on the string}
\label{sec:oscillator}


\subsection{Boost Killing vectors}
\label{ssc:boostKV}

In the previous sections, we used the static frame adjusted to the global timelike Killing vector $\ts{\partial}_\tau$, and we also explored the rotation symmetry generated by $\ts{\partial}_\ph$. However, the full group of symmetry $SO(2,2)$ of the global AdS spacetime is six-dimensional. We call the remaining generators of the isometry group the boost Killing vectors since they have, in some regions, a structure similar to the boost Killing vectors in Minkowski spacetime.

The boost Killing vector is first characterized by a bifurcation line, or more precisely, a system of periodically repeating bifurcation lines separated by a time shift $\Delta\tau=\pi$. The second characteristic is the two-dimensional axis orthogonal to the bifurcation lines. 

\begin{figure}[t]\centering
    \includegraphics[]{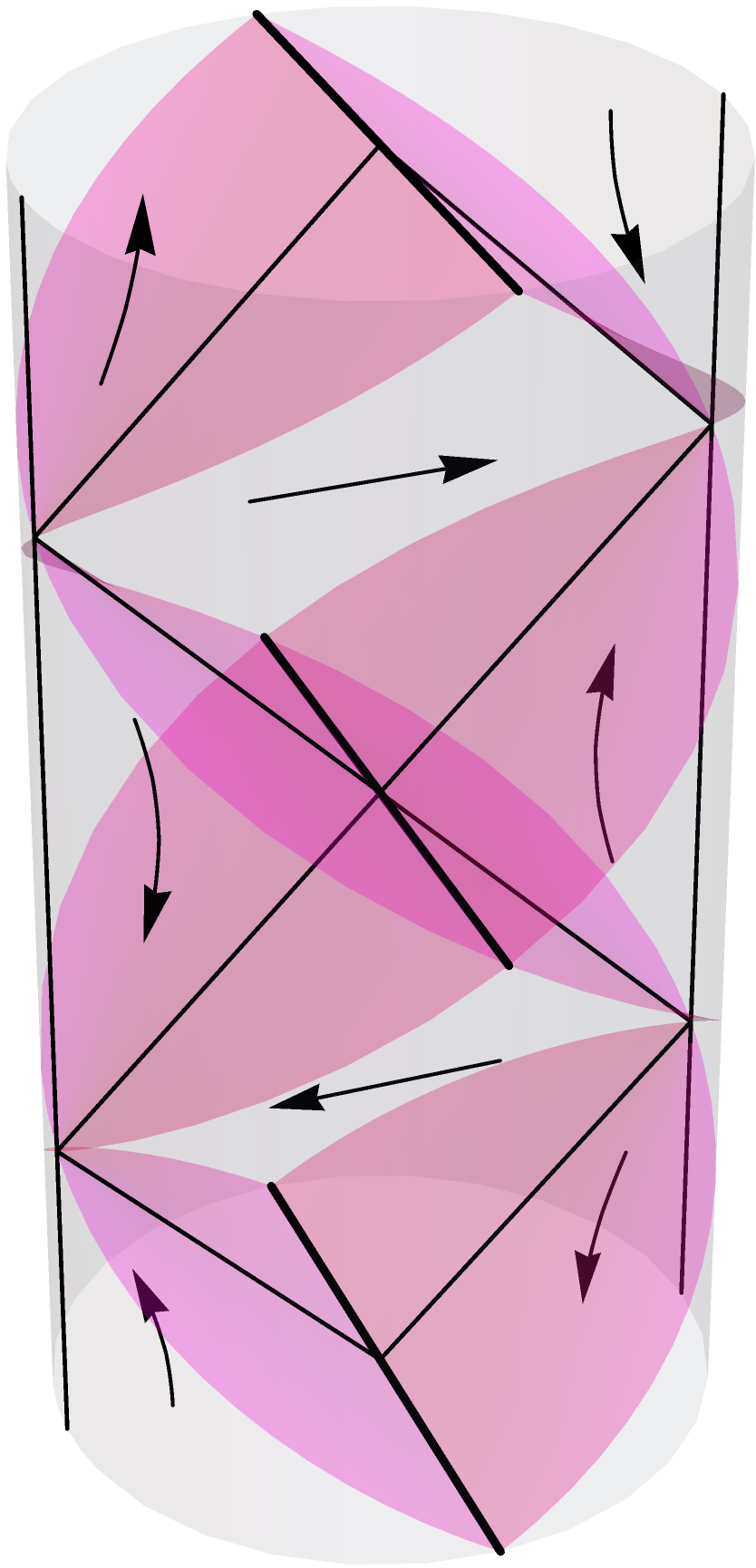}
    \caption{\label{fig:bKVx3D}%
    Boost Killing vector $\bKV_x$. The causal character of the boost Killing vector divides the spacetime into static and non-static domains. They are separated by acceleration horizons -- null surfaces (magenta) intersecting in bifurcation lines. Arrows indicate the direction of the Killing vector.}
\end{figure}

We describe an example of such a Killing vector, the vector $\bKV_x$, in the global static frame \eqref{eq:AdStauchiph}, cf.~\Cref{fig:bKVx3D}. Its bifurcation lines are space-like geodesics given by $\ph=\pm\frac\pi2$ and $\tau=m\pi$ with $m\in\intn$. The axis is given by the $\tau\text{-}\chi$ plane, i.e., $\ph=0$ and $\ph=\pm\pi$ in spherical coordinates. The Killing vector $\bKV_x$ changes its causal character on null surfaces that intersect along the bifurcation lines. They are called acceleration horizons since they form horizons of observable events for uniformly accelerated observers moving along timelike orbits of $\bKV_x$. These are supercritical observers mentioned earlier: their acceleration $a$ is larger than the cosmological one, $a>\frac1\ell$.

It will be more convenient to prescribe the boost Killing vector $\bKV_x$ in the equidistant coordinates $\tau,\rho,\zeta$ aligned to the axis of $\bKV_x$. These coordinates form the same static frame as the spherical coordinates (the time coordinate is not changed), but the spatial coordinate $\rho$ parametrizes the distance from the axis, and $\zeta$ the distance along the axis. These coordinates are closely related to the rotated coordinates $\btau, \bchi, \bph$ for the parameter $\chi_\oi=-\frac\pi2$, cf.~\eqref{eq:S2rotation}, namely
\begin{equation}\label{eq:equidistcoors}
    \begin{aligned}
        \zeta &= \bchi -\frac{\pi}{2}\;,&&&\hskip-1.5ex\sin\zeta &= \sin\chi\,\cos\ph\;,&\hskip-0.5ex\cos\chi &= \cos\zeta\cos\rho\\[-2ex]
        &&&\hskip-2ex\Rightarrow\\[-2ex]
        \rho &= \bph\;,                 &&&\hskip-1.5ex\tan\rho  &= \tan\chi\sin\ph\;,&\hskip-0.5ex\tan\ph &= \cot\zeta\sin\rho\;.
    \end{aligned}
\end{equation}
We also mention a useful relation
\begin{equation}
{\cos\zeta}\,{\sin\rho} = {\sin\chi}\,{\sin\ph}\;.
\end{equation}
The ranges of coordinates are ${\tau\in\realn}$, ${\zeta\in(-\frac\pi2,\frac\pi2)}$, ${\rho\in(-\frac\pi2,\frac\pi2)}$. The spatial infinity is reached for ${\rho=\pm\frac\pi2}$. The AdS metric reads
\begin{equation}
	\label{eq:AdStaurhozeta}
	\ts{g}_{\AdS} = \frac{\ell^2}{\cos^2\zeta\cos^2\rho}\left(
      -\dif \tau^2 + \cos^2\!\zeta\,\dif\rho^2 + \dif\zeta^2
      \right)\,.
\end{equation}
The spatial geometry ${\tau=\const}$ is, of course, hyperbolic. Lines ${\zeta=\const}$ are geodesics orthogonal to the axis ${\rho=0}$. Lines ${\rho=\const}$ are equidistant from the axis and exocycles -- curves of a constant undercritical (${\alpha<\frac1\ell}$) curvature. The distance from the axis is $\ell\arcsinh\tan\rho$, the distance along the line $\rho=\const$ runs as $\frac{\ell}{\cos\rho}\,\arcsinh\tan\zeta$.

In these coordinates, the boost Killing vector $\bKV_x$ has the form
\begin{equation}
	\label{eq:bKVx}
	\bKV_x = \cos\tau\sin\zeta\,\ts{\partial}_\tau+\sin\tau\cos\zeta\,\ts{\partial}_\zeta\,.
\end{equation}
We see that its action leaves $\rho$ unchanged. The shift along $\bKV_x$ parameterized by $\Delta\theta$ can be written explicitly,
\begin{equation}\label{eq:boosttransf}
\begin{aligned}
\tan\tau' 
  &= \frac{\cosh\Delta\theta\,\sin\tau+\sinh\Delta\theta\,\sin\zeta}{\cos\tau}\;,\\
\tan\zeta' 
  &= \frac{\sinh\Delta\theta\,\sin\tau+\cosh\Delta\theta\,\sin\zeta}{\cos\zeta}\;,\\
\rho'&=\rho\;,
\end{aligned}
\end{equation}
where $\tau',\rho',\zeta'$ are coordinates of a shifted point. Let us also mention a useful identity.
\begin{equation}
\frac{\cos\zeta'}{\cos\tau'} = \frac{\cos\zeta}{\cos\tau}\;.
\end{equation}

\begin{figure}[t]\centering
    \includegraphics[]{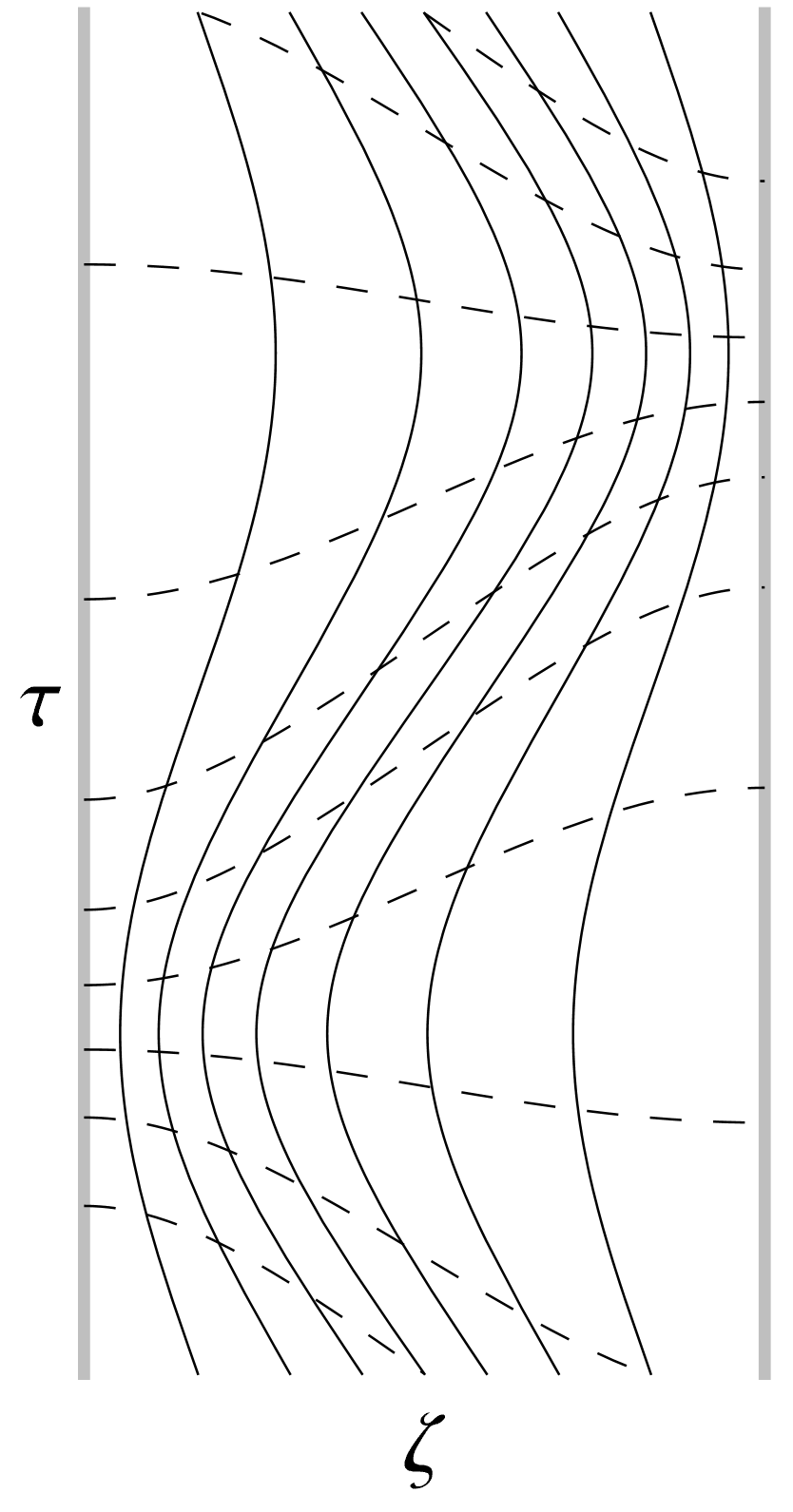}\qquad
    \includegraphics[]{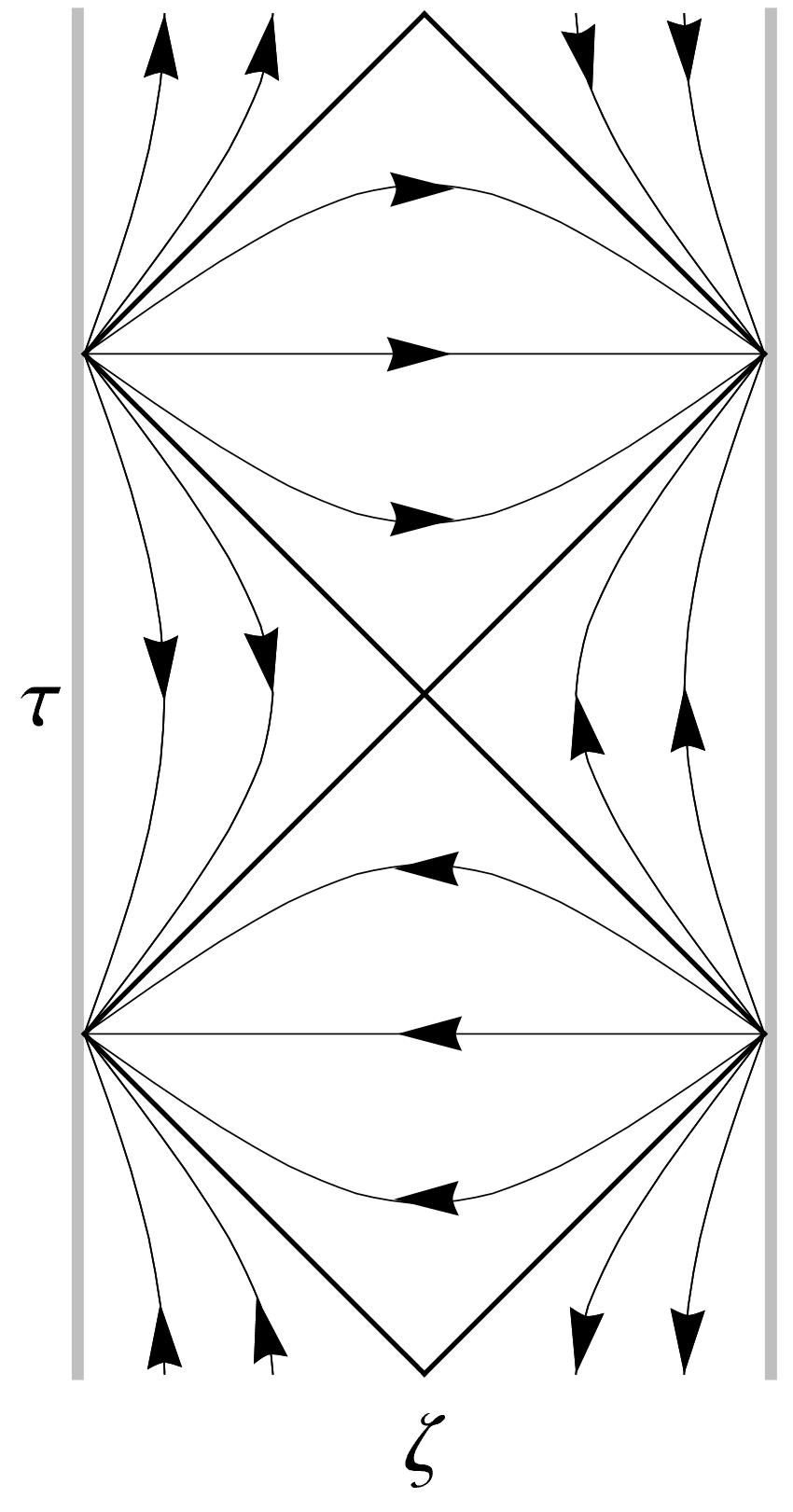}
    \caption{\label{fig:bKVx2D}%
    The boost Killing vector in the surface $\rho=\const$. The action of the Killing vector $\bKV_x$ leaves coordinate $\rho$ unchanged. The left diagram shows the deformation of the coordinate lines $\tau=\const$ (dashed) and $\zeta=\const$ (solid) in the surface $\rho=\const$ under the action of the Killing vector. The right diagram shows horizons (bold lines) and orbits of the Killing vector in the surface $\rho=\const$.}
\end{figure}

The other three boost Killing vectors generating the full isometry group can be obtained by rotating $\bKV_x$ by $\frac\pi2$ in $\ph$ direction (the boost $\bKV_y$) and shifting them by $\frac\pi2$ in $\tau$ direction (boosts $\hat\bKV_x$ and $\hat\bKV_y$). Any Killing vector can thus be written as a linear combination of $\ts{\partial}_\tau$, $\ts{\partial}_\ph$, $\bKV_x$, $\bKV_y$, $\hat\bKV_x$, and $\hat\bKV_y$. 

In particular, we are interested in the boost Killing vector $\bKV_{\ph_*}$ obtained by rotating $\bKV_x$ by the angle $\ph_*$ in the global static frame. Clearly, $\bKV_{\ph_*}=\cos\ph_*\,\bKV_x + \sin\ph_*\,\bKV_y$. The action of such a Killing vector is again given by relations \eqref{eq:boosttransf}, only the equidistant coordinates $\tau,\rho,\zeta$ have to be adjusted to the axis in the direction $\ph=\ph_*$.

Coordinate lines $\tau$ and $\zeta$ deformed by the action of the boost Killing vector are depicted in \Cref{fig:bKVx2D}. The action of the boost Killing vector looks the same in all surfaces $\rho=\const$, since $\rho$ remains unchanged by the action. Deformed timelike $\tau$-lines can be understood as worldlines of test observers. The original orbits of the Killing vector $\ts{\partial}_\tau$ represent subcritically accelerated static observers in the original static frame. The congruence of orbits obtained by the action of \eqref{eq:boosttransf} is isometric, so these orbits also represent subcritically uniformly accelerated observers. However, they are not static in the original static frame. In the 2+1 view, we describe them as periodically moving. For example, although the central orbit of deformed congruence is geodesic, static observers see this orbit in oscillating motion under the influence of the cosmological attraction. All other orbits are moving under the influence of a constant force combined with the cosmological attraction.


\subsection{Single particle oscillating on the string}
\label{ssc:oneboostedpart}

\begin{figure}[t!]\centering
    \includegraphics[]{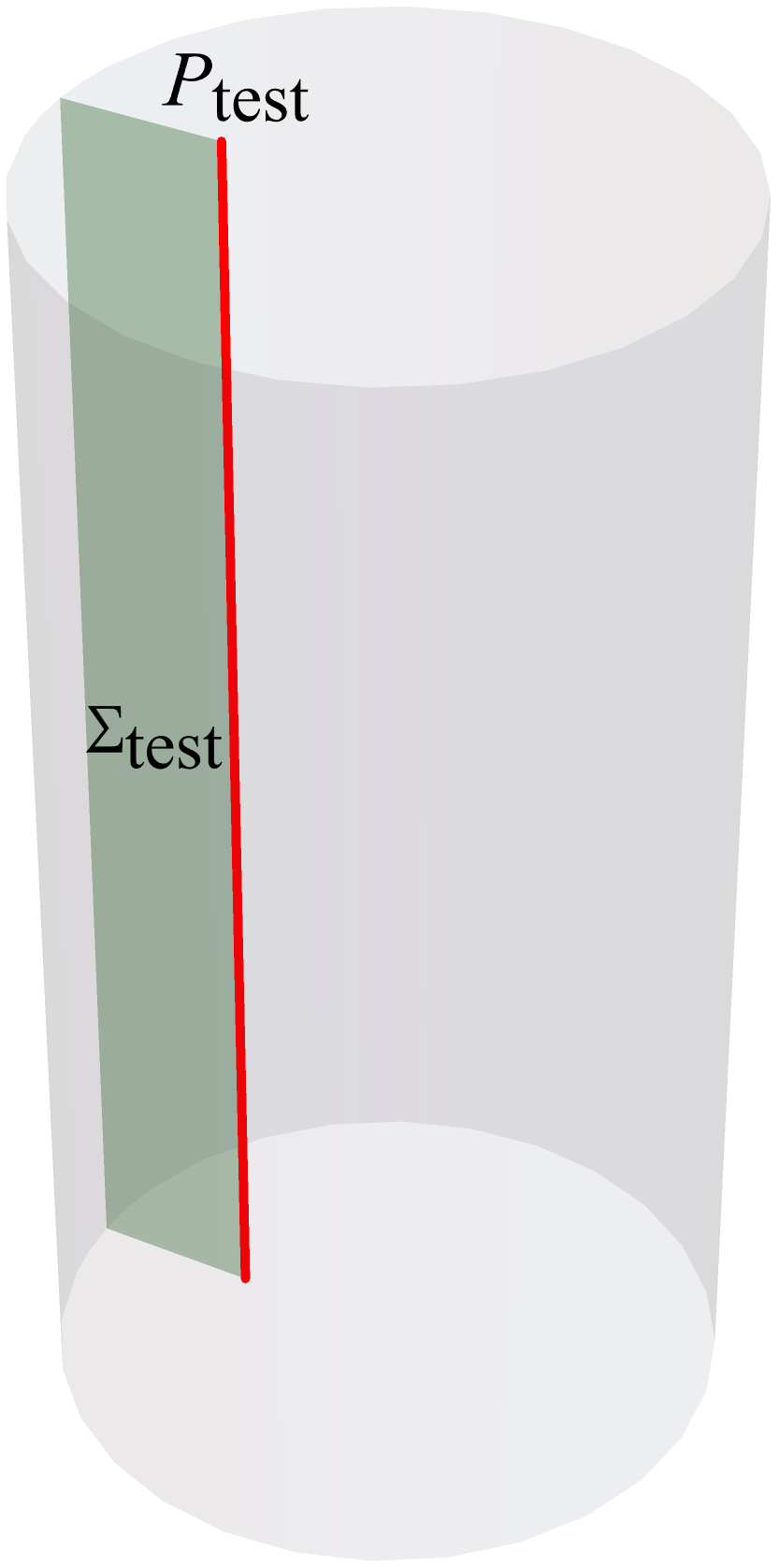}\qquad
    \includegraphics[]{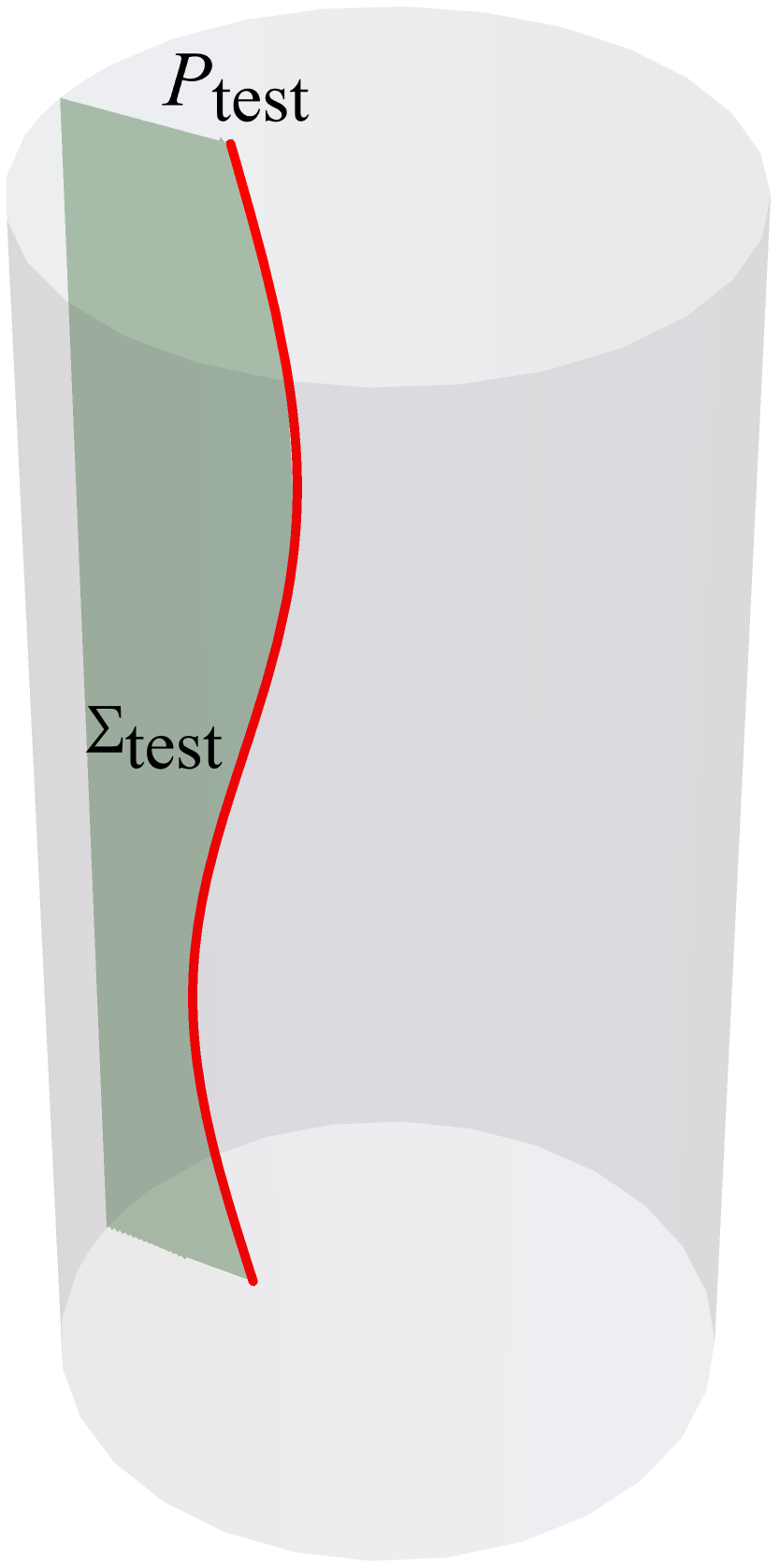}
    \caption{\label{fig:3d_test}%
    Spacetime diagram of the AdS universe with test particle $P_\tst$ (red) hanging on the test string $\Sigma_\tst$ (green) in the static frame $S_\oi$ (left) and the same situation with respect to a boosted frame $S$ (right). In the boosted frame, the particle exhibits movement periodic in time (exactly one period is depicted).}
\end{figure}

Now we want to promote the oscillating accelerated test observer into a fully interacting massive particle hanging on the string, oscillating with respect to the static frame. The case of the test particle is shown in \Cref{fig:3d_test}. The oscillating test particle on a test string is obtained by the boost along $\bKV_x$ of the static test particle.

\begin{figure}[t!]\centering
    \includegraphics[]{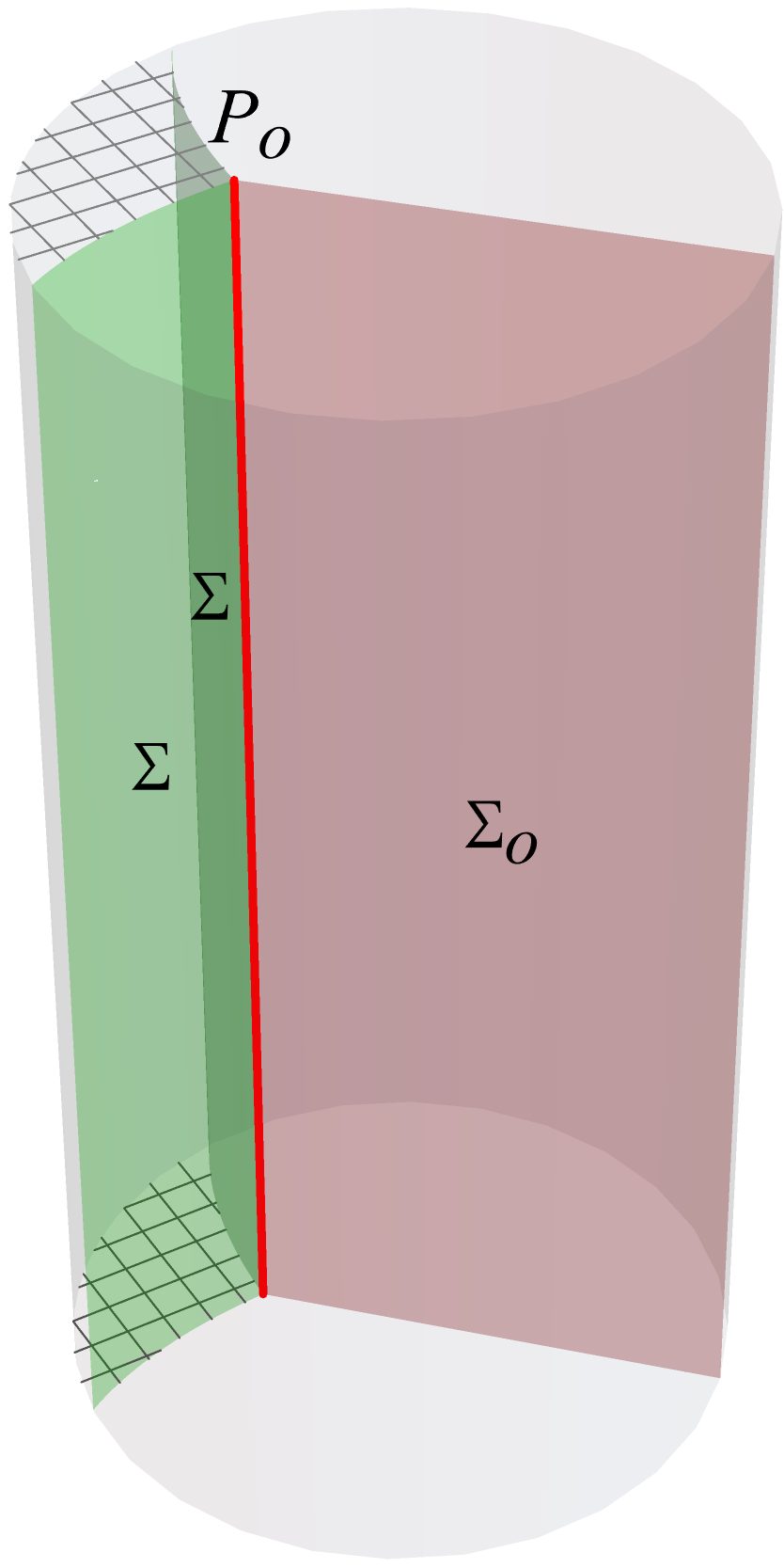}\qquad
    \includegraphics[]{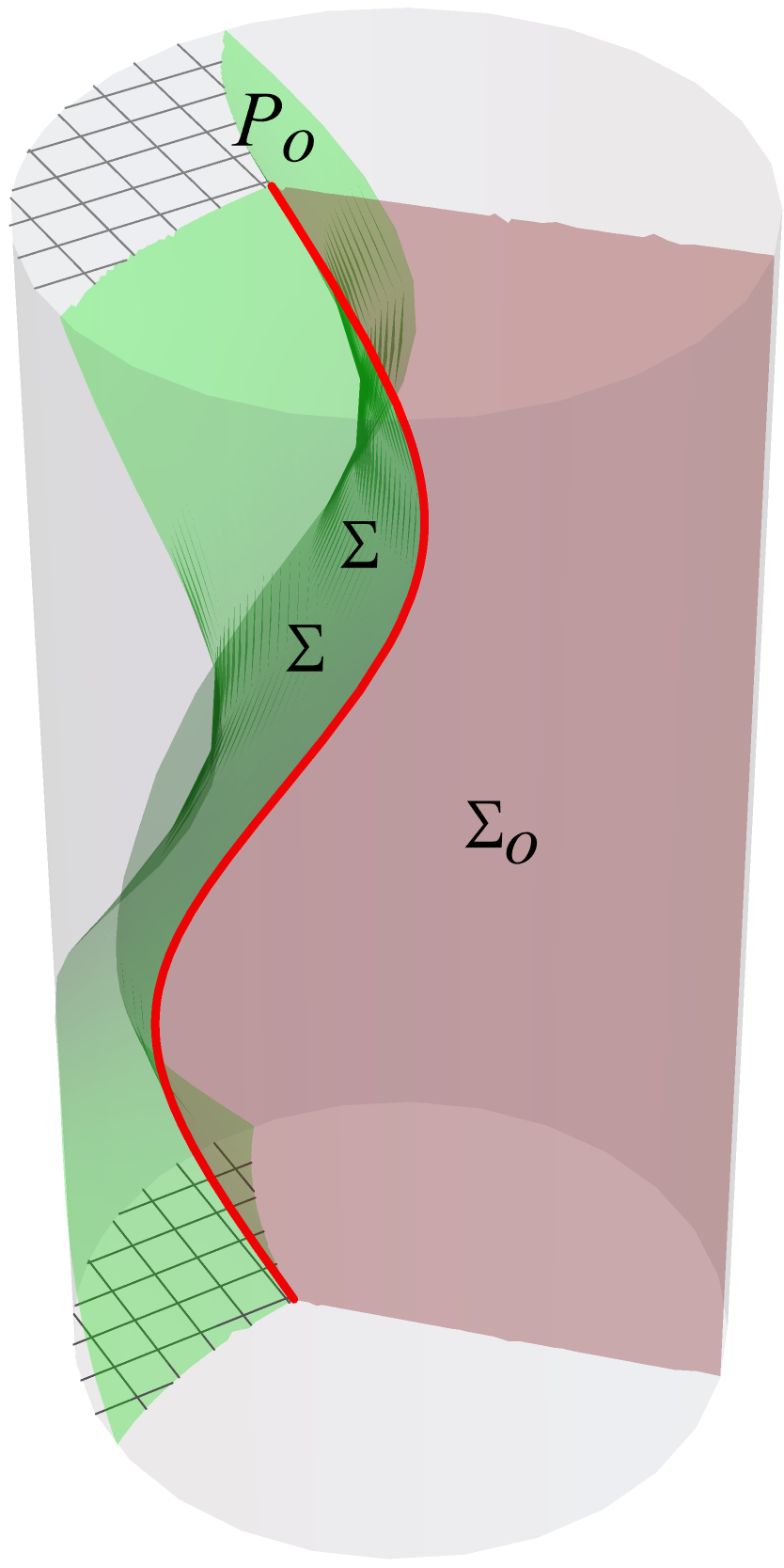}
    \caption{\label{fig:3d_oscpartstring}%
    Left: Spacetime diagram of the massive particle~$P_\oi$ (red) hanging on the string $\Sigma$ (green) in the static frame $S_\oi$ (cf.~\Cref{fig:string_strut}). Right:~The same situation with respect to a boosted frame $S$. The green surfaces should be identified; the hatched region does not belong to the spacetime. In the boosted frame~$S$, the particle exhibits oscillating movement.}
\end{figure}

Similarly, we can boost the static massive particle hanging on the string in the same direction discussed in \Cref{ssc:string}. The resulting oscillating particle attached to a properly moving string is depicted in \Cref{fig:3d_oscpartstring}. 

However, we should stress that this does not generate a qualitatively new system. Essentially, we only view the static particle on the string from a different static frame~$S$, which is related to the original frame~$S_\oi$ by the boost along $\bKV_x$. The AdS cosmological attraction causes the motion of the boosted particle to remain bounded in space and looks like an oscillating motion.

\begin{figure}
    \centering
        \includegraphics[]{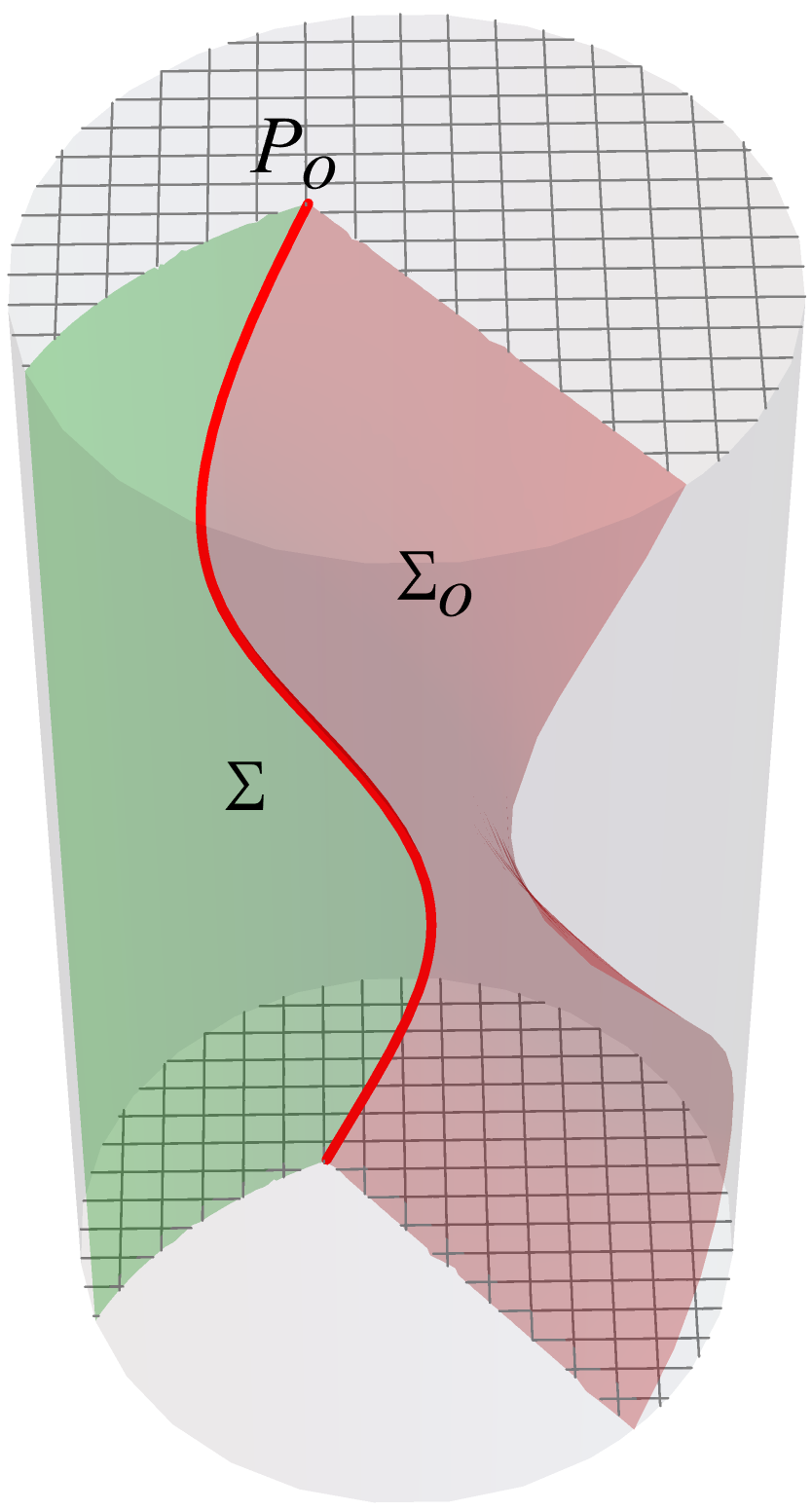}\qquad
        \includegraphics[]{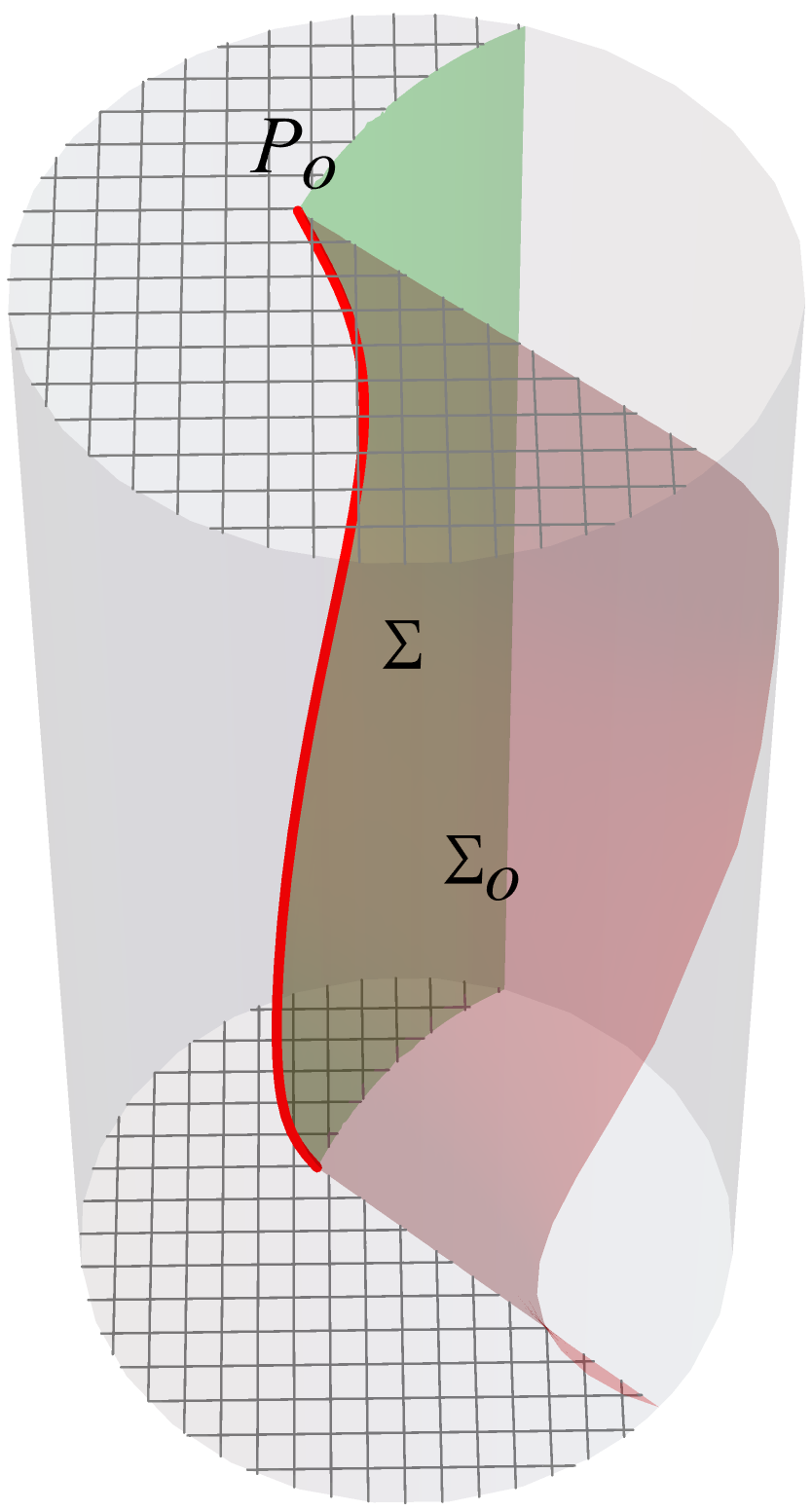}
    \vspace{-2ex}
    \caption{\label{fig:3d_one_particle}%
    Spacetime of the massive particle $P_\oi$ (red curve) hanging on the string $\Sigma$ (green surface) is shown in boosted frames $S_+$ and $S_-$. Two halves of the spacetime ${\ph>0}$ (left) and ${\ph<0}$ (right) are boosted along different boost Killing vectors. Regions indicated by hatching are excluded from the spacetime. The boundaries of both halves, composed by the axis $\Soax$ (red) and the string $\Sigma$ (green), should be identified. On the left, the region ${\ph>0}$ is shown in the frame $S_+$ obtained by the boost along $\bKV_{\ph_+}$ with respect to the original static frame $S_\oi$. The right diagram depicts the region ${\ph<0}$ in the frame $S_-$ obtained by the boost along $\bKV_{\ph_-}$. The surfaces of the string (green) remain unchanged under the corresponding boost. The axis $\Soax$ of the frame $S_\oi$ is moving in frames $S_+$ and $S_-$. However, it is of vanishing extrinsic curvature, and, therefore, the gluing at the $\Soax$ is transparent.}
    \vspace*{2ex}
        \includegraphics[]{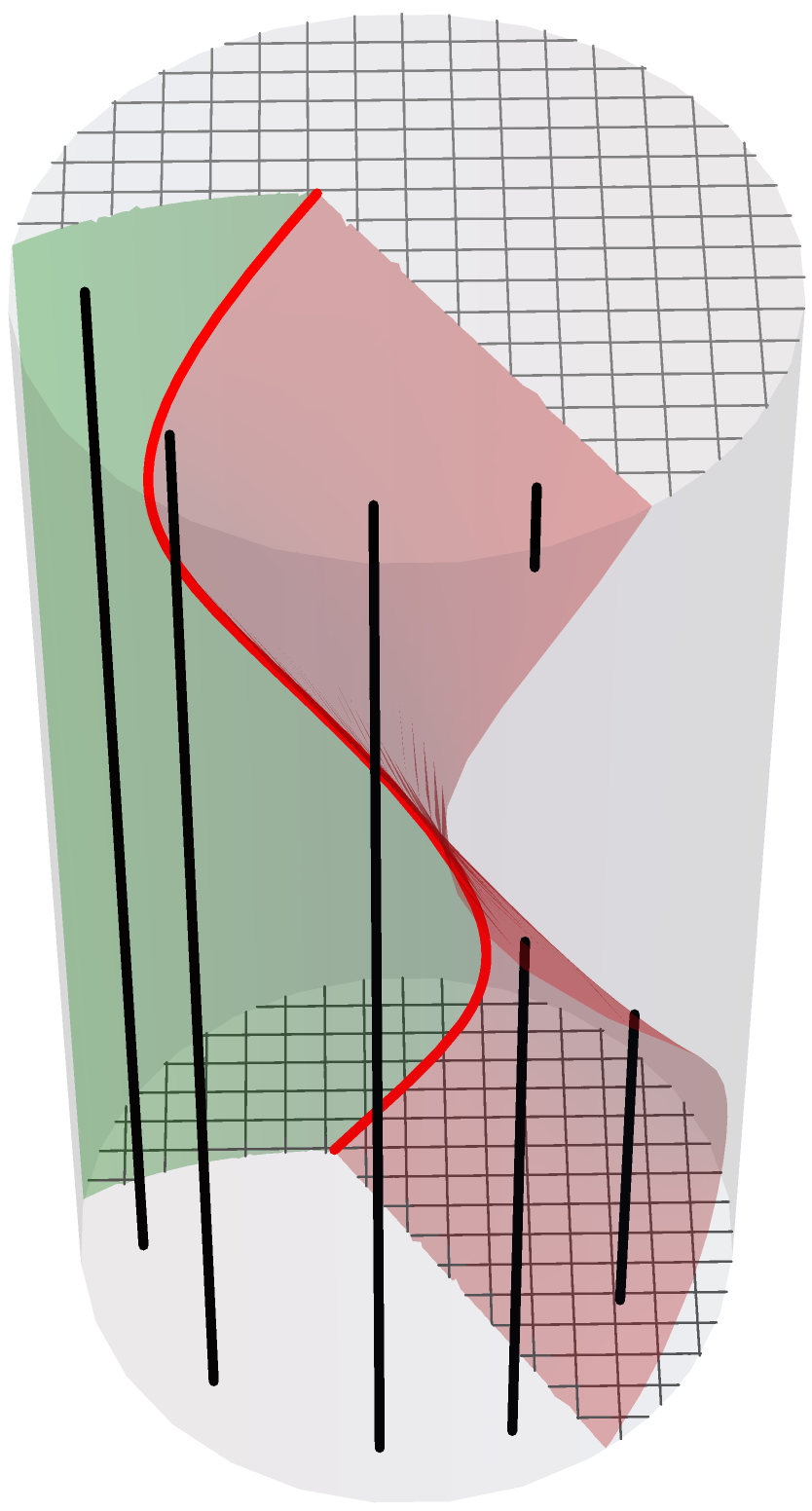}\qquad
        \includegraphics[]{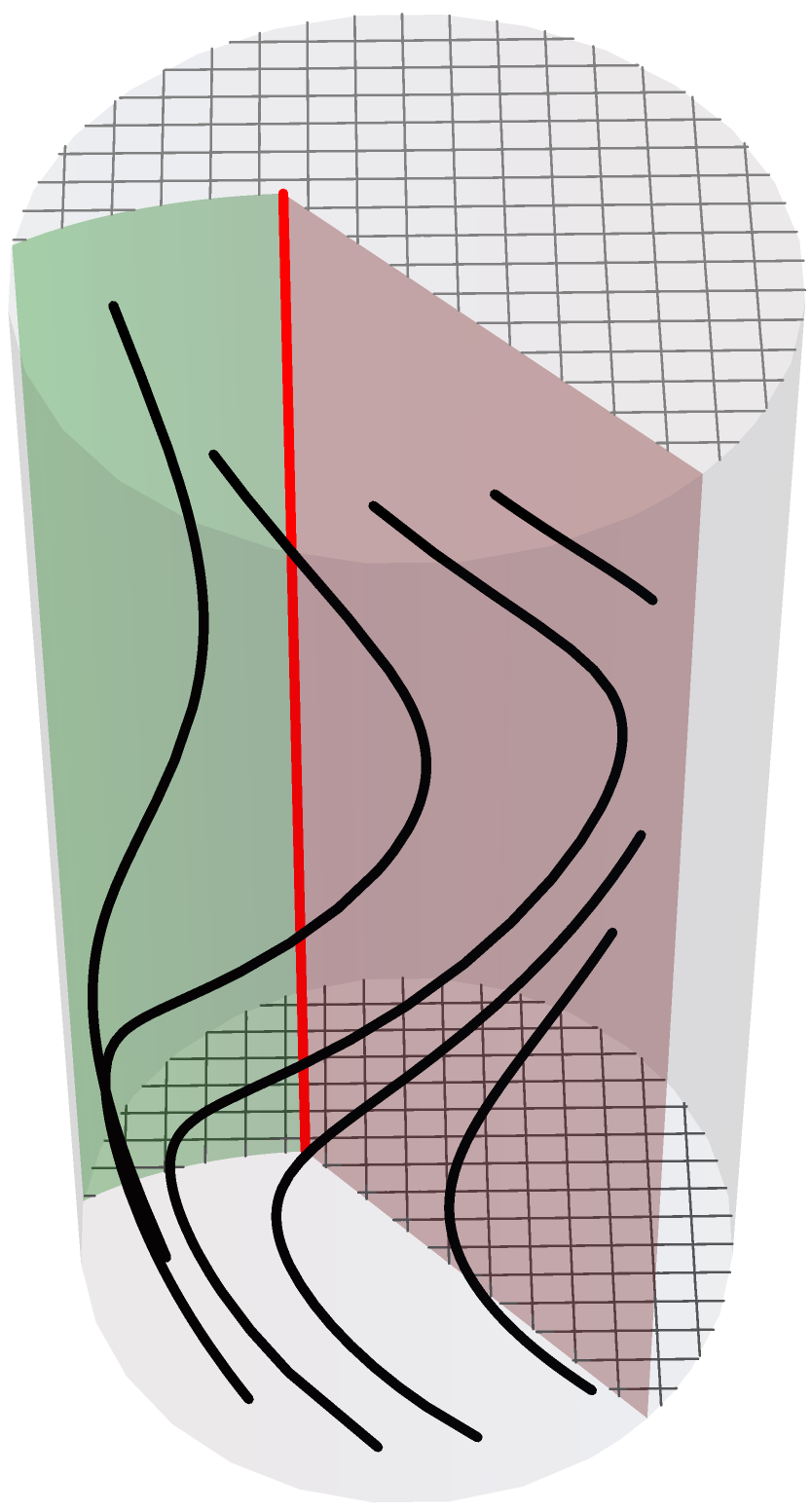}
    \vspace{-2ex}
    \caption{\label{fig:3d_boosted_observers}%
    The half of spacetime of the massive particle $P_\oi$ (red curve) hanging on the string $\Sigma$ (green surface) is depicted in the boosted static frame $S_+$ (left) and in the original static frame $S_\oi$ (right). The hatched region is excluded. On the right, in the frame $S_\oi$, the particle, the string, and the axis (red surface) are static, not moving. On the left, in the frame $S_+$, the axis of $S_\oi$ (red surface), as well as the particle, are oscillating in time. The surface of the string remains static in $S_+$ except for its boundary, where it is attached to the particle. The static observers of the frame $S_+$ are shown in black in both diagrams. Of course, in frame $S_\oi$ (on the right), they appear moving and crossing the axis $\Soax$.} 
\end{figure}

Next, we explore other boosts of the same system. As we discussed in \Cref{ssc:string}, spatial curves representing the static string are exocycles. The curves $\bph=\pm\ph_\oi$, which after gluing form the string, are equidistant to the radial geodesics $\ph=\ph_\pm$, where 
\begin{equation}
    \label{eq:axis_exocycle_rotation_angle}
    \tan\ph_\pm=\pm{\tan\ph_\oi\cos\chi_\oi}\;,
\end{equation}
respectively, see \eqref{eq:asymptotic_angle} in \Cref{apx:string_curvature}. Let us focus first on the half of spacetime ${\ph>0}$. The string curve $\bph=+\ph_\oi$ is equidistant to the line $\ph=\ph_+$. We perform the boost $\bKV_{\ph_+}$ along this direction in the chosen half of spacetime. It leaves the string surface $\Sigma$ (given by $\bph=\ph_\oi$) unchanged. However, the boost deforms the worldline of the particle and the axis $\Soax$ ($\ph=0,\pm\pi$). The resulting region is depicted in \Cref{fig:3d_one_particle}. Similarly, we perform the boost (with the same boost parameter) along the Killing vector $\bKV_{\ph_-}$ in the other half ${\ph<0}$ of the spacetime. We glue both halves together in the same way as before the boosts.

Again, we can understand the resulting situation as merely a different visualization of the original system. Only this time, we view the particle with the string from two different static frames. After performing the boost along $\bKV_{\ph_+}$, we represent half of spacetime ${\ph>0}$ in the frame $S_+$; similarly, we view half of spacetime ${\ph<0}$ in the frame $S_-$. The relation of frames $S_\oi$ and $S_+$ is shown in \Cref{fig:3d_boosted_observers}, and a similar relation is between $S_\oi$ and $S_-$. However, since the identification of both halves of the spacetime remains the same, independent of the boosts, the resulting spacetime is unchanged.


\subsection{Two particles on a finite string}
\label{ssc:twoboostedpart}

\begin{figure}[t]\centering
    \includegraphics[]{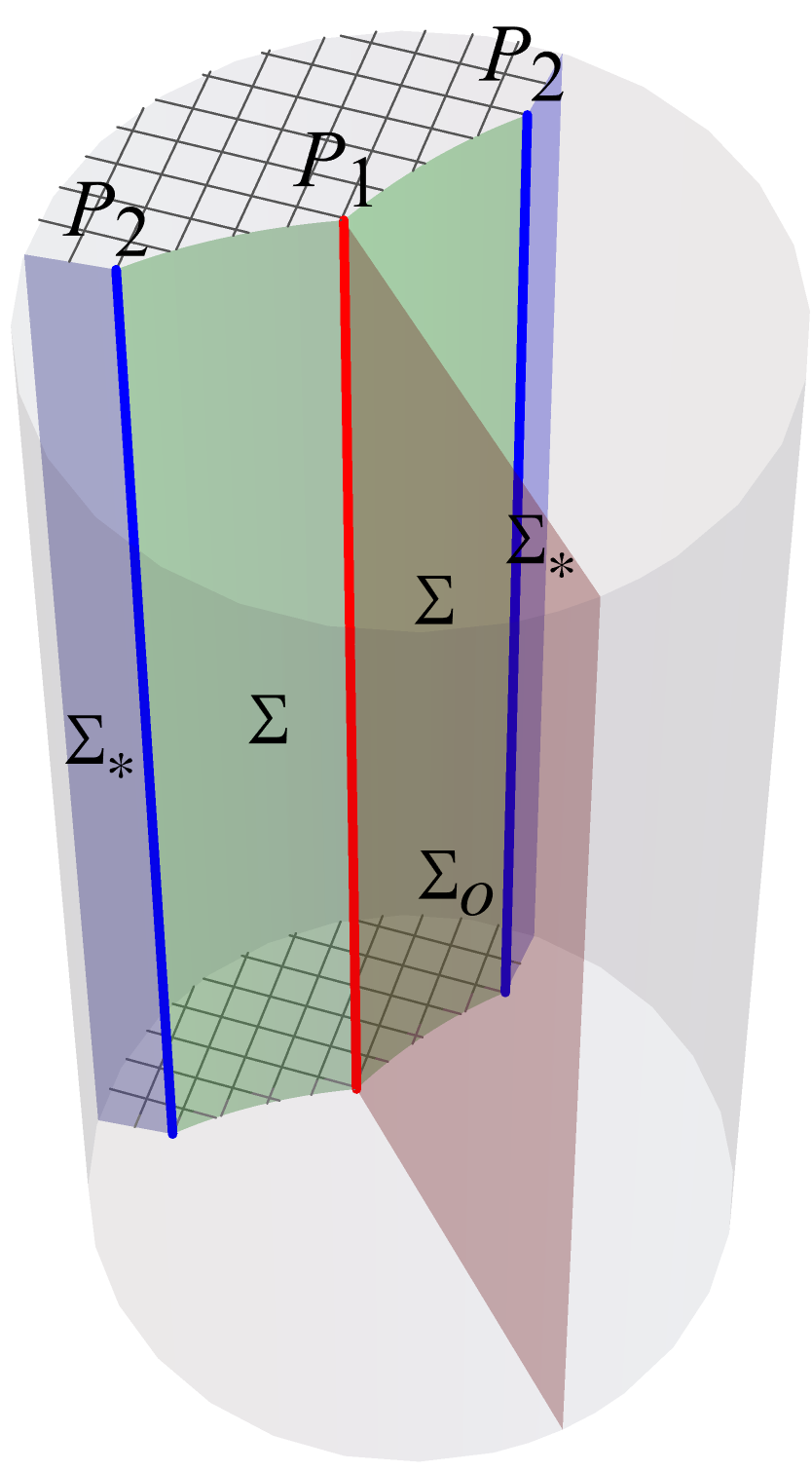}
    \caption{\label{fig:3d_twostatparts}%
    The spacetime diagram of two particles $P_1$ and $P_2$ connected by a string $\Sigma$ corresponding to the spatial diagram in \Cref{fig:two_particles_string/strut} (left). The red curve represents the worldline of the positive massive particle $P_1$, the blue line that of particle $P_2$ with a negative mass. The green surfaces $\Sigma$ represent the string, the red surface $\Soax$ ($\ph=0,\pm\pi$), and the blue surfaces $\Sbax$ ($\ph=\pm\ph_\tot$) depict flat radial surfaces of vanishing extrinsic curvature. The hatched region is excluded from the spacetime. Both string surfaces $\Sigma$ and both surfaces $\Sbax$ should be identified, respectively. Everything is static with respect to the static frame $S_\oi$.\\[-5ex]}   
\end{figure}

Until now, we have done just exercises in viewing the same system from different static frames. Now we finally construct a qualitatively different system: two particles oscillating relative to each other.

First, \Cref{fig:3d_twostatparts} depicts the spacetime of two static particles connected by the string discussed in \Cref{ssc:two_particles}, cf.~also \Cref{fig:two_particles_string/strut} (left). Particle~$P_1$ lying on the axis~$\Soax$ of the frame~$S_\oi$ is attached to the string~$\Sigma$. The string, starting at particle~$P_1$, is terminated by particle~$P_2$ along the intersection with the radial flat plane~$\Sbax$.

Now we perform the same operation -- the intersection of the string $\Sigma$ with a radial plane $\Sbax$ -- however, applied to the system boosted along the Killing vectors $\bKV_{\ph_\pm}$. More precisely, we take half of spacetime ${\ph>0}$ boosted along $\bKV_{\ph_+}$, \Cref{fig:3d_one_particle} right. The equidistant surface of the string worldsheet remains unchanged by the boost. We intersect this surface by the flat plane $\Sbax$ going through the origin, but in the boosted frame $S_+$, \Cref{fig:3d_boosted_1}. Next, we perform the same procedure in the second half ${\ph<0}$ of the spacetime and glue both halves together. 

\begin{figure}[t]\centering
    \includegraphics[]{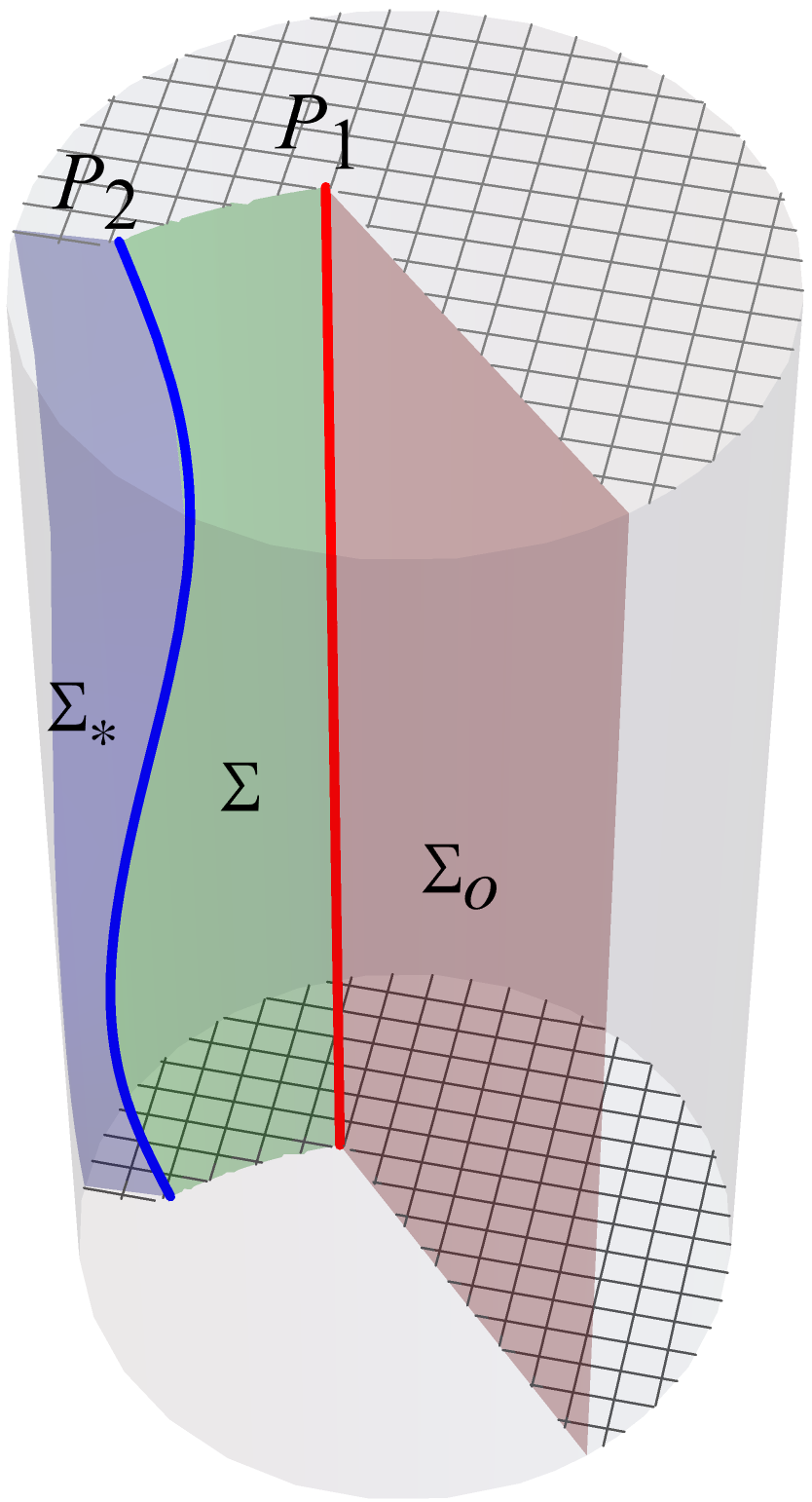}\qquad
    \includegraphics[]{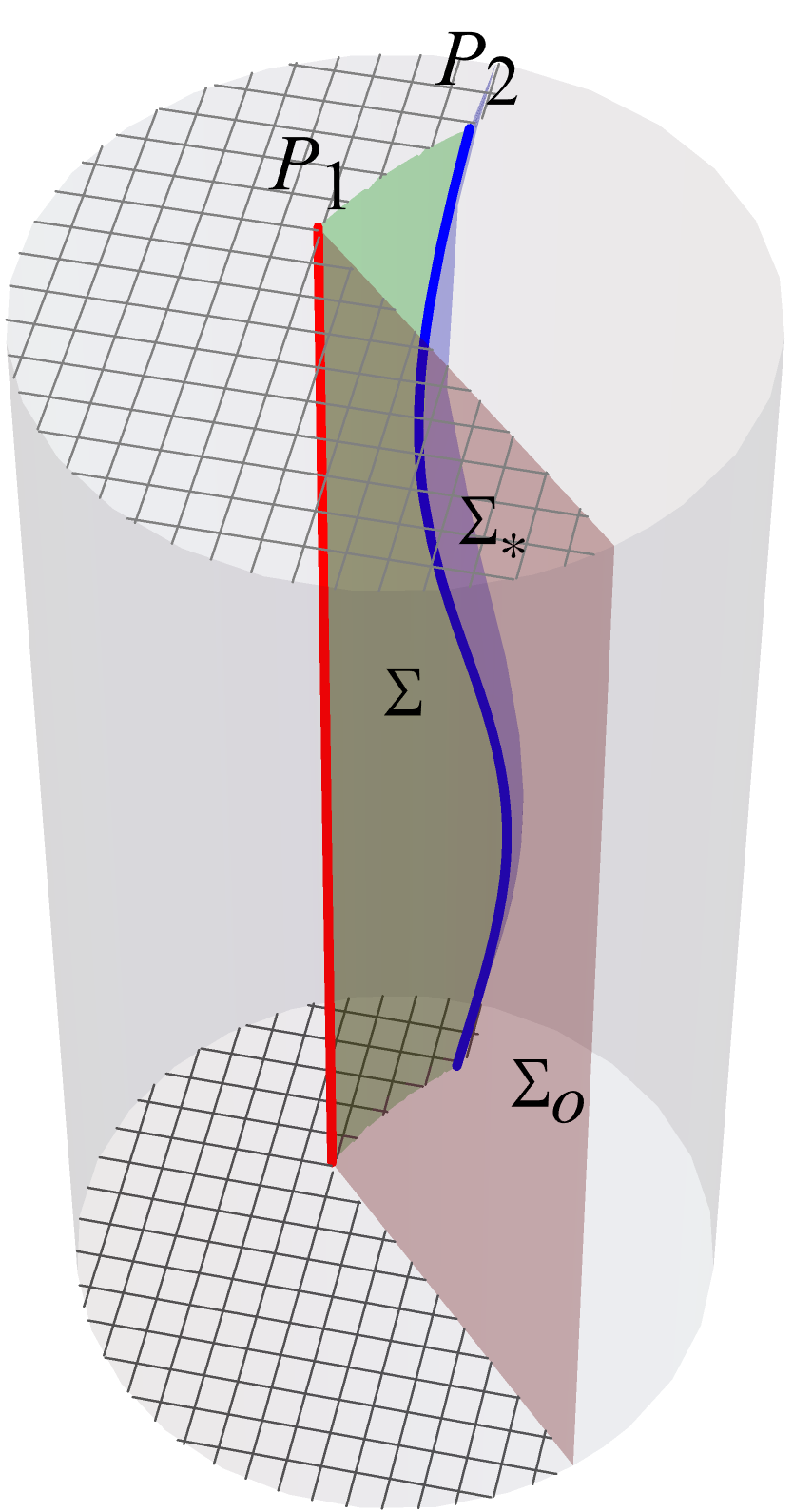}
    \vspace{-1ex}
    \caption{\label{fig:3d_boosted_0}%
    Spacetime diagrams of static particle $P_1$ (red curves) connected by the string $\Sigma$ (green surfaces) to moving particle $P_2$ (blue curves). Half of the spacetime ${\ph>0}$ is shown on the left, another half ${\ph<0}$ on the right. Both diagrams are depicted in the original static frame $S_\oi$. The hatched regions are excluded from the spacetime, and both halves should be identified along the \mbox{$S_\oi$-axis}~$\Soax$ (red surfaces), string~$\Sigma$ (green surfaces), and \mbox{$S_\bst$-axis}~$\Sbax$ (blue surfaces). Particle~$P_1$ and \mbox{$S_\oi$-axis}~$\Soax$ are at rest in frame $S_\oi$. Particle~$P_2$ and \mbox{$S_\bst$-axis}~$\Sbax$ are moving in frames $S_\oi$ but at rest in $S_\bst$. The string is static in $S_\oi$, as well as in $S_\bst$. The system represents two particles oscillating with respect to each other under the influence of string tension and cosmological attraction. The same situation from the point of view of the boosted frame $S_\bst$ is shown in \Cref{fig:3d_boosted_1}.}
\end{figure}

\begin{figure}\centering
    \includegraphics[]{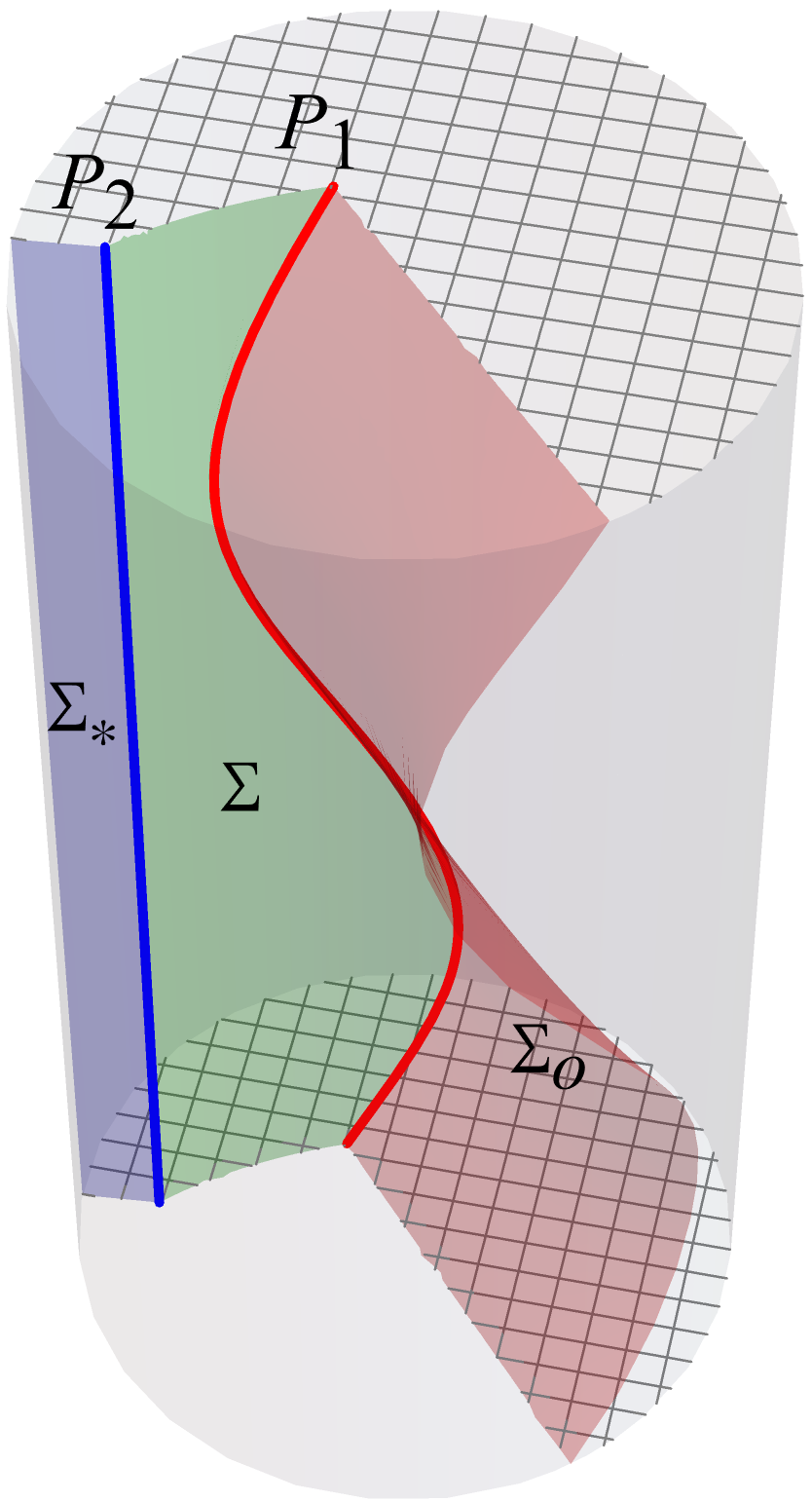}\qquad
    \includegraphics[]{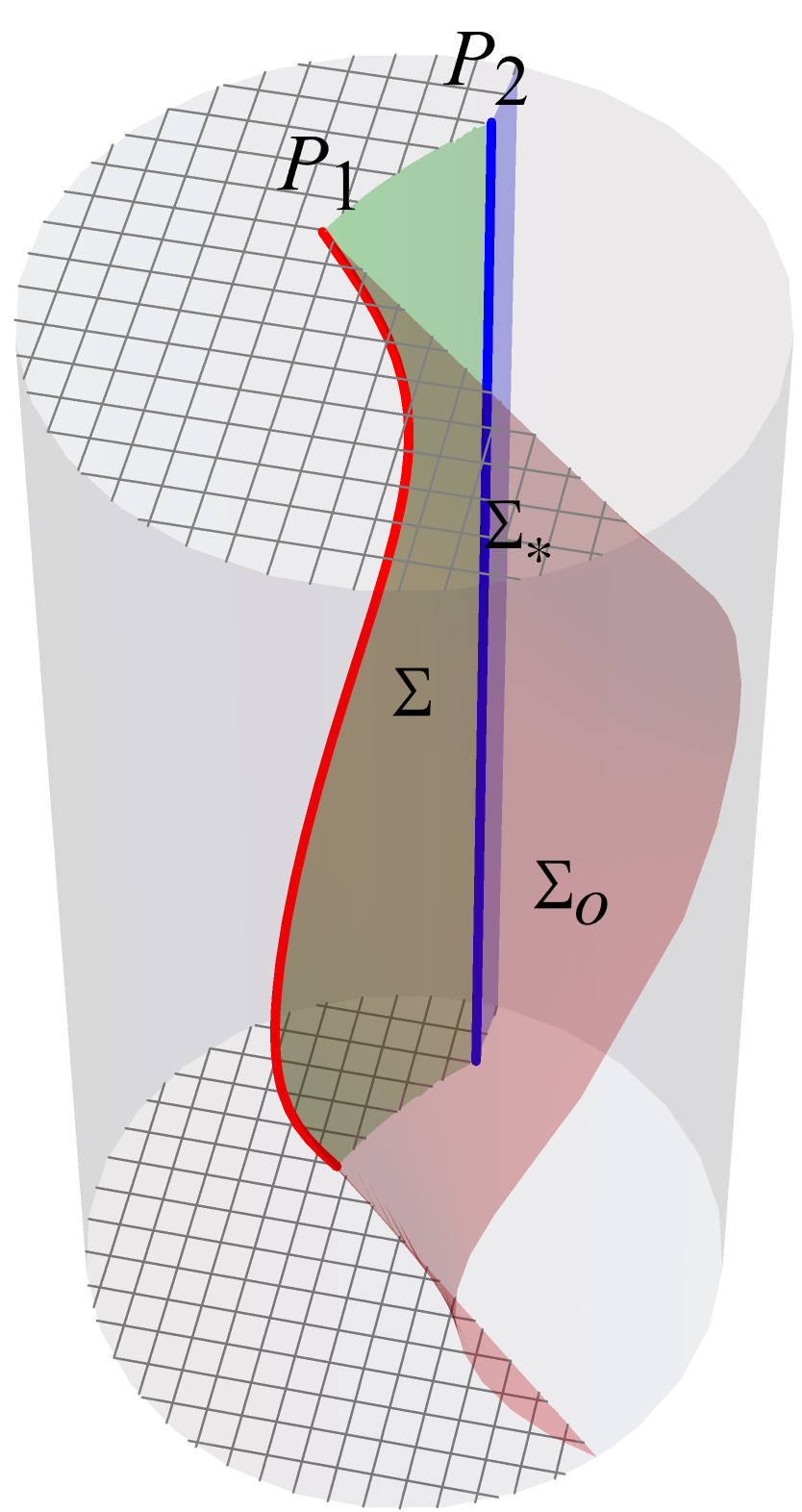}
    \vspace{-1ex}
    \caption{\label{fig:3d_boosted_1}%
    The same situation as in \Cref{fig:3d_boosted_0} shown in boosted frames $S_+$ and $S_-$. The left diagram shows half of the spacetime ${\ph>0}$ in the boosted frame $S_+$. The right diagram shows half of the spacetime ${\ph<0}$ in the boosted frame $S_-$. Compare with one particle on the string in \Cref{fig:3d_one_particle}. The hatched regions are excluded from the spacetime, and both halves should be identified along the \mbox{$S_\oi$-axis}~$\Soax$ (red surfaces), string~$\Sigma$ (green surfaces), and \mbox{$S_\bst$-axis}~$\Sbax$ (blue surfaces). Particle $P_1$ (red curves) and $S_\oi$-axis $\Soax$ are moving in frames $S_+$ and $S_-$. Particle $P_2$ (blue curves) and surfaces $\Sbax$ are static in respective frames $S_+$ and $S_-$. These frames can be smoothly joined into one frame $S_\bst$ through the surface $\Sbax$. The surface $\Sbax$ can be considered as the axis of the frame~$S_\bst$. The visualization in the original static frame $S_\oi$ is shown in \Cref{fig:3d_boosted_0}.}
    \vspace*{1ex}
        \includegraphics[]{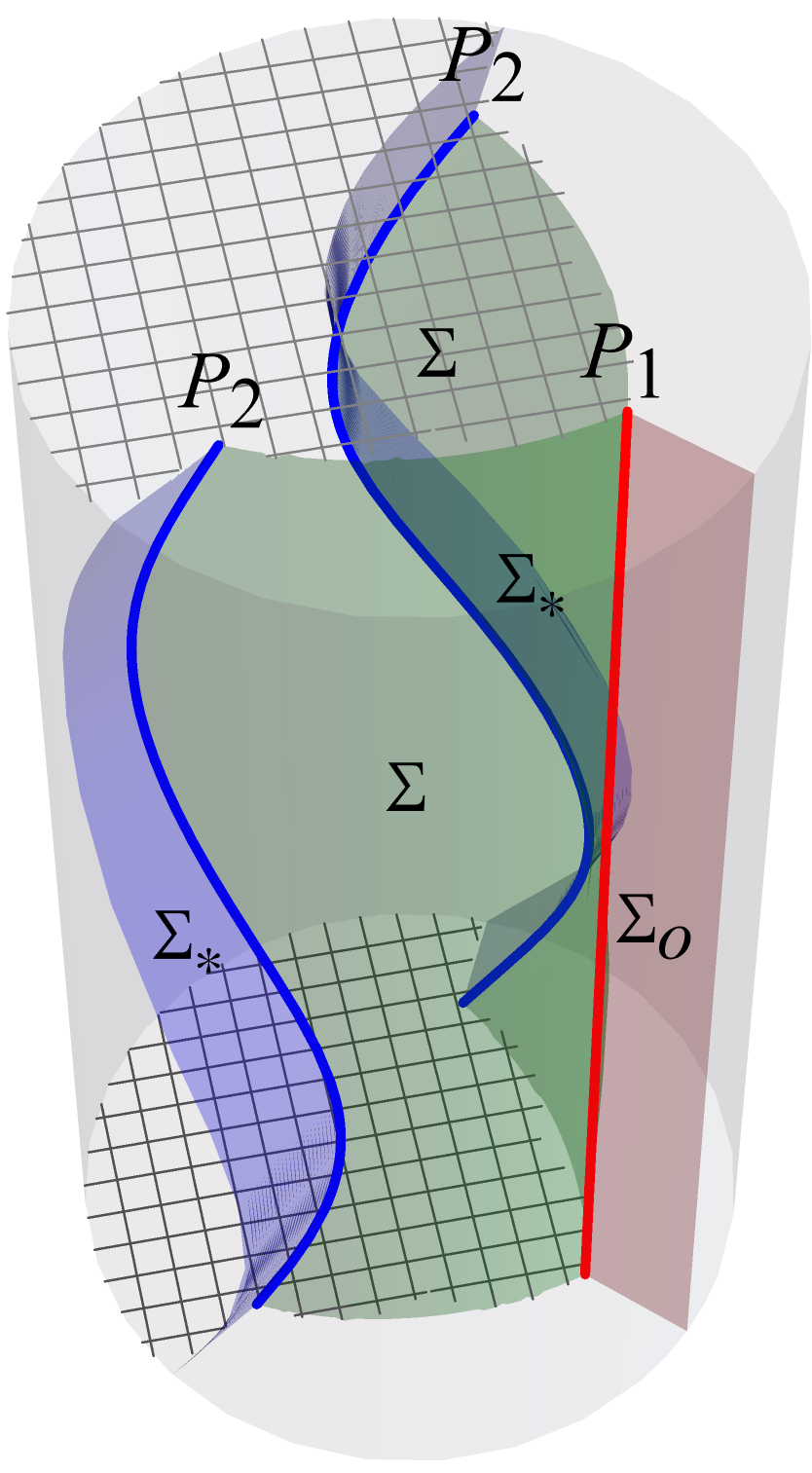}\qquad
        \includegraphics[]{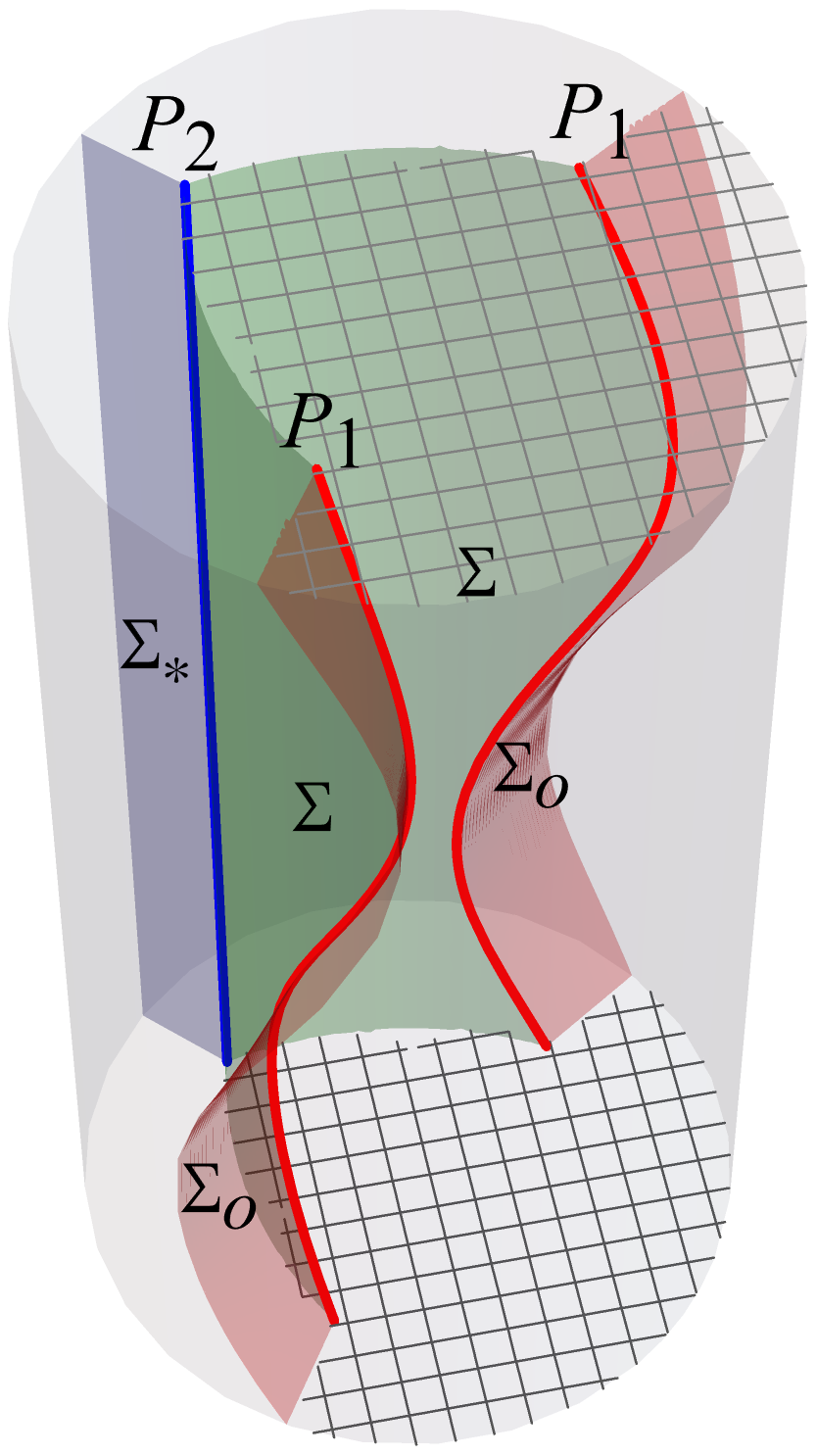}
    \vspace{-2ex}
    \caption{\label{fig:3d_oscpart}%
    The spacetime of two oscillating positively massive particles separated by the strut. It describes a similar system as \Cref{fig:3d_boosted_0,fig:3d_boosted_1} in which, however, particle $P_2$ has negative mass. Left:~The spacetime depicted in the static frame~$S_\oi$. It corresponds to \Cref{fig:3d_boosted_0}, only with both diagrams glued together along the $S_\oi$-axis $\Soax$ (red surface). Right:~The same spacetime depicted in the static frame $S_\bst$. It shows diagrams analogous to \Cref{fig:3d_boosted_1} glued along the $S_\bst$-axis $\Sbax$ (blue surface). The hatched region is excluded. The string surfaces (green) should be identified, as well as surfaces $\Sbax$ (on the left) and $\Soax$ (on the right). Both static frames $S_\oi$ and $S_\bst$ cover the whole spacetime, but they are not smooth along the string and one of the axes. The frame $S_\oi$ fails to be smooth on $\Sbax$, the frame $S_\bst$ is not well defined at~$\Soax$.}
\end{figure}

We can also depict the same situation in the original static frame $S_\oi$, see~\Cref{fig:3d_boosted_0}. In this frame, the axis $\Soax$ looks static, and the first particle is at rest (but remember, the particle is subcritically accelerated in the spacetime sense). The second particle oscillates (since it is at rest in frames $S_+$ and $S_-$; nevertheless, it is also subcritically accelerated in the spacetime sense). Although the planes $\Sbax$ continuing from the second particle up to infinity look moving in the frame $S_\oi$, they are geometrically flat with vanishing extrinsic curvature (cf.~\Cref{fig:3d_boosted_1}); therefore, their identification is also transparent.

Since both identified surfaces $\Soax$ and $\Sbax$ are transparent, these surfaces are just geometrical planes in the resulting spacetime, without any matter contributions. On the other hand, the surface~$\Sigma$ represents the string with energy and tangential tension. The surface $\Soax$ is static in the original static frame $S_\oi$. The surface $\Sbax$ is static in both frames $S_-$ and $S_+$. Thus, these two frames can be joined into one common static frame $S_\bst$, which is smooth through $\Sbax$. The worldsheet of the string $\Sigma$ is static in both frames if one ignores its boundary, where the string is attached to the particles. Particle~$P_1$ is static in the frame $S_\oi$, particle~$P_2$ in $S_\bst$.


However, neither frame $S_\oi$ nor $S_\bst$ is globally smooth. Frame $S_\oi$ is non-smooth at $\Sbax$, frame $S_\bst$ fails to be smooth at $\Soax$ -- even though the spacetime is smooth both at $\Soax$ and $\Sbax$. It demonstrates that the resulting spacetime is dynamical, without static asymptotics. We cannot thus compare this spacetime with canonical static spacetimes \eqref{eq:BTZ_metric2} discussed in \Cref{ssc:PointParticleAdS}.

It opens the question of a general asymptotic character of locally AdS spacetimes. It is clear that the asymptotics of canonical static rotationally symmetric spacetimes do not cover, for example, rotating systems. However, the spacetime of oscillating particles fails to be static for other reasons than rotation. It has an intrinsically dynamical character. It may be surprising because one would expect that the system bounded in space and periodic in time should be asymptotically static in a kind of rest frame. But it seems that the dynamical character propagates up to infinity and destroys the global static asymptotic. Therefore, one also needs to consider asymptotics other than the plain canonical monopole, even in the non-rotating case.

Finally, let us mention that a similar construction can be done for two particles of positive mass connected by a strut, see \Cref{fig:3d_oscpart}. The resulting system can be interpreted as two oscillating particles attached to a `spring' in a cosmological gravitational field.

    \begin{figure*}\centering
    \begin{minipage}[c]{.66\linewidth}
        \includegraphics[]{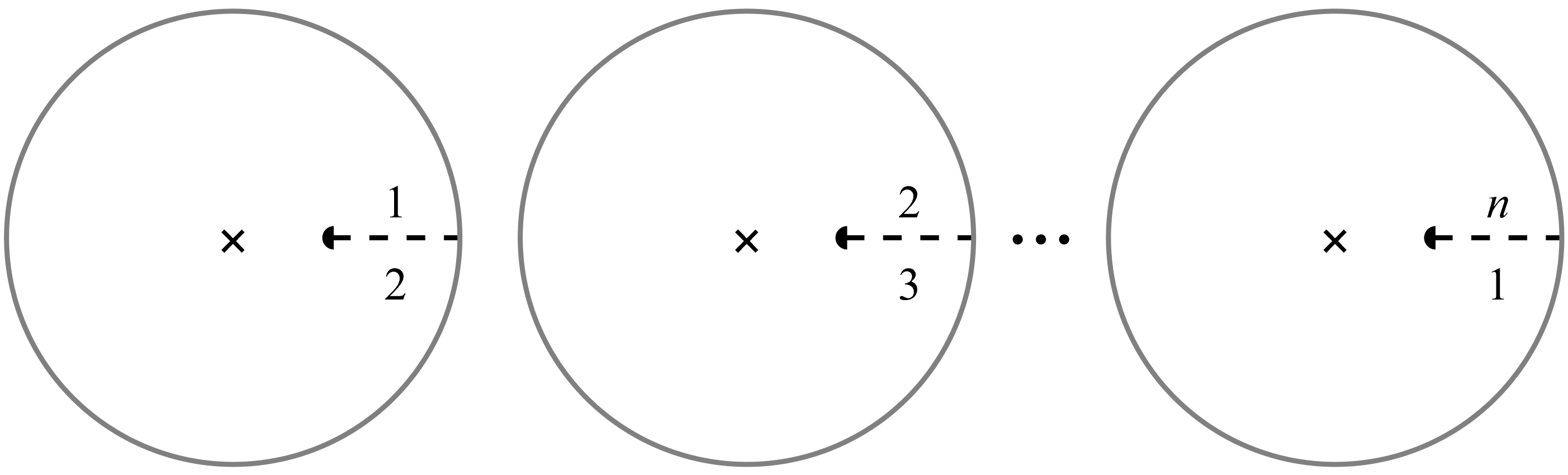}
    \end{minipage}\hfill
    \begin{minipage}[c]{.33\linewidth}
        \includegraphics[]{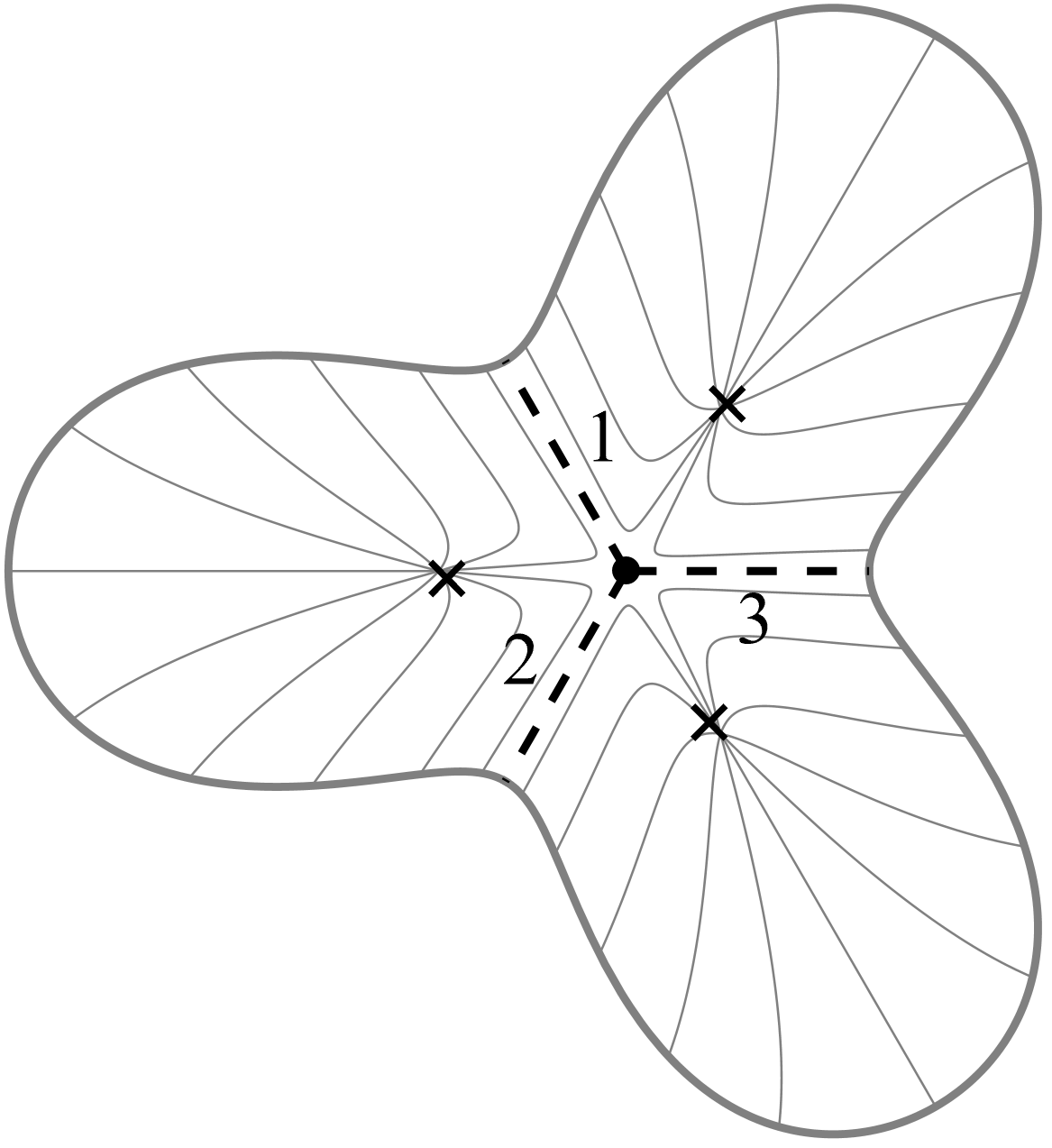}
    \end{minipage}
\caption{\label{fig:negmassinst}%
    Left: Spatial sections of $n$ empty AdS spacetimes. Cutting them along the axis $\ph=0$ on the interval $\chi\in(\chi_\oi,\tfrac\pi2)$ and identifying them as indicated by numbers creates the spacetime with a particle of mass $m_\oi = -n\frac{2\pi}{\kap}$. Since all identifications are of vanishing extrinsic curvature, they are transparent, and the spacetime does not contain any other matter than the particle. Although the particle is not interacting with any external agent, it is surrounded by the static observers of a nontrivial acceleration $a_\oi$. We call such a particle the negative mass instanton. 
    Right: Rescaling the accelerated angular coordinate $\bph$ allows us to draw all copies of AdS spacetime into one diagram (the case $n=3$ is shown). The diagram demonstrates that the particle is in equilibrium among $n$ symmetrically placed gravity centers. 
    }
\end{figure*}

\section{Interesting spacetimes}
\label{sec:spacetimes}

In this section, we discuss a couple of spacetimes constructed by the cut and glue method. We do not present an exhaustive description of all possible cases since the landscape of possibilities of the method is enormous. We want to point out only some interesting features that one can encounter.


\subsection{Negative mass instantons}
\label{ssc:negmassinst}

In \Cref{ssc:string}, we discussed the particle of mass $m_\oi$ hanging on the semi-infinite string of linear energy density $\mu_\oi$ at position $\chi_\oi$ above the gravity center. The mass $m_\oi$ is directly related to the deficit angle $\Delta\ph_\oi$ around the particle, cf.~\eqref{eq:mascondef}, the energy density to the string curvature $\scur_\oi$, cf.~\eqref{eq:surface_stress_energy}, and the position $\chi_\oi$ to the acceleration $a_\oi$ of the particle \eqref{eq:aodef}. All these quantities are related by the equilibrium condition \eqref{eq:meancurvaopho}. Two of these parameters can be varied freely, provided that \eqref{eq:meancurvaopho} holds.

Inspecting the equilibrium condition, we find an interesting special case of the particle staying at rest with a nontrivial acceleration $a_\oi$ without any interaction with a string. Namely, for the masses 
\begin{equation}
    \label{eq:negmassinst}
    \begin{aligned}
        m_\oi &= - (n-1) \frac{2\pi}{\kap}\,,\\
        \Delta\ph_\oi &= -(n-1) \pi\,,\\
    \end{aligned}
        \qquad n=1,2,3,\dots\;,
\end{equation}
the string curvature $\scur_\oi$ vanishes and the string becomes transparent, disappearing completely. Of course, since the relevant masses are negative, the situation is not very physical.

The system of such a negative mass instanton can be directly obtained by the cut-and-glue method. We start with $n$ copies of empty AdS spacetimes and cut them along surfaces $\chi\in(\chi_\oi,\frac\pi2)$, $\ph=0$. Next, we identify these cuts between various copies as indicated in \Cref{fig:negmassinst}. We obtain locally AdS spacetime with a conical singularity with the vertex angle $2\pi n$, i.e., precisely with the deficit angle $\Delta\ph_\oi=\pi-n\pi$.

Squeezing the angular coordinate $\bph$ of accelerated coordinates \eqref{eq:S2rotation}, cf.~\Cref{fig:coords} right, by a factor $\frac1n$ and including all $n$ glued copies of AdS spacetime into one picture, see \Cref{fig:negmassinst} right, we can understand why a string is not needed to keep the particle at a nontrivial acceleration. The particle is in equilibrium between $n$ gravity centers symmetrically placed around it. We also realize that although the magnitude of acceleration seems to be given by $a_\oi$, the direction of the acceleration is not well defined. Not only is the tangent structure at the conical singularity undefined, but even intuitively, there is no single natural candidate for the direction of acceleration. Static observers near the particle have the acceleration of magnitude close to $a_\oi$, but with directions pointing towards various gravity centers around the particle.  

Let us finally mention that the instanton can be added to other solutions -- for example, to a generic particle hanging on the string -- shifting the particle's mass by a discrete multiple of $-\frac{2\pi}{\kap}$. Indeed, one can imagine that one of the copies of AdS in \Cref{fig:negmassinst} is replaced by the spacetime with a particle on the strut, cf.~\Cref{fig:string_strut} right.


\subsection{Spacetimes of various topology}
\label{ssc:vartopology}

\begin{figure}[t]\centering
        \includegraphics[]{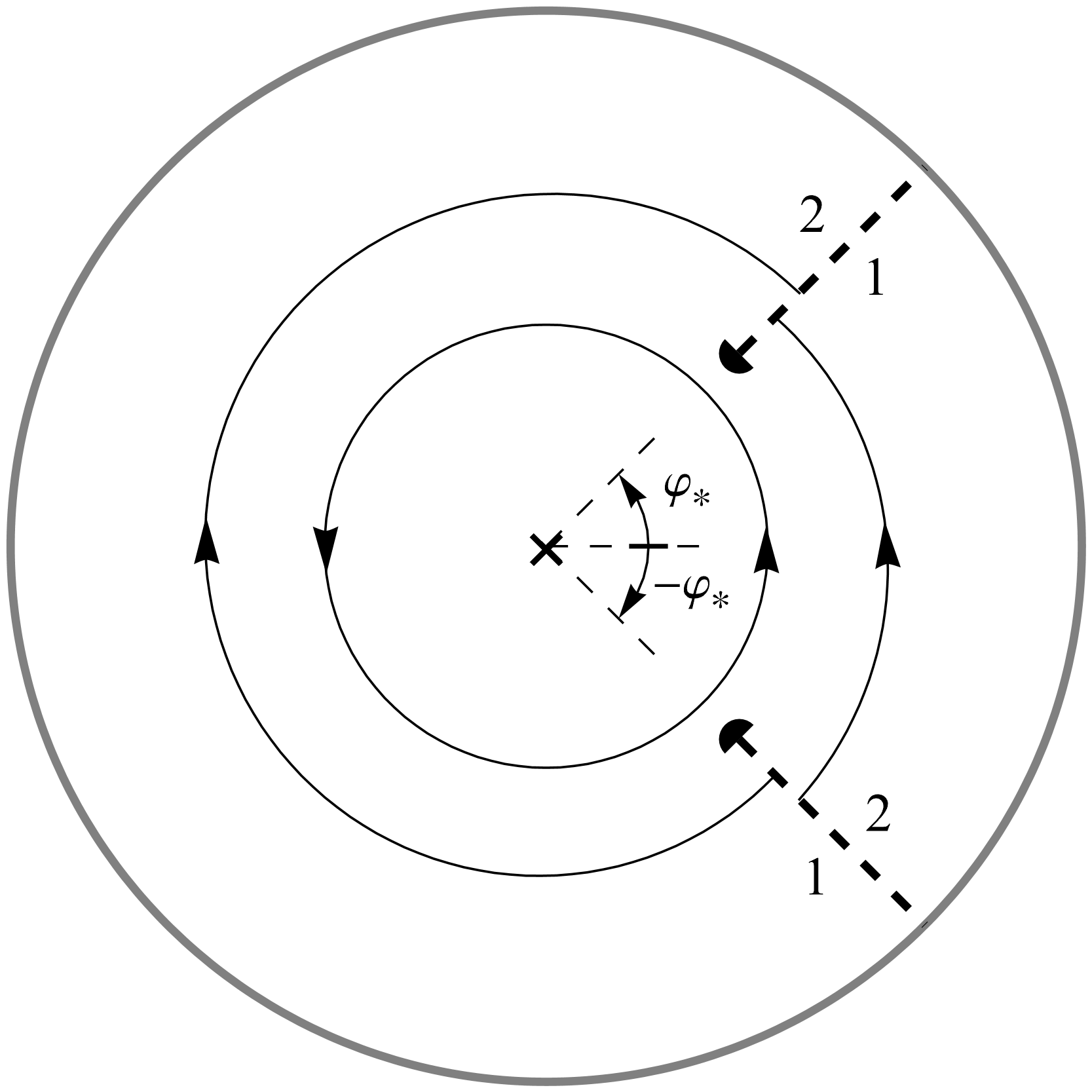}
        \caption{\label{fig:one_asymptote_non-ori}%
        The spatial section of the static spatially non-orientable spacetime is obtained by the identification along two radial semi-infinite cuts. The identification is indicated by numbers. The asymptotic region is identical to the empty AdS spacetime. The spacetime contains the conical singularity at the end of cuts representing the particle of mass $-\frac{2\pi}{\kap}$. The spacetime is non-orientable, the topology of the spatial section is that of the M\"obius strip. Examples of spatial trajectories of two observers circulating around the gravity center are indicated: one on the radius below the cuts (a simple circle), another moving in the region with the cuts (with slightly varying radius, merely for clarity of the diagram). We see that although both observers move in the same direction near $\ph=0$, they move in an opposite direction near $\ph=\pm\pi$.}        
\end{figure}

\begin{figure}[t]\centering
        \includegraphics[]{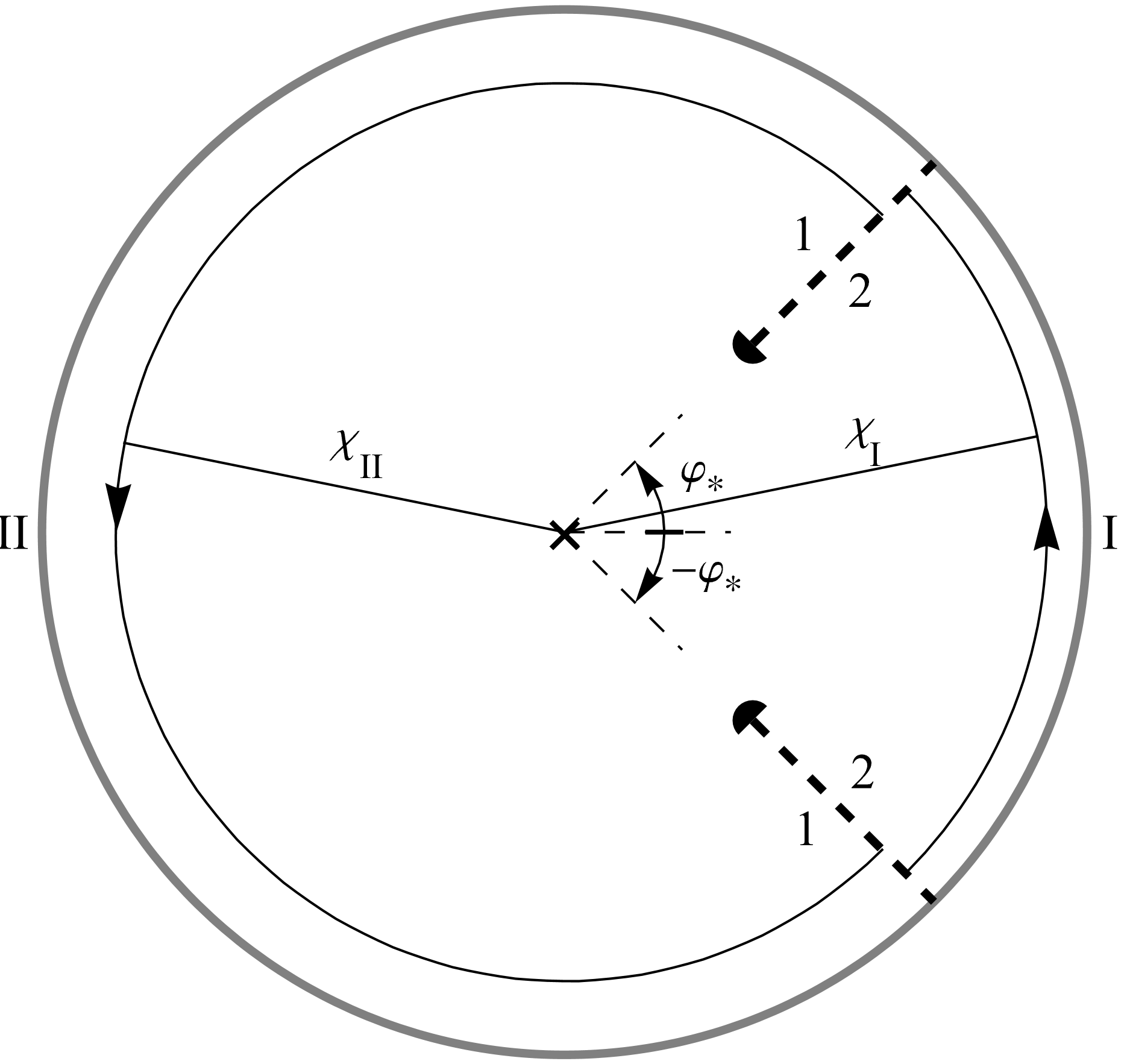}
        \caption{\label{fig:one_asymptote_wh}%
         The spatial section of the spacetime containing a wormhole between two AdS asymptotics. It is obtained by identifying two radial semi-infinite cuts as indicated by numbers. The spacetime contains two separate AdS asymptotic regions $\asA$ and $\asB$, and the topology of the spatial section is that of a cylinder. The spacetime contains the conical singularity at the end of cuts representing the particle of mass $-\frac{2\pi}{\kap}$. 
         The diagram shows spatial trajectories of two observers circulating in the asymptotic regions at radii $\chi_\asA$ and $\chi_\asB$. Each asymptotic region can be compared with the canonical conical space with the masses of the central particles $m_\asA=\frac{2(\pi-\ph_*)}{\kap}$ and $m_\asB=\frac{2\ph_*}{\kap}$, respectively.}
\end{figure}

Similarly to the just discussed case, we can identify semi-infinite cuts $\chi\in(\chi_\oi,\tfrac\pi2)$ along radial directions $\ph_j$, $j=1,2,\dots,n$ in one AdS spacetime. Let us focus on the simplest nontrivial case $n=2$ with $\ph_1=\ph_*$ and $\ph_2=-\ph_*$. There are two qualitatively different cases of identifying the cuts, as indicated in \Cref{fig:one_asymptote_non-ori,fig:one_asymptote_wh}. One leads to non-orientable spacetime with spatial topology of the M\"obius strip (\Cref{fig:one_asymptote_non-ori}), the other represents the wormhole between two asymptotic AdS spacetimes (\Cref{fig:one_asymptote_wh}).


In the first case, the asymptotic region is identical to the empty AdS spacetime: ``switching'' one part between cuts in the asymptotic region does not change the geometry of this region. However, the spacetime contains the conical singularity at the end of cuts representing the particle of mass $m_\oi=-\frac{2\pi}{\kap}$. The spacetime has a non-orientable spatial section of genus $\frac12$ with one $S^1$ boundary (the infinity), i.e., the topology of M\"obius strip. Although the spacetime is rather trivial and restricted, it may be instructive to analyze the local mass formula for this case. We compare the ``ball'' ${}^\nor\!\dom{B}_{\chi_\ai}$ (i.e., the domain $\chi<\chi_\ai$) in the non-orientable spacetime with the ball ${}^\can\!\dom{B}_{\chi_\ai}$ in the conical spacetime of the same asymptotic:
\begin{equation}
	\label{eq:locmass_nor}
\begin{aligned}
	\locmass_\nor[{}^\nor\!\dom{B}_{\chi_\ai}] &= m_\oi + \eps_\Lambda\, {}^\nor\!A_{\chi_\ai}\\
       &= \frac{2\pi}{\kap}\chi[{}^\nor\!\dom{B}_{\chi_\ai}] + \frac1\kap \int_{\chi=\chi_\ai}\alpha\,ds
       \;,\\
	\locmass_\can[{}^\can\!\dom{B}_{\chi_\ai}] &= m_\tot + \eps_\Lambda\,\, {}^\can\!A_{\chi_\ai}\\
       &= \frac{2\pi}{\kap}\chi[{}^\can\!\dom{B}_{\chi_\ai}] + \frac1\kap \int_{\chi=\chi_\ai}\alpha\,ds
       \;.
\end{aligned}
\end{equation}
Here we first write the contributions to the local mass integral \eqref{eq:locmass}, including the contribution of the point particles; next, we use the formula \eqref{eq:locmaspardef} to transform the mass to the boundary integral. The contributions of the boundary integrals are the same. However, the Euler characteristics of the ``balls'' in both spaces differ. For the ball in the canonical conical space, it is just standard $\chi[{}^\can\!\dom{B}_{\chi_\ai}]=1$, but for a non-orientable ball of genus $\frac12$, it is $\chi[{}^\nor\!\dom{B}_{\chi_\ai}]=0$. It is consistent with the fact that the areas ${}^\nor\!A_{\chi_\ai}$ and ${}^\can\!A_{\chi_\ai}$ of both balls are the same and that the canonical spacetime is, in fact, the empty spacetime with $m_\tot=0$. The mass $m_\oi = -\frac{2\pi}{\kap}$ of the particle in the non-orientable space thus corresponds to the difference in the Euler characteristics, and instead of \eqref{eq:diflocmass} we obtain
\begin{equation}
	\label{eq:diflocmass_nor}
    \Delta\locmass = \locmass_\nor[{}^\nor\!\dom{B}_{\chi_\ai}] - \locmass_\can[{}^\can\!\dom{B}_{\chi_\ai}] = m_\oi =  -\frac{2\pi}{\kap}\;.
\end{equation}

\begin{figure}[b]\centering
  \includegraphics[]{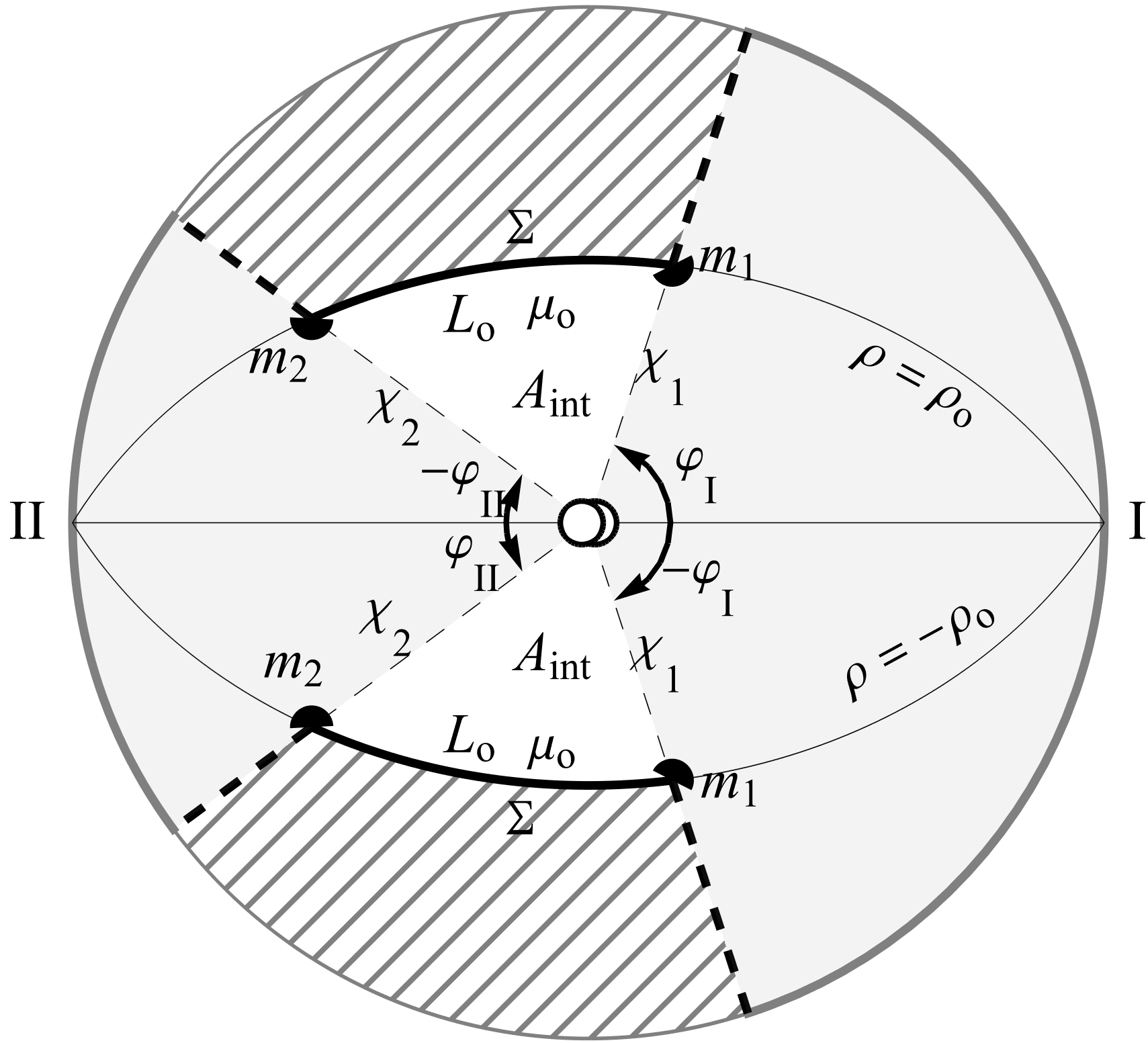}
  \caption{\label{fig:symwormhole}%
    Spatial section of the wormhole spacetime formed by symmetric exocycles. It contains two particles connected by a string. The radial cuts $\ph=\ph_\asA$ and $\ph=-\ph_\asA$ (thick dashed lines) starting from the exocycle $\Sigma$ up to infinity are identified, as well as the cuts $\ph=\pm(\pi-\ph_\asB)$. The symmetric exocycles $\Sigma$ between the cuts are also identified, forming the string of the length $L_\oi$ and the linear energy density $\mu_\oi$. Spacetime has two separate asymptotics $\asA$ and $\asB$. Domains corresponding to the canonical conical spaces are in gray. Areas $A_\itr$ reflect the difference between the whole wormhole spacetime and the canonical spaces.\\[-4ex]}
\end{figure}

Similarly, we can analyze the wormhole case in \Cref{fig:one_asymptote_wh}. Here we have to compare the spacetime with two canonical conical spaces of the corresponding asymptotics. In the wormhole spacetime, we choose the domain ${}^\wh\dom{B}_{\chi_\asA,\chi_\asB}$ between two circles, each in separate asymptotic regions. We compare it with the standard balls ${}^{\asA}\!\dom{B}_{\chi_\asA}$ and ${}^{\asB}\!\dom{B}_{\chi_\asB}$ in respective canonical spaces. The central masses of these canonical conical spaces are $m_\asA=2\frac{\pi-\ph_*}{\kap}$ and $m_\asB=2\frac{\ph_*}{\kap}$. The local mass balances read
\begin{equation}
	\label{eq:locmass_wh}
\begin{aligned}
	\locmass_\wh[{}^\wh\!\dom{B}_{\chi_\asA,\chi_\asB}] &= m_\oi + \eps_\Lambda\,\, {}^\wh\!A_{\chi_\asA,\chi_\asB}\\
       &= \frac{2\pi}{\kap}\chi[{}^\wh\!\dom{B}_{\chi_\asA,\chi_\asB}] + \frac1\kap \int_{\chi=\chi_\asA,\chi_\asB}\mspace{-25mu}\alpha\,ds\;,\\
	\locmass_{\asA}[{}^{\asA}\!\dom{B}_{\chi_\asA}] &= m_\asA + \eps_\Lambda\, {}^{\asA}\!A_{\chi_\asA}\\
       &= \frac{2\pi}{\kap}\chi[{}^{\asA}\!\dom{B}_{\chi_\asA}] + \frac1\kap \int_{\chi=\chi_\asA}\mspace{-25mu}\alpha\,ds \;,\\
	\locmass_{\asB}[{}^{\asB}\!\dom{B}_{\chi_\asB}] &= m_\asB + \eps_\Lambda\, {}^{\asB}\!A_{\chi_\asB}\\
       &= \frac{2\pi}{\kap}\chi[{}^{\asB}\!\dom{B}_{\chi_\asB}] + \frac1\kap \int_{\chi=\chi_\asB}\mspace{-25mu}\alpha\,ds\;.
\end{aligned}
\end{equation}
The boundary integrals in both canonical spaces together are equivalent to the boundary contribution in the wormhole space. A straightforward inspection shows that the areas ${}^{\asA}\!A_{\chi_\asA}$ and ${}^{\asB}\!A_{\chi_\asB}$ of the standard balls give together the area ${}^\wh\!A_{\chi_\asA,\chi_\asB}$ of the domain ${}^\wh\!\dom{B}_{\chi_\asA,\chi_\asB}$. The Euler characteristics are $\chi[{}^{\asA}\!\dom{B}_{\chi_\asA}]=\chi[{}^{\asB}\!\dom{B}_{\chi_\asB}]=1$ and $\chi[{}^\wh\!\dom{B}_{\chi_\asA,\chi_\asB}]=0$. Putting together, we obtain
\begin{equation}
	\label{eq:diflocmass_wh}
    \begin{aligned}
        \Delta\locmass &\equiv \locmass_\wh[{}^\wh\!\dom{B}_{\chi_\asA,\chi_\asB}] 
        - \locmass_{\asA}[{}^{\asA}\!\dom{B}_{\chi_\asA}] - \locmass_{\asB}[{}^{\asB}\!\dom{B}_{\chi_\asB}]\\ 
        &= m_\oi - m_\asA - m_\asB 
        = -2\frac{2\pi}{\kap}\;,
    \end{aligned}
\end{equation}
Taking into account the masses of the canonical spaces, the mass $m_\oi$ of the particle 
\begin{equation}
	\label{eq:parmass_wh}
    m_\oi = 2\frac{\pi-\ph_*}{\kap} + 2\frac{\ph_*}{\kap} -2\frac{2\pi}{\kap} = - \frac{2\pi}{\kap} \;.
\end{equation}
consistently gives the negative mass instanton.

\begin{figure}[b]\centering
  \includegraphics[]{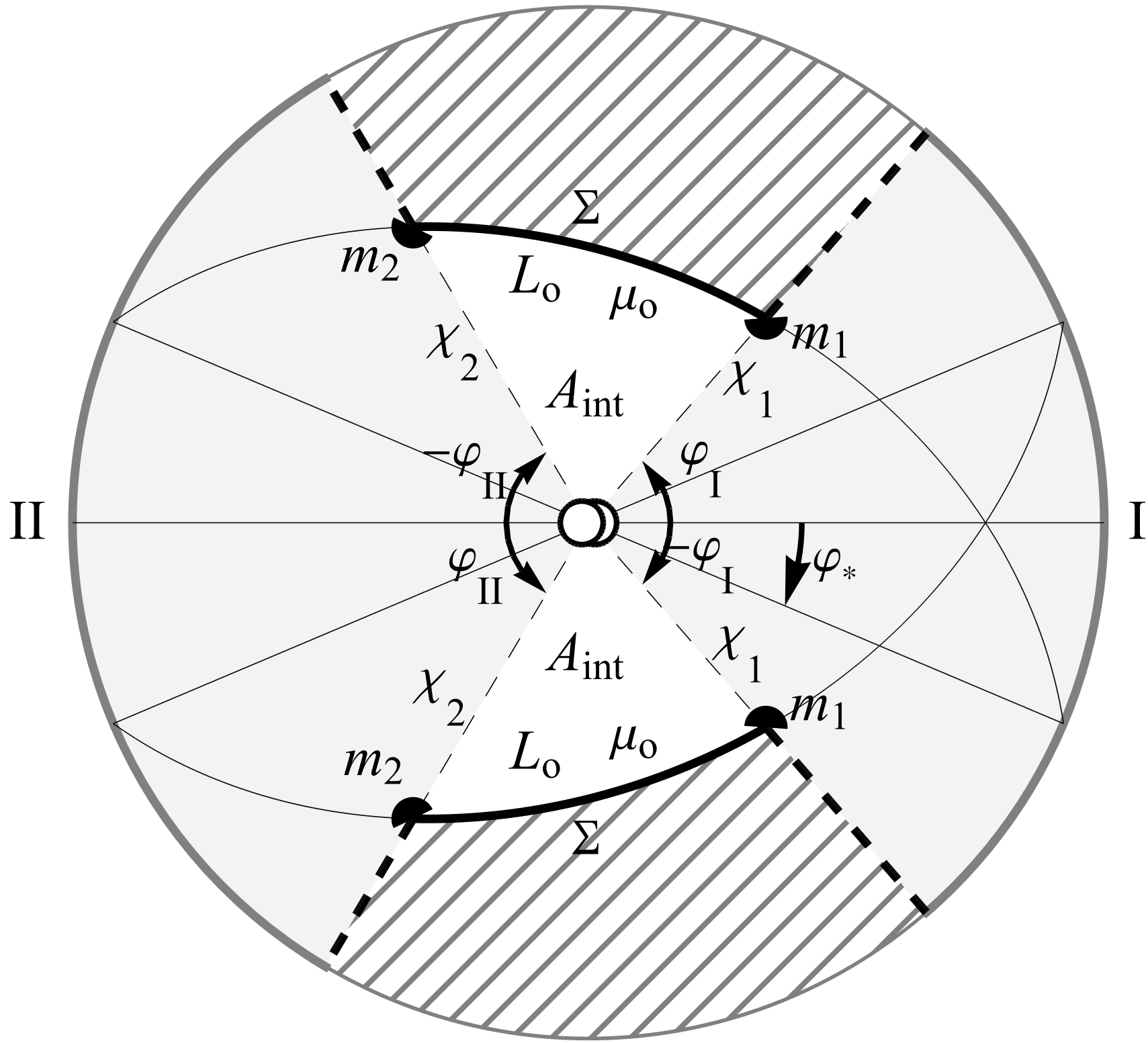}
  \caption{\label{fig:genwormhole}%
   The wormhole obtained by the same construction as in \Cref{fig:symwormhole}, only with the exocycles and all related structures rotated by angle $\ph_*$ and $-\ph_*$, respectively. The spacetime can be obtained from that of \Cref{fig:symwormhole} by adding angle $2\ph_*$ to the asymptotic $\asA$ and removing the same angle from the asymptotic $\asB$. The figure corresponds to negative $\ph_*$. The geometry and masses of the system of two particles connected by the string remain the same, independent of the parameter $\ph_*$. The system can thus be placed at a different position inside the wormhole; of course, the chosen position affects the asymptotics of the wormhole\\[-4ex]}
\end{figure}


\subsection{Wormhole spacetime with point particles}
\label{ssc:wormhole}

The just discussed wormhole spacetime is generated by the negative mass instanton. It is not surprising since the static wormholes usually require an exotic (read: badly behaved) energy distribution supporting them \cite{ThorneMorris1988}. Instead of a particle of a large special negative mass, we can support the wormhole by a system of two particles attached to the string. Both particles still have a negative energy distribution, so we do not escape the curse of the wormhole exotic energy. However, the system has freedom in the choice of the masses of the particles and the tension of the string, which determine the asymptotic behavior of both sides of the wormhole. We first discuss a special type of such wormhole, cf.~\Cref{fig:symwormhole}, and later add one more degree of freedom, cf.~\Cref{fig:genwormhole}.

We construct the spacetime in a similar way as we did in \Cref{ssc:two_particles}. Let us describe the spatial section of the static spacetime depicted in \Cref{fig:symwormhole}. We start with two symmetric exocycles $\Sigma$ given in the equidistant coordinates \eqref{eq:equidistcoors} by $\rho=\pm\rho_\oi$. We intersect them by radial cuts $\ph=\pm\ph_\asA$ and $\ph=\pm(\pi-\ph_\asB)$. We identify the cuts $\ph=\ph_\asA$ and $\ph=-\ph_\asA$ starting from the exocycle up to the space infinity, $\chi\in(\chi_1,\tfrac\pi2)$. Similarly, we identify cuts $\ph=\pm(\pi-\ph_\asB)$ on the interval $\chi\in(\chi_2,\tfrac\pi2)$. Finally, we identify both exocycles $\Sigma$ in the interval between the cuts. The domain outside the exocycles between the cuts (hatched) is removed.

We thus obtain the wormhole space with two separate asymptotics $\asA$ and $\asB$. These asymptotics can be compared with the canonical conical spaces of masses $m_\asA=\frac2\kap(\pi-\ph_\asA)$ and $m_\asB=\frac2\kap(\pi-\ph_\asB)$. The wormhole contains two particles of masses $m_1$ and $m_2$ located at the intersections of the cuts with the exocycle. The particles are connected by the string of linear energy density $\mu_\oi=-\frac{2\scur_\oi}{\kap}$ between both particles. The string curvature $\scur_\oi$ is related to the coordinate $\rho_\oi$ of the exocycle as
\begin{equation}
    \ell\scur_\oi = \sin\rho_\oi\;,
    \label{eq:curexocycle}
\end{equation}
cf.~\eqref{eq:scurrhoo} in \Cref{apx:string_curvature}. The equilibrium conditions \eqref{eq:meancurvaopho} between particles and the string thus read\footnote{In the equilibrium condition \eqref{eq:meancurvaopho}, the angle total vertex $\ph_\oi$ differs from the total vertex angle in the present case: they are complementary. The deficit angles thus differ by a sign, and one has to substitute $\Delta\ph_\oi\to-\frac{\kap m_{1,2}}2$, which leads to an additional minus sign.}
\begin{equation}
\begin{aligned}
    \sin\rho_\oi &= -\sin\chi_1\, \sin\frac{\kap m_1}{2}\;,\\
    \sin\rho_\oi &= -\sin\chi_2\, \sin\frac{\kap m_2}{2}\;,
\end{aligned}
    \label{eq:equilibriumcond}
\end{equation}
where we expressed the accelerations $a_i$ in terms of the radii $\chi_i$ using \eqref{eq:aodef}. The coordinate definition \eqref{eq:equidistcoors} gives the relation between $\chi$ and $\ph$ for points at the exocycle ${\rho=\rho_\oi}$. Evaluating it at the position of the first particle $\ph=\ph_\asA$ and the second particle $\ph=\pi-\ph_\asB$ gives
\begin{equation}
    \tan\rho_\oi = \tan\chi_1\, \sin\ph_\asA\;,\qquad 
    \tan\rho_\oi = \tan\chi_2\, \sin\ph_\asB\;.
    \label{eq:parsonexocycle}
\end{equation}
This allows us to relate asymptotic masses $m_\asA$, $m_\asB$ and the particle masses $m_1$, $m_2$ 
\begin{equation}
    \cos\frac{\kap m_\asA}{2} = \frac{\cos\frac{\kap m_1}{2}}{\cos\rho_\oi}
      \;,\quad
    \cos\frac{\kap m_\asB}{2} = \frac{\cos\frac{\kap m_2}{2}}{\cos\rho_\oi}
      \;,
    \label{eq:whasmasses}
\end{equation}
where $\cos\rho_\oi=\sqrt{1 - \Bigl(\frac{\kap\mu_\oi\ell}{2}\Bigr)^2}$ is given by the string energy density $\mu_\oi$.

Relations  \eqref{eq:equidistcoors} determine also the coordinates $\zeta_1$ and $\zeta_2$ of the particles, 
\begin{equation}
    \tan\zeta_1 = \cot\ph_\asA\, \sin\rho_\oi\;,\hskip2ex
    \tan\zeta_2 = -\cot\ph_\asB\, \sin\rho_\oi\;,
    \label{eq:zetaonexocycle}
\end{equation}
and the length of the string is
\begin{equation}
\begin{split}
    L_\oi &= \frac{\ell}{\cos\rho_\oi}
      \Bigl(\arcsinh\tan\zeta_1-
      \arcsinh\tan\zeta_2\Bigr)\\
    &=-\frac{\ell}{\cos\rho_\oi}
      \biggl(\arctanh\frac{\cot\frac{\kap m_1}{2}}{\cot\rho_\oi}+
      \arctanh\frac{\cot\frac{\kap m_2}{2}}{\cot\rho_\oi}\biggr)
      \;.
    \label{eq:Lexocycle}
\end{split}\raisetag{10.5ex}
\end{equation}

It is interesting that for fixed string energy density $\mu_\oi$, the asymptotic mass $m_\asA$ depends only on the mass $m_1$ of the first particle, and $m_\asB$ only on $m_2$. Of course, varying particle masses also change the length $L_\oi$ of the string.

The just-constructed wormhole can be modified by a simple cut-and-glue operation. We can add an angle $2\ph_*$ to the asymptotic $\asA$ and remove the same angle at the asymptotic $\asB$. Added and removed angles meet at the common center of gravity and compensate each other. Therefore, the center of gravity remains a regular point in spacetime. This operation only changes angular characteristics of the asymptotics as $\ph_\asA\to\ph_\asA+\ph_*$ and $\ph_\asB\to\ph_\asB-\ph_*$. The parameter $\ph_*$ thus allows a shift of the rigid system of two particles connected by the string with respect to the wormhole, modifying accordingly both asymptotics.

Effectively, the resulting spacetime corresponds to \Cref{fig:genwormhole}. It depicts the spatial section, in which all structures related to the exocycle $\rho=\rho_\oi$ are rotated by $\ph_*$ and structures related to the exocycle $\rho=-\rho_\oi$ are rotated by $-\ph_*$. Beware that the angle $\ph_*$ is chosen negative in \Cref{fig:symwormhole}.

\begin{figure*}\centering
\vspace*{-6ex}
    \begin{minipage}[c]{.66\linewidth}
        \includegraphics[]{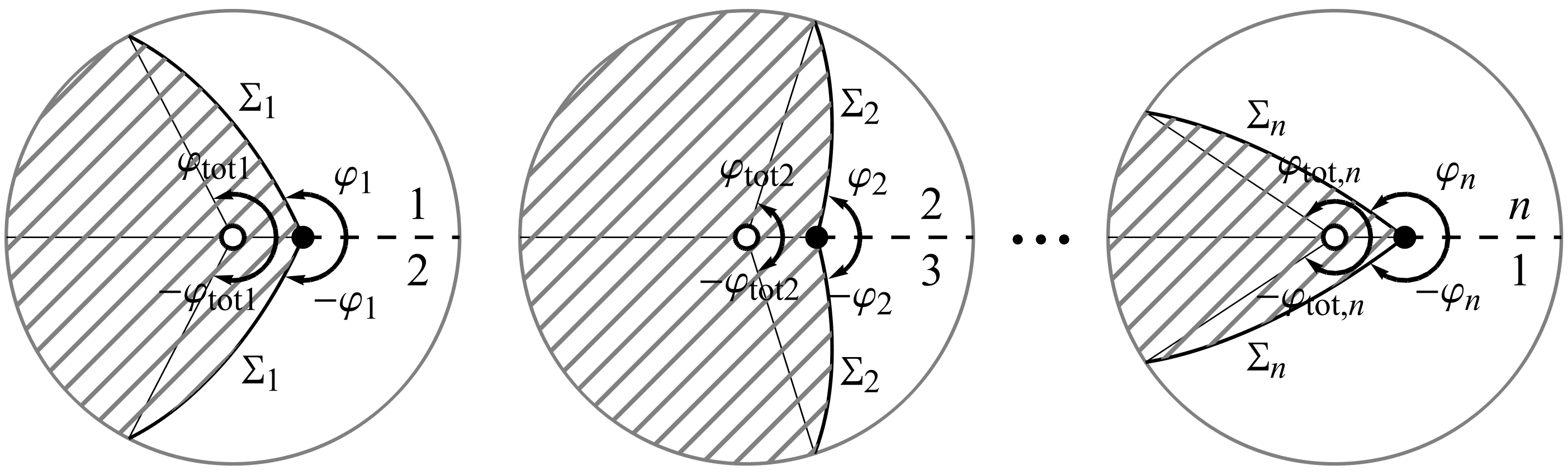}
    \end{minipage}
    \begin{minipage}[c]{.33\linewidth}\centering
        \includegraphics[]{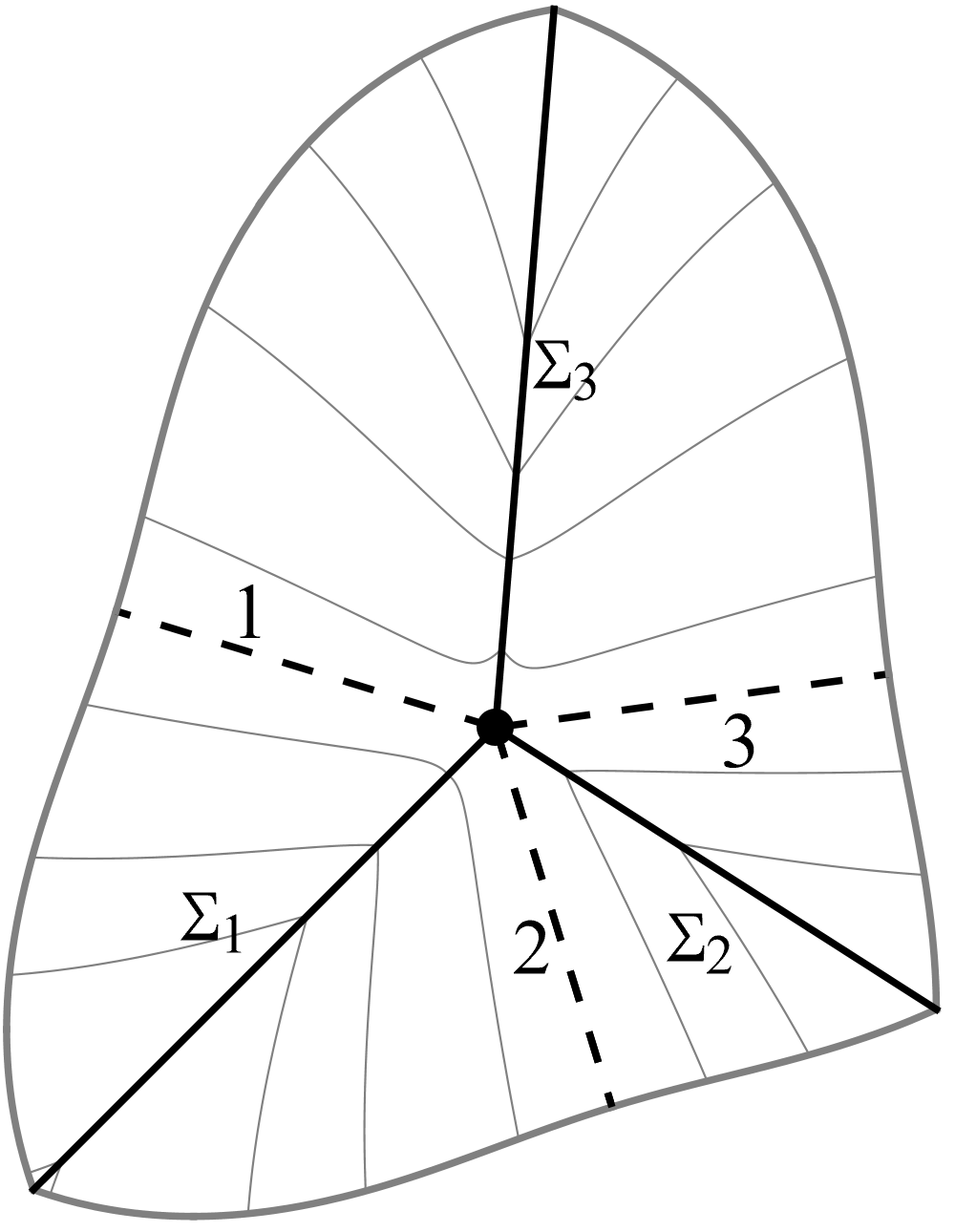}
    \end{minipage}
    \caption{\label{fig:multistring}%
        Left: Spatial sections of $n$ spacetimes with a semi-infinite string (cf.~\Cref{fig:string_strut}). Cutting them along the axis $\ph=0$ on the interval $\chi\in(\chi_\oi,\tfrac\pi2)$ and identifying them as indicated by numbers creates the spacetime with a particle of mass $m_\oi$ attached to $n$ strings reaching up to infinity. The particle is kept at rest in the static frame by the equilibrium of string forces.  
        Right: All spaces squeezed into one diagram by rescaling the accelerated angular coordinate $\bph$ (the case $n=3$ is shown). 
    }
\vspace*{2ex}
    \begin{minipage}[c]{.66\linewidth}
        \includegraphics[]{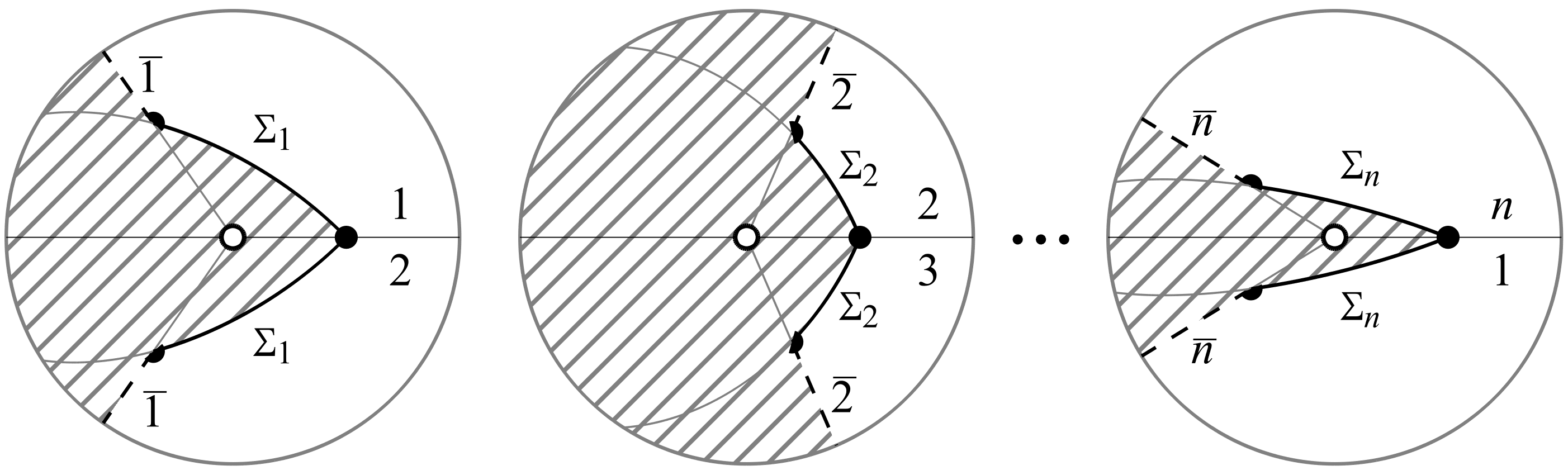}
    \end{minipage}
    \begin{minipage}[c]{.33\linewidth}\hfill
        \includegraphics[]{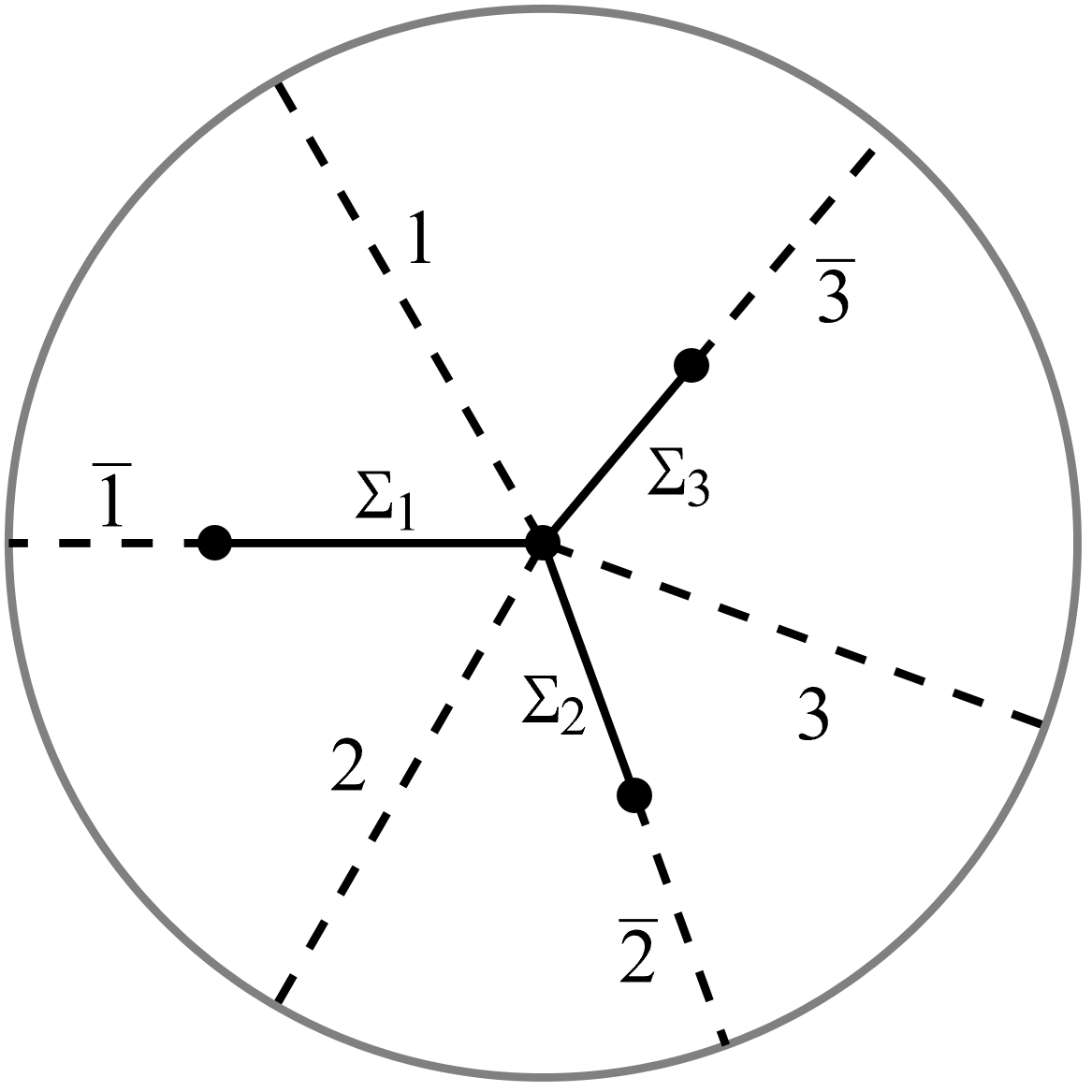}
    \end{minipage}
    \caption{\label{fig:multistring_fin}%
        Left: Spatial sections of $n$ spacetimes, each with a finite string between two particles (cf.~\Cref{fig:two_particles_string/strut}). Cutting these spaces along the axis $\ph=0$ on the interval ${\chi\in(\chi_\oi,\tfrac\pi2)}$ and identifying them as indicated by numbers creates the spacetime with a central particle of mass attached to $n$ strings, each reaching up to another particle at a finite distance. The central particle is kept at rest in the static frame by the equilibrium of string forces.  
        Right: Intuitive visualization of the situation with angular coordinate $\bph$ rescaled and radial coordinate properly deformed to match AdS asymptotic character of the spacetime. The case $n=3$ is shown. 
    }
\end{figure*}

The asymptotics of the spacetime are equivalent to the canonical conical spacetimes of the central masses $m_\asA$ and $m_\asB$ given now by
\begin{equation}
\begin{aligned}
    \cos\Bigl(\frac{\kap m_\asA}{2}-\ph_*\Bigr) &= \frac{\cos\frac{\kap m_1}{2}}{\cos\rho_\oi}
      \;,\\
    \cos\Bigl(\frac{\kap m_\asB}{2}+\ph_*\Bigr) &= \frac{\cos\frac{\kap m_2}{2}}{\cos\rho_\oi}
      \;,
\end{aligned}
    \label{eq:genwhasmasses}
\end{equation}
cf.~\eqref{eq:whasmasses}. Comparing similar domains as in the wormhole case in the previous section, we obtain
\begin{equation}
	\label{eq:locmass_parwh}
\begin{aligned}
	\locmass_\wh[{}^\wh\!\dom{B}_{\chi_\asA,\chi_\asB}] &= m_1 + m_2 + \mu_\oi L_\oi + \eps_\Lambda\,\, {}^\wh\!A_{\chi_\asA,\chi_\asB}\\
       &= \frac{2\pi}{\kap}\chi[{}^\wh\!\dom{B}_{\chi_\asA,\chi_\asB}] 
          + \frac1\kap \int_{\chi=\chi_\asA,\chi_\asB}\mspace{-25mu}\alpha\,ds
       \;,\\
	\locmass_{\asA}[{}^{\asA}\!\dom{B}_{\chi_\asA}] &= m_\asA + \eps_\Lambda\, {}^{\asA}\!A_{\chi_\asA}\\
       &= \frac{2\pi}{\kap}\chi[{}^{\asA}\!\dom{B}_{\chi_\asA}] 
          + \frac1\kap \int_{\chi=\chi_\asA}\mspace{-25mu}\alpha\,ds
       \;,\\
	\locmass_{\asB}[{}^{\asB}\!\dom{B}_{\chi_\asB}] &= m_\asB + \eps_\Lambda\, {}^{\asB}\!A_{\chi_\asB}\\
       &= \frac{2\pi}{\kap}\chi[{}^{\asB}\!\dom{B}_{\chi_\asB}] 
          + \frac1\kap \int_{\chi=\chi_\asB}\mspace{-25mu}\alpha\,ds
       \;.
\end{aligned}
\end{equation}
Here, the boundary integrals again match, but the area of ${}^\wh\!\dom{B}_{\chi_\asA,\chi_\asB}$ differs from the sum of areas of ${}^{\asA}\!\dom{B}_{\chi_\asA}$ and ${}^{\asB}\!\dom{B}_{\chi_\asB}$ by $2A_\itr= {}^\wh\!A_{\chi_\asA,\chi_\asB} - {}^{\asA}\!A_{\chi_\asA} - {}^{\asB}\!A_{\chi_\asB}$, cf.~\Cref{fig:genwormhole}.
The difference of the local masses of the wormhole and canonical spacetimes thus yields
\begin{equation}
	\label{eq:diflocmass_parwh}
    \Delta\locmass 
    = m_1 + m_2 + \mu_\oi L_\oi - m_\asA - m_\asB + 2\eps_\Lambda A_\itr
    = -2\frac{2\pi}{\kap}\;,
\end{equation}
Substituting for masses and the string length, we can, in principle, express the interaction area $A_\itr$. Notice that it does not depend on $\ph_*$ as can be also seen by comparison of \Cref{fig:symwormhole,fig:genwormhole}.


\subsection{Multiple strings attached to the particle}
\label{ssc:multistring}

As the last example, we discuss the possibility of attaching multiple strings to one particle. We can glue together $n$ spaces in a similar way as in \Cref{ssc:negmassinst}, only now each space contains the particle attached to a string~$\Sigma_j$, see \Cref{fig:multistring}. Spaces are cut along the interval $\chi\in(\chi_\oi,\frac\pi2)$ of the axis $\ph=0$ and identified as indicated by numbers. The space contains one conical singularity of vertex angle $\ph_\oi = \sum_j \ph_j$ and $n$ strings of linear energy densities $\mu_j$. Recall that it is related to the string curvature $\scur_j$ and exocycle coordinate $\rho_j$ as
\begin{equation}
\scur_j =-\frac{\kap\mu_j}{2}= \frac1\ell\sin\rho_j\;.    
\end{equation}

The spacetime has one asymptotic, which, however, cannot be matched to a canonical space since we have strings extending up to infinity. Nevertheless, we can calculate the total angle $\ph_\tot=\sum_j \ph_{\tot j}$ composed from the asymptotic angle contributing $\ph_{\tot j}$ for each string. These angles are related to the exocycle parameter of the string $\rho_j$ and the position of the particle $\chi_\oi$ as
\begin{equation}
 \tan\rho_j = \tan\chi_\oi\,\sin\ph_{\tot j}\;,
\end{equation}
(cf.\ similar relation \eqref{eq:parsonexocycle}). 
The equilibrium condition \eqref{eq:meancurvaopho} for each string, rewritten in terms of variables $\rho_j$ and $\chi_\oi$, reads
\begin{equation}
 \sin\rho_j = \sin\chi_\oi\,\sin\ph_j\;.
\end{equation}
Again, combining these equations allows us to relate $\ph_\oi$ and $\ph_\tot$, with the help of string parameters~$\rho_j$ or $\mu_j$.

It is possible to make the system finite, without strings reaching infinity. We just replace the spaces with a particle hanging on the string (cf.~\Cref{fig:string_strut}) by spaces with two particles connected by the string (cf.~\Cref{fig:two_particles_string/strut}) as indicated in \Cref{fig:multistring_fin}. The total spacetime has a single conical asymptotic. It contains a central particle connected by multiple strings to other particles. Combining the steps presented in the previous cases, we could express the total mass $m_\tot$ in terms of the parameters of the strings and individual particles. We leave this tedious exercise to the interested reader.


\section{Summary}
\label{sec:summary}

In this work, we have discussed 2+1-dimensional locally vacuum AdS spacetimes with lower-dimensional deficits corresponding to point particles and cosmic strings or struts. We focused on the static systems of accelerated point particles attached to a linear agent, keeping them in equilibrium. The simplest case of one particle attached to a semi-infinite string is described by the C-metric. We have reviewed the relation between the famous BTZ metric and the C-metric and discussed the importance of the proper choice of the coordinate intervals. The choice of the definition interval of the angular coordinate introduces a conical defect describing the particle and a 2-dimensional defect describing the cosmic string. 

We have also constructed more general examples using a cut-and-glue method, obtaining, for example, the pair of self-accelerating particles joined by a string or strut. Such a system is bound to a finite domain and thus it has a reasonable asymptotic.

For static systems, we introduced a notion of additive mass located in a domain. This mass can be evaluated by an integral over the boundary of the domain. It contains the dark energy contribution, and therefore, it diverges for infinite domains. It can be compared with the mass of canonical spacetimes with the identical asymptotic. It allows for the introduction of an interaction area that characterizes the difference between spacetime with generic static matter content and simple canonical spacetime. We will discuss applications of this concept for examples of systems composed of more realistic matter distributions in \cite{2+1_mass}. 

Next, we discussed the boost symmetry and realized that it can preserve the worldsheet of the string or strut. Taking advantage of this behavior, we constructed a dynamic spacetime of two particles oscillating with respect to each other, connected by a string of varying length. We believe that such a spacetime goes beyond the scope of the canonical asymptotically static conical spaces. It requires a new type of asymptotic behavior that encodes features that are more complicated than just a monopole character.

Finally, exploring the cut-and-glue method and combining previously discussed cases, we presented examples of several new spacetimes. These can have multiple centers of gravity, they can have wormholes or non-orientable topology, or they can contain several strings and particles. Comparing such spacetimes to corresponding canonical spacetimes revealed that the topology is relevant in considerations of mass assessment.

We have concentrated mainly on the discussion of point particles and mostly avoided the discussion of black holes, which are physically even more interesting and relevant. We plan to discuss the interaction of BTZ-like black holes with strings and particles in a subsequent contribution~\cite{BHs_particles}.

\begin{acknowledgments}
The authors thank the Czech Science Foundation grant GA\v{C}R~22-14791S. P.L. acknowledges the support from the Charles University Research Center grant UNCE24/SCI/016 and the Charles University Student Science Project SVV260833.
\end{acknowledgments}

    \appendix
\section*{Appendix}
\renewcommand\theequation{A\arabic{equation}}
\renewcommand\thesubsection{A.\arabic{subsection}}

\subsection{Extrinsic curvature and umbilical hypersurfaces}
\label{apx:K_propto_h}

Let us assume a $d$-dimensional manifold equipped with metric $\ts{g}$ with an embedded hypersurface $\Sigma$ of codimension 1 with a unit normal vector field $\ts{n}$. Assuming $\ts{n}$ is nowhere null, with constant normalization
\begin{equation}
  s = \ts{n}\cdot \ts{g} \cdot \ts{n} = \pm1\;,
\end{equation}
the embedding is characterized by intrinsic metric $\ts{h}$ and the extrinsic curvature $\ts{K}$. The intrinsic metric $\ts{h}$ is an orthogonal projection of the full metric $\ts{g}$ onto $\Sigma$,
\begin{equation}\label{eq:metricsplit}
  \ts{g}=s\, \ts{n}\ts{n}+\ts{h}\;.
\end{equation}
Extrinsic curvature 2-form $\ts{K}$ is defined as
\begin{equation}\label{eq:ex_curv}
  \ts{K}=\ts{h}\cdot\!\ts{\nabla} \ts{n}\;,
\end{equation}
where $\ts\nabla$ is the Levi-Civita covariant derivative adapted to~$\ts{g}$. It can be shown that $\ts{K}$ is symmetric.

A very specific case of the embedding is the so-called umbilic hypersurface, for which the extrinsic curvature is proportional to the metric, 
\begin{equation}\label{eq:ex_curv_umbilic}
    \ts{K} = \scur\, \ts{h}\;,
\end{equation}
with the mean curvature $\scur$ being a scalar function on the hypersurface.

In this paper, we study anti-de~Sitter spacetime, the maximally symmetric spacetime of constant negative curvature. The curvature defines the length scale $\ell$. The curvature tensors are then given by\footnote{Here, $\doublewedge$ denotes the Kulkarni-Nomizu product -- a bilinear form on symmetric tensors of rank two, whose result has symmetries of the Riemann tensor. Using indices,
\[\frac12\,(A\doublewedge B)_{abcd} = A_{ac}B_{bd}-A_{ad}B_{bc}\;.\]}
\begin{equation}\label{eq:MaxSymST}
    \begin{gathered}
        \Riem(\ts{g}) = -\frac{1}{2} \frac{1}{\ell^2}\; \ts{g}\doublewedge \ts{g}\;,\\
        \Ric(\ts{g}) = -\frac{d{-}1}{\ell^2}\; \ts{g}\;,\\
        \scR(\ts{g}) = -d(d{-}1)\,\frac1{\ell^2}\;. 
    \end{gathered}
\end{equation}

The embedding of the umbilic hypersurface in such a space is very special, e.g., \cite{umbilical}. Namely, the hypersurface is also a maximally symmetric space with both intrinsic and extrinsic curvature constant. The Gauss and Gauss-Codazzi equations imply that the curvature tensors of the intrinsic hypersurface geometry are
\begin{equation}\label{eq:MaxSymHS}
    \begin{gathered}
        \Riem(\ts{h}) = \frac{1}{2}(-\ell^{-2}{+}s\scur^2)\; \ts{h}\doublewedge \ts{h}\;,\\
        \Ric(\ts{h}) = (d-2)(-\ell^{-2}{+}s\scur^2)\; \ts{h}\;,\\
        \scR(\ts{h}) = (d{-}2)(d{-}1)(-\ell^{-2}+s\scur^2) \;.
    \end{gathered}
\end{equation}
Half of the scalar curvature in the last formula of \eqref{eq:MaxSymST} and \eqref{eq:MaxSymHS} is also called the Gauss curvature. Note that $\lvert\scur\rvert$ defines the geometry of the hypersurface fully; the sign of $\scur$ reflects just a hypersurface orientation given by the orientation of the normal $\ts{n}$. 

For a timelike hypersurface, $s=+1$, the hypersurface describes a worldsheet of a domain wall in the AdS universe. In dimension $d=3$, it can also be interpreted as the worldsheet of a string. The magnitude of the mean curvature has a critical value $\lvert\scur\rvert = \ell$. For $\lvert\scur\rvert< \ell$, the hypersurface has the geometry of $d{-}1$ dimensional AdS spacetime; for $\lvert\scur\rvert > \ell$, it is $d{-}1$ dimensional de~Sitter spacetime and it is intrinsically flat for $|\scur|=\ell$; cf.\ the sign of the scalar curvature in \eqref{eq:MaxSymHS}.  

Another case of interest is a spacelike hypersurface of constant static time. In this case, $s=-1$, and thanks to the staticity, $\scur=0$. We immediately obtain that the hypersurface has a hyperbolic geometry of the curvature given by the AdS scale $\ell$. In this case, we denote the spatial metric $\ts{q}$, and the future-oriented normalized normal vector, being the velocity of the static observers, $\ts{u}$.

\subsection{Israel's junction condition}
\label{apx:IJC}

Here we quickly summarize Israel's junction conditions \cite{Israel1966}.

Let us consider two $d$-dimensional spacetimes $\dom{M}_+$ and $\dom{M}_-$, each with a timelike boundary $\Sigma_\pm$ with the inner metric $\ts{h}_\pm$. If the inner geometries of both boundaries are isometric,
\begin{equation}
    \ts{h}\equiv\ts{h}_- = \ts{h}_+ \;,
\end{equation}
we can identify these boundaries and understand them as a single hypersurface $\Sigma$ embedded into a manifold $\dom{M}$ formed by the union of $\dom{M}_-$ and $\dom{M}_+$.

Let us denote inner normals of the boundaries $\Sigma_\pm$ as $\ts{n}_\pm$. After identification of the boundaries, we can also define a common normal 
\begin{equation}
    \ts{n}\equiv - \ts{n}_- = + \ts{n}_+ \;.
\end{equation}
We define the extrinsic curvature $\ts{K}_\pm$ using this common normal $\ts{n}$,
\begin{equation}
    \ts{K}_\pm = \ts{h}\cdot\ts{\nabla}_{\!\pm} \ts{n} \;,
\end{equation}
with Levi-Civita derivatives $\ts{\nabla}_{\!-}$ and $\ts{\nabla}_{\!+}$ defined in $\dom{M}_-$ and  $\dom{M}_+$, respectively.

Since the extrinsic curvatures $\ts{K}_-$ and $\ts{K}_+$ may differ, the resulting geometry does not have to be smooth at $\Sigma$. It turns out that the jump in the extrinsic curvature,
\begin{equation}\label{eq:Kjump}
[\ts{K}] \equiv \ts{K}_+ - \ts{K}_-\;,
\end{equation}
characterizes a distributional curvature localized on the hypersurface $\Sigma$. Plugging into the Einstein equations, W.~Israel showed that the hypersurface describes a thin massive shell with a distributional stress-energy tensor $\Terg=\ts{S}\delta_\Sigma$, where
\begin{equation}
   \kap \ts{S} =  - [\ts{K}] + \ts{h}\,\Tr([\ts{K}])\;.
   \label{eq:IJC}
\end{equation}

Note that the direction of the normal $\ts{n}$ determines the sign of the extrinsic curvatures and also the direction from $\dom{M}_-$ to $\dom{M}_+$. So not only would the change $\ts{n}\to-\ts{n}$ switch the signs of the extrinsic curvatures, but it would also switch the roles of $\ts{K}_+$ and $\ts{K}_-$ in \cref{eq:Kjump}. Therefore, the quantity $[\ts{K}]$ is invariant under a change of the orientation of the normal.

\subsection{Umbilic surfaces and static exocycles}
\label{apx:string_curvature}

For completeness, here we present several computations mentioned in the main text.

We start with the computation of the extrinsic curvature of the surface $\bph = \ph_\oi$ given in \eqref{eq:meancurvaopho}. Extrinsic curvature $\ts{K}$ of the surface with the normal $\ts{n}$ and inner metric $\ts{h}$ is given by \eqref{eq:metricsplit}. Let us assume coordinates of \eqref{eq:AdSAcctauchiph}. The inner normal is oriented along decreasing $\bph$,
\begin{equation}\label{eq:surfnorm}
    \ts{n} = -\ell\,\frac{\sin \bchi}\Omega\, \dif\bph\,,
\end{equation}
with $\Omega = {\cos\bchi\cos\chi_\oi{-}\sin\bchi\sin\chi_\oi\cos\bph}$. 
The relevant nonvanishing Christoffel symbols $\Gamma^{\bph}_{ij}$ are straightforward to compute due to the staticity and diagonality of the metric,
\begin{equation}
\begin{gathered}
  -\Gamma^{\bph}_{\btau\btau} = \Gamma^{\bph}_{\bchi\bchi} = \frac{1}{\Omega}\,\frac{\sin\chi_\oi}{\sin\bchi}\,\sin\bph 
     \,,\\
  \Gamma^{\bph}_{\bchi\bph} = \Gamma^{\bph}_{\bph\bchi} = \frac{1}{\Omega}\,\frac{\cos\chi_\oi}{\sin{\bchi}}
     \,,\\
  \Gamma^{\bph}_{\bph\bph} = -\frac{1}{\Omega}\sin\bchi\sin\chi_\oi\sin\bph
     \,.
\end{gathered}
\end{equation}
Evaluating the extrinsic curvature \eqref{eq:ex_curv} at $\bph=\ph_\oi$ yields
\begin{equation}
    \label{eq:ex_curv_comp}
    \ts{K} =
    \frac{\sin\chi_\oi\,\sin\ph_\oi}{\ell}\,\ts{h}.
\end{equation}
Recalling $\sin\ph_\oi = \sin\Delta\ph_\oi$, we identify the surface curvature $\scur_\oi$ as given in \ref{eq:meancurvaopho}.

The extrinsic curvature \eqref{eq:ex_curv_comp} shows that the surface ${\bph = \ph_\oi}$ is umbilic with subcritical mean curvature ${\scur_\oi<\frac{1}{\ell}}$. We will now show that the spatial projection of such a surface in its static frame is always an exocycle.

Let us introduce a triad $\{\ts{u},\ts{e},\ts{n}\}$ where $\ts{n}$ is the unit normal \eqref{eq:surfnorm}, ${\ts{u} \propto \ts{\partial}_\tau}$ is a timelike unit tangent vector parallel to the time flow of the static frame, and $\ts{e}$ is the spacelike unit vector tangent to the spatial projection of the surface. We can write
\begin{equation}
    \ts{g}_{\AdS} = -\ts{uu}+\ts{ee}+\ts{nn}
\end{equation}
and
\begin{equation}
    \label{eq:app_ex_curv}
    \ts{K} = \scur_\oi (-\ts{uu}+\ts{ee})
\end{equation}
on the surface.

For the curvature vector ${\ts{e}\cdot \ts{\nabla} \ts{e}}$ of the spatial projection, the staticity implies ${\ts{e}\cdot (\ts{\nabla}\ts{e}) \cdot \ts{u} = 0}$, and thanks to normalization ${\ts{e}\cdot (\ts{\nabla}\ts{e}) \cdot \ts{e} = 0}$. Therefore, it must hold
\begin{equation}
    \ts{e}\cdot \ts{\nabla}\ts{e} \propto \ts{n}.
\end{equation}
Let us evaluate the proportionality factor. Thanks to orthogonality ${\ts{e}\cdot\ts{n}=0}$ we have
\begin{equation}
    \ts{e}\cdot (\ts{\nabla}\ts{e}) \cdot \ts{n} = - \ts{e}\cdot (\ts{\nabla}\ts{n}) \cdot \ts{e} = - \ts{e}\cdot\ts{K}\cdot\ts{e} = - \scur_\oi\,.
\end{equation}
Therefore,
\begin{equation}
    \ts{e}\cdot \ts{\nabla}\ts{e} = -\scur_\oi\ts{n}\,,
\end{equation}
and the curvature of the spatial projection is given by $\scur_\oi$, the same as the string curvature in \eqref{eq:meancurvaopho}. It is constant and subcritical. The spatial projection of the surface in the static frame is thus an exocycle (the curve of constant sub-critical curvature) of the spatial hyperbolic geometry.

Next, we show that the exocycle $\bph=\ph_\oi$ has the axis passing through the center of gravity of its static frame. For that, we find the asymptotic angle $\ph_*$ of the exocycle in the unaccelerated coordinates. From \eqref{eq:S2rotation} it follows
\begin{equation}
    \tan{\ph_\oi} = \frac{\sin{\chi}\sin{\ph}}{\sin{\chi}\cos{\chi_\oi}\cos{\ph}-\cos{\chi}\sin{\chi_\oi}}.
\end{equation}
The value $\ph_*$ should be achieved at infinity, $\chi=\frac\pi2$. We find
\begin{equation}\label{eq:asymptotic_angle}
    \tan{\ph_*} = \tan{\ph_\oi}\cos{\chi_\oi}\,.
\end{equation}
It has two solutions corresponding to antipodal angles $\ph_*$ and $\ph_*-\pi$. The axis of the exocycle thus runs through the gravity center. The static worldsheet of this axis forms the axis referred to in \Cref{sec:oscillator}.

The exocycles with the axis $\ph=0,\pi$ and the corresponding static umbilic surfaces are naturally described in coordinates $\{\tau,\rho,\zeta\}$ of \eqref{eq:AdStaurhozeta}. The lines $\rho=\const$ are equidistant from the axis $\rho=0$. Since these coordinates are a special case of coordinates $\{\btau,\bchi,\bph\}$ (for parameter $\chi_\oi=\frac\pi2$, cf.~\eqref{eq:equidistcoors}), the lines of $\rho=\const$ are equivalent to $\bph-\const$ and thus they are exocycles. They enter infinity $\chi-\frac\pi2$ at the same improper points $\ph=0$ and $\pm\pi$. Setting $\chi_\oi=\frac\pi2$ and $\sin\Delta\ph_\oi = \sin\rho_\oi$, cf.~\eqref{eq:equidistcoors}, we obtain that the curvature of the exocycles $\rho=\rho_\oi$ is
\begin{equation}\label{eq:scurrhoo}
    \scur_\oi=\frac{\sin\rho_\oi}{\ell}\,.
\end{equation}

All umbilic surfaces in a maximally symmetric spacetime with the same mean curvature are isometric. Consider the Lie algebra of symmetries generated by static time shift, $\ts{\partial}_\tau$, rotation $\ts{\partial}_\ph$, and four boosts. Transformation along $\ts{\partial}_\tau$ leaves the globally static surface invariant. The rotation along $\ts{\partial}_\ph$ changes it. Of the boosts, two independent can be chosen so that they leave the surface invariant, namely those generated by $\bKV_{\ph_*}$ and $\hat\bKV_{\ph_*}$, where $\ph_*$ defines the axis of the umbilic surface (we used such a boost in \Cref{sec:oscillator}). The remaining boosts transform the surface into a non-static one.


    \bibliography{ZZ__Bibliography}
    \label{sec:bib}

\end{document}